\documentclass[11pt]{article}
\usepackage[letterpaper, left=1in, top=1in, right=1in, bottom=1in,nohead,includefoot]{geometry}
\usepackage{color,charter,graphicx,natbib,here,setspace,multirow,amsmath,amssymb,amsfonts,enumitem}
\usepackage[dvipsnames]{xcolor}
\usepackage{pdflscape}
\usepackage{adjustbox}
	\usepackage[pdftex]{hyperref}
	\hypersetup{colorlinks,
	citecolor=NavyBlue  
	}

\bibpunct{(}{)}{;}{a}{}{,}

\def\figfile{jpg}
\def\figdir{.} 

\def\a{\mbox{\boldmath$a$}}
\def\h{\mbox{\boldmath$h$}}
\def\H{\mbox{\boldmath$H$}}
\def\Y{\mbox{\boldmath$Y$}}
\def\A{\mbox{\boldmath$A$}}
\def\C{\mbox{\boldmath$C$}}

\def\D{\mbox{\boldmath$D$}}
\def\F{\mbox{\boldmath$F$}}
\def\f{\mbox{\boldmath$f$}}
\def\Q{\mbox{\boldmath$Q$}}
\def\V{\mbox{\boldmath$V$}}

\def\R{\mbox{\boldmath$R$}}
\def\I{\mbox{\boldmath$I$}}
\def\zero{\mbox{\boldmath$0$}}
\def\one{\mbox{\boldmath$1$}}
\def\x{\mbox{\boldmath$x$}}
\def\y{\mbox{\boldmath$y$}}
\def\z{\mbox{\boldmath$z$}}

\def\bnu{\mbox{\boldmath$\nu$}}

\def\bPhi{\mbox{\boldmath$\Phi$}}

\def\btheta{\mbox{\boldmath$\theta$}}

\def\bUpsilon{\mbox{\boldmath$\Upsilon$}}

\def\X{\mbox{\boldmath$X$}}
\def\W{\mbox{\boldmath$W$}}

\def\bomega{\mbox{\boldmath$\omega$}}

\def\e{\mbox{\boldmath$e$}}
\def\m{\mbox{\boldmath$m$}}

\def\h{\mbox{\boldmath$h$}}

\def\mA{\mathcal{A}}
\def\mD{\mathcal{D}}
\def\mH{\mathcal{H}}
\def\seq#1#2{#1{:}#2}
\def\eqno#1{eqn.~(\ref{eq:#1})}

\begin{document}

\title{Multivariate Bayesian Predictive Synthesis \\in Macroeconomic Forecasting\thanks{\footnotesize{The views expressed are those of the authors and do not necessarily reflect those of the Norges Bank or  of the
Bank for International Settlements.}}}  

\author{Kenichiro McAlinn$^{1,2}$\thanks{Corresponding author.
\newline \indent\, {\em E-mail}:
\href{mailto:kenichiro.mcalinn@chicagobooth.edu}{kenichiro.mcalinn@chicagobooth.edu},
\href{mailto:knut-are.aastveit@norges-bank.no}{knut-are.aastveit@norges-bank.no},
\href{mailto:jouchi.nakajima@bis.org}{jouchi.nakajima@bis.org}  \newline\indent\qquad\qquad and
\href{mailto:mike.west@duke.edu}{mike.west@duke.edu}
}
\,\, Knut Are Aastveit$^{3,4}$, Jouchi Nakajima$^5$, \& Mike West$^2$\\
\small $^1$Booth School of Business, University of Chicago\\
\small $^2$Department of Statistical Science, Duke University\\
\small $^3$Norges Bank\\
\small $^4$BI Norwegian Business School\\
\small $^5$Bank for International Settlements
}

\maketitle\thispagestyle{empty}\setcounter{page}0

\begin{abstract}
We develop the methodology and a detailed case study in use of a class of Bayesian predictive synthesis (BPS) models
for multivariate time series forecasting. This extends the recently introduced foundational  framework of
BPS to the multivariate setting, with detailed application in the topical and challenging
context of multi-step macroeconomic forecasting in a monetary policy setting.
BPS evaluates-- sequentially and adaptively over time-- varying forecast biases and facets of miscalibration of individual forecast densities, and-- critically-- of time-varying inter-dependencies among them over multiple series.
We develop new BPS methodology for a specific subclass of
the dynamic multivariate latent factor models implied by BPS theory.  Structured dynamic latent factor BPS is here
motivated by  the application context-- sequential forecasting of multiple US macroeconomic time series with forecasts
generated from several traditional econometric time series models.  The case study highlights the potential of BPS   to improve of forecasts of
multiple series at multiple forecast horizons,  and its use in learning   dynamic relationships among forecasting
models or agents.

\end{abstract}

\newpage

\section{Introduction \label{sec:intro} }

The societal importance of sequential decision making in many areas has promoted developments in time series modeling and forecasting to feed into decisions. In complex dynamic systems with multiple, inter-related time series,
the dependencies across series can critically impact on decisions, policies, and their outcomes.
In economic policy making, dependencies among macroeconomic time series provide fundamental insights into the state of economy. The interest is to use such relationships to  improve forecasts over multiple horizons, to
help to guide the policy decisions and understand their impact.
Central banks set national target interest rates based on (implicit or explicit) utility/loss considerations that
weigh  future outcomes of inflation and other measures of the real economy.
Understanding the (time-varying) dependencies of these measures-- especially the dynamics over multiple horizons-- is simply critical.
Driven by this, multivariate models, ranging from vector autoregressive models (VAR) to dynamic stochastic generalized equilibrium models (DSGE), have been developed and used by researchers and policy makers.
A huge literature reflects the critical nature of the field, from the early works of, for example,~\citet{sims1993nine}, \citet{stock1996evidence}, and~\citet{sims1998bayesian}, to more
recent advances in dynamic Bayesian models
in~\citet{Cogley2005}, \citet{Primiceri2005}, \citet{benati2008evolving}, \citet{Koop2009}, \citet{koop2010bayesian}, \citet{nakajima2011time}, \citet{Nakajima2010, Nakajima2011a}, and \citet{Zhou2012}, among others.

Concerned with accurate and useful forecasts, policy makers routinely rely on multiple sources, employing multiple models, forecasters, and economists, to produce forecasts.
To ensure appropriate normative decision making as well as reflecting increased uncertainty into the future, it has become popular, particularly for central banks, to provide probabilistic (density) forecasts.
For example, forecasts reported in the monetary policy reports of the Bank of England, Norges Bank, Swedish Riksbank, and recently also for the Federal Reserve Bank, have reflected this change.
To respond to this increased usage of density forecasts, there has been a recent resurgence in interest in forecast comparison, calibration, and combination of density forecasts in macroeconomics, econometrics, and statistics.
These new developments range from combining predictive densities using weighted linear combinations of prediction models, evaluated using various scoring  rules~\citep[e.g.][]{HallMitchell2007,Amisano2007,JoreMitchellVahey2010,Hooger2010,Kascha2010,Geweke2011,Geweke2012,Aastveit2014}, to more complex combination approaches that allows for time-varying weights with possibly both learning and model set
incompleteness~\citep[e.g.][]{Billio2013,Casarin2015,Aastveit2015,Pettenuzzo2015,Negro2016}.

The extensive literature on forecast combination has, for the most part, focused on forecasting a single series.  This is true from the seminal paper by~\cite{BatesGranger1969} to applications in business, economics, technology, meteorology, management science, military intelligence, seismic risk, and environmental risk, among other areas~\citep[e.g.][]{Clemen1989,  ClemenWinkler1999, Timmermann2004, ClemenWinkler2007}, as it is to the recent developments reported above.
In contrast, the literature on multivariate forecasting  is dominated by traditional statistical model comparison and variable selection~\citep[e.g.][]{chan2012time,korobilis2013var,Nakajima2010}.
Little attention  has yet been given to forecast comparison, calibration, and combination in the context of forecasting multiple series.
A few exceptions~\citep[e.g.][]{andersson2008bayesian,amendola2015model,AmisanoGeweke2017}  recognize this, but
restrict attention to direct extensions of univariate methods, with models   combined linearly using one metric for overall performance.
This is potentially    limiting in several ways: in ignoring inter-dependencies among series that
can be detrimental in   informing decisions,  in ignoring the reality that some models  might be good at forecasting one series but poor in another, and for the fact that some or all models
maybe be poor overall.
Partly reflecting the lack of formal statistical frameworks for holistic multivariate forecast model assessment and combination,
economic policy makers use ``ad hoc" strategies, which either rely on the policy maker's ``favorite" model, or ignore inter-dependencies all together.
The need for coherent methodology that gives policy makers flexibility in incorporating multivariate density forecasts from multiple sources cannot be understated.

The developments of the current paper   address the above issues, challenges, and needs.  The new methodology and case study presented
builds on theory and methods of dynamic {\it Bayesian Predictive Synthesis} (BPS) recently introduced in a univariate forecasting
setting~\citep{McAlinnWest2017}.
BPS is a coherent Bayesian  framework for evaluation, calibration, comparison, and context- and data-informed combination of multiple forecast densities. The approach applies whether forecast densities arise from sets of models, forecasters, agencies, or institutions.
As detailed in~\cite{McAlinnWest2017} the framework includes, as special cases, a range of
existing univariate density forecasts combination methods.  Our multivariate extensions here naturally allow modeling and estimation of varying forecast biases and facets of miscalibration of individual forecast densities,  time-varying inter-dependencies among models or forecasters over multiple series, and addresses the above noted problems in multivariate settings.

Section~\ref{sec:dynamicBPS} summarizes the BPS framework and implications in terms of the broad class of implied theoretical  models
for dynamic multivariate problems.   Methodological details are developed for one specific subclass of models-- flexible
dynamic latent factor models with seemingly-unrelated regression structure (DFSUR models). In this setting,
each individual model generating multivariate forecast densities  is linked to a set of multivariate dynamic latent factor processes--
the relationships across each set of latent factors are then a key focus in understanding and leading to forecast combination that
addresses interdependencies.
In Section~\ref{sec:Inf}, analyses of U.S macroeconomic time series  illustrates and highlights the benefits of the new framework
using the class of DFSUR models in the BPS context.
Additional comments in Section~\ref{sec:summary} conclude the main paper. An appendix of detailed supplementary
material summarizes Bayesian computational methods (MCMC) for fitting and using DFSUR models in the BPS context,
and contains more extensive graphical and tabular summaries of results of the multiple BPS forecasting analyses from the case study.

\paragraph{Some notation:} We use lower case bold font for vectors and upper case bold font for matrices. Vectors are columns by default. Distributional notation
$y\sim N(f,v),$ $\x\sim N(\a,\A)$ and $k\sim G(a,b)$ are   for the
univariate normal, multivariate normal and gamma distributions, respectively.
We use, for example, $N(y|f,v)$ to denote the actual density function of $y$ when $y\sim N(f,v).$
Index sets $s{:}t$ stand for $s,s+1,\ldots,t$ when $s<t$, such as in   $y_{\seq1t} = \{ y_1,\ldots,y_t\}.$

\section{Dynamic Multivariate BPS\label{sec:dynamicBPS}}

The new  developments forming the methodological core of this paper adapt and extend the basic BPS framework of \cite{McAlinnWest2017} to multivariate density forecast synthesis with practical decision goals in mind.
 \cite{McAlinnWest2017} defined formal,  coherent methodology for integrating density forecasts from multiple, potentially competing
 statistical model-- or forecasters, or institutions-- in a univariate time series setting.   The dynamic BPS approach there
 has a foundation in coherent Bayesian reasoning with predictive and decision analytic goals, based on historical developments in
assessing and combining subjective probabilities~\citep{DVL1979,West1984,GenestSchervish1985,West1988,West1992c,West1992d,APDetal1995,French2011}.
Drawing on key theoretical results from that Bayesian \lq\lq agent/expert opinion analysis" literature,~\cite{McAlinnWest2017}
define a class of time-varying parameter, latent factor models in which each of the univariate latent factors relates to one of the
set of models or forecasters generating predictions.  The models are developed methodologically and shown to have promise in
understanding relationships among forecasting models, their biases and inter-dependencies over time, and can improve
short and medium term forecasting for univariate time series.

We now develop the new, multivariate extension of dynamic BPS, beginning with a brief summary of the key background theory
free from the time series context.

\subsection{BPS Background\label{sec:BPSbackground}}

Consider forecasting a $q{\times}1-$ dimensional vector of outcomes $\y$.  Outcomes are typically real-valued, as is the case in
our applications below, though the foundational theory is general.
A Bayesian decision maker  $\mD$ is to receive forecast distributions for $\y$ from each of $J$ {\em agents}; in our
application, the agents are different Bayesian time series models, while in other contexts they may include professional forecasters, or forecasting agencies, etc.,  labelled $\mA_j$, $(j=\seq1J).$  Then $\mD$ aims to  incorporate the information
provided by the set of agent forecast distributions in her thinking and forecasting $\y$ and any resulting decisions.
Agent $\mA_j$ provides a probability density function $h_j(\y).$
These forecast densities represent the individual inferences from the  agents, and define the information set $\mH = \{ h_1(\cdot), \ldots, h_J(\cdot) \} $ now available to $\mD.$
Formal subjective Bayesian analysis dictates that, $\mD$ will then use the information set $\mH$ to predict $\y$ using the implied posterior $p(\y|\mH)$ from a full Bayesian prior-to-posterior analysis.

To obtain a full Bayesian prior-to-posterior analysis, \cite{West1992d} extended prior theory~\citep{GenestSchervish1985,West1992c}
to show that there is a subset of all Bayesian models
in which $\mD$'s posterior has the mathematical form
\begin{equation}\label{eq:theorem1}
	p(\y|\mH)=\int_{\X} \alpha(\y|\X)\prod_{j=\seq1J}h_j(\x_j)d\x_j
\end{equation}
where each $\x_j$ is a latent $q{\times}1-$dimensional vector,
$\X = [\x_1,\ldots,\x_J]'$   collects these latent vectors in a $J{\times}q-$dimensional matrix,
  and $\alpha(\y|\X)$ is a conditional p.d.f. for $\y$ given $\X.$
The interpretation is as follows. First, in the subjective view of $\mD$ there {\em must}
exist latent factors $\x_j$ potentially related to $\y$ and  such that
agent $\mA_j$'s forecast density is that of $\x_j$. Second, {\em conditional on learning $\mH$},  the $\mD$ regards the
latent factors as conditionally independent with $\x_j\sim h_j(\x_j). $ Note that this does not imply that $\mD$ regards the
forecasts as independent, since under her prior the $\h_j(\cdot)$ are uncertain and likely highly inter-dependent, and the
key element $\alpha(\y|\X)$ is how $\mD$ expresses her views of dependencies. Third, this key element $\alpha(\y|\X)$
is $\mD$'s regression model relating the $\x_j$ as a collective to the outcome $\y.$  Refer to $\alpha(\y|\X)$ as the BPS
{\em synthesis function}, and to the $\x_j$ as {\em latent agent states}.

The key \eqno{theorem1}  does define the functional form of $\alpha(\y|\X)$.
\cite{McAlinnWest2017} show that, for scalar outcomes $\y=y,$  many forecast and model combination methods \citep[e.g.][among others]{Geweke2011,Fawcett2014,Pettenuzzo2015,Aastveit2015} can be considered as special cases of \eqno{theorem1},  realized via different
choices of the form of the BPS synthesis function $\alpha(\cdot|\cdot)$.
For vector outcomes $\y$, \eqno{theorem1} similarly allows flexibility for $\mD$  to specify $\alpha(\y|\X)$ to reflect decision goals and incorporate views and historical information about, for example,  agent-specific biases, patterns of miscalibration, inter-dependencies among agents, and their relative expertise and expected forecast accuracy.   Any specific BPS model will be created by assuming a specific
model form for the synthesis p.d.f.

\subsection{Dynamic Sequential Setting}

For a $q-$vector time series $\y_t, t=1,2, \ldots,$   decision maker $\mD$  receives forecast densities from each
agent sequentially over time.
At   time $t-1,$ $\mD$ receives current forecast densities
$\mH_t = \{ h_{t1}(\y_t),\ldots, h_{tJ}(\y_t) \}$ from the set of agents and aims to forecast $\y_t$. The
full information set used by $\mD$ at time $t$ is thus  $\{ \y_{1:t-1}, \ \mH_{1:t} \}.$
As $\mD$ observes more information, her views of the agent biases and calibration characteristics, as well as of
inter-dependencies among agents are repeatedly updated.  A formal, parametrized Bayesian dynamic model is the vehicle for
structuring this sequential learning in a general state-space context.  This defines the dynamic BPS framework.

The time series extension of  \eqno{theorem1} implies that $\mD$ has a time $t-1$ distribution for $\y_t$ as
\begin{equation}\label{eq:theorem}
	p(\y_t|\bPhi_t,\y_{\seq1{t-1}},\mH_{\seq1t})
	\equiv p(\y_t|\bPhi_t,\mH_t)=\int \alpha_t(\y_t|\X_t,\bPhi_t)
		\prod_{j=\seq1J}h_{tj}(\x_{tj})d\x_{tj}
\end{equation}
where $\X_t= [\x_{t1},\ldots,\x_{tJ}]'$ is a $J{\times}q-$dimensional matrix of latent agent states at time $t$,
the conditional p.d.f. $\alpha_t(\y_t|\X_t,\bPhi_t)$ is $\mD$'s   synthesis p.d.f. for $\y_t$ given $\X_t,$
and involves time-varying parameters $\bPhi_t$
for which $\mD$ has current beliefs represented in terms of her
(time $t-1$) posterior $p(\bPhi_t|\y_{\seq1{t-1}},\mH_{\seq1{t-1}}).$

This general framework defines
the $\x_{tj}$ as realizations of inherent dynamic latent factors-- the {\em latent agent states} at time $t$--
and synthesis is achieved by relating these latent factor processes to the time series $\y_t$ via
models of the time-varying synthesis function $\alpha_t(\y_t|\X_t,\bPhi_t).$  The foundational theory does
not specify this p.d.f.,  and methodology is based on specific chosen forms.
For the multivariate extension of~\cite{McAlinnWest2017}, we look to a specific class of models that
extends the traditional seemingly unrelated regression model~\citep[SUR; ][]{Zellner1962} to a dynamic Bayesian framework, as a first approach to defining a  computationally accessible yet flexible framework for dynamic multivariate BPS.

\subsection{Multivariate Latent Factor Dynamic  Models \label{sec:dynamic}}

Consider a dynamic multivariate BPS synthesis function
\begin{equation}\label{eq:BPSnormalalphadynamic}
\alpha_t(\y_t|\X_t,\bPhi_t) = N(\y_t|\F_t\btheta_t,\V_t)
\end{equation}
with
\begin{equation}\label{eq:BPSalphadynamicFtheta}
\F_t=\left(
    \begin{array}{cccccccc}
      1 &  \f_{t1}'  & 0 & \zero' & \cdots & \cdots  & 0 & \zero'  \\
      0 & \zero' &  1& \f_{t2}' &  &  & &  \vdots \\
      \vdots &  &  &  & \ddots &  &  & \vdots  \\
      0 &  \zero' &  \cdots & \cdots & \cdots  & \cdots & 1& \f_{tq}'
    \end{array}
  \right) \quad\textrm{and}\quad
\btheta_t= \left(
    \begin{array}{c} \btheta_{t1}\\ \btheta_{t2}\\ \vdots \\ \btheta_{tq}\end{array} \right)
\end{equation}
where for each series $r=\seq1q,$  the $J{\times}1-$vector $\f_{tr}=(x_{tr1},x_{tr2},...,x_{trJ})'$ is a  realization of the set of $J$ latent
agents states for series $r$, and the $(J+1){\times}1-$vector
 $\btheta_{tr}=(1,\theta_{tr1},\theta_{tr2},...,,\theta_{trJ})'$ contains an intercept and coefficients representing time-varying bias/calibration
weights of the $J$ latent agent states for  series $r$.   Note that $\F_t$ has $q$ rows and $(J+1)q$ columns, and
$\btheta_t$ is a column $(J+1)q-$vector. Observation noise is reflected in the-- likely volatile-- residual
$q{\times}q$ variance matrix $\V_t$, and the general time-varying parameter of~\eqno{theorem} is set as
$\bPhi_t = \{ \btheta_t, \V_t\}$.

This defines the first component of a conditionally linear, conditionally normal dynamic multivariate model-- a subclass of
multivariate dynamic linear models but with latent factors as predictors.   Modeling time evolution of the
parameter processes $\bPhi_t = (\btheta_t,\V_t)$  is needed to complete model specification. We do this using the first step
into dynamic models, with traditional random walk models to allow for-- but not anticipate direction in--
stochastic changes over time in both regressions $\btheta_t$ and matrix volatilities $\V_t$, as is traditional in
Bayesian time series literatures; see, for example,~\citet{WestHarrison1997book2} (chap. 16) and \citet{Prado2010} (chap. 10).
Thus we take
\begin{subequations}
\label{DLM}
\begin{align}
	\y_t&=\F_t\btheta_t+\bnu_t, \quad \bnu_t\sim N(\zero,\V_t), \label{eq:DLMa} \\
	\btheta_t&=\btheta_{t-1}+\bomega_t, \quad \bomega_t\sim N(\zero, \W_t)\label{eq:DLMb}
\end{align}
\end{subequations}
where  $\btheta_t$ evolves  in time according to a linear/normal random walk with
innovations variance matrix $\W_t$ at time $t$,  and $\V_t$ is the residual variance in predicting
$\y_t$ based on past information and the set of agent forecast distributions.

Model specification is completed using standard discount methods.
As with the univariate DLM, the time-varying intercept and agent coefficients $\btheta_t$
follow the random walk evolution of \eqno{DLMb} where $\W_t$ is defined via a standard,
single discount factor specification (\citealt[][Chap 10]{Prado2010}).
The residual variance matrix $\V_t$ follows a standard inverse Wishart random walk volatility model,
also based on discounting with a second discount factor.

We now have a class of dynamic, multivariate latent factor models in which latent factors are realized as   draws from the set of agent densities $h_{tj}(\cdot)$, becoming available to $\mD$ at $t-1$ for forecasting $\y_t$.
Thus, coupled with eqns.~(\ref{eq:DLMa},\ref{eq:DLMb}), we have the  time $t$ {\em prior}  for the latent states-- conditional on
$\mH_{\seq1t},$ as
\begin{equation}\label{eq:dfmh}
 p(\X_t| \bPhi_t,\Y_{\seq1{t-1}},\mH_{\seq1t}) \equiv p(\X_t|\mH_t) = \prod_{j=\seq1J} h_{tj}(\x_{tj})
\end{equation}
with $\X_t,\X_s$ conditionally independent for all $t\ne s.$
Again it is important to stress that the {\em conditional} independence of the $\x_{tj} $ given the $h_{tj}(\cdot)$
must not be confused with the $\mD$'s modeling and estimation of the dependencies among agents.
This dependence is central and integral, and is reflected through the effective dynamic parameters $\bPhi_t = (\btheta_t,\V_t)$.

\subsection{Bayesian Analysis and Computation \label{sec:comp}}

At any current time $t,$    $\mD$ has historical information $\{ \y_{\seq1t}, \mH_{\seq1t}\}$ and the history of the BPS analysis up until that point. This will have defined inferences on past  latent agent states $\X_\ast$ and the dynamic BPS model parameters $\bPhi_\ast = (\btheta_\ast,\V_\ast)$.
The former, importantly, provides insight into the dependencies, biases, and other characteristics pertaining to  $\y_{\seq1t}$, among agents and individual agents.
Posterior summaries for  $\X_t$ over time inform on this-- a key feature of BPS.
This inference is topical, as issues of herding (overlap and redundancies) among groups of agents (either models or individuals) is of practical importance, and understanding how these characteristics change over time and across series is key.

Posterior analysis is enabled by Markov chain Monte Carlo (MCMC) methods, followed by forecasting from time $t$ onward utilizing   theoretical and simulation-based extrapolation of the model.
$\mD$ is interested in the inference on the full set of past latent agent states   and dynamic parameters   $\{ \X_{\seq1t}, \bPhi_{\seq1t}\}$, as well as   forward filtering to update posteriors for current values
  $\{ \X_t, \bPhi_t\}$.
Posterior MCMC-based computation uses nowadays traditional methods, and extends the MCMC method used in \cite{McAlinnWest2017} for the univariate case with several   modifications.

\paragraph{Posterior Computations via MCMC.}     At a given current time $t,$ the multivariate dynamic latent factor model structure of eqns.~(\ref{eq:DLMa},\ref{eq:DLMb},\ref{eq:dfmh}) leads easily to a three-component block Gibbs sampler for the latent agent states, dynamic coefficient parameters, and dynamic volatility parameters.
The components are iteratively resampled from the three conditional posteriors noted below, initialized given agent states drawn independently from priors $h_\ast(\ast).$

First, conditional on the agent states and residual volatility, the MCMC step draws new dynamic coefficient parameters from $  p( \btheta_{\seq1t} |  \X_{\seq1t}, \V_{\seq1t}, \y_{\seq1t}).$
This is the full (normal) posterior for the sequence of states in the implied
conditional multivariate DLM, and is efficiently sampled using an extension of the
traditional forward filtering, backward sampling (FFBS) algorithm (e.g.~\citealt[][chap 10]{Prado2010}).

Second,  the MCMC step draws new dynamic volatility  matrices $\V_t$ from the full joint conditional posterior
$ p( \V_{\seq1t}|  \X_{\seq1t},\btheta_{\seq1t}, \y_{\seq1t} )$-- conditional on the agent states and dynamic coefficient parameters.
This employs the standard FFBS algorithm for inverse Wishart discount volatility models~\citep[][chap. 10]{Prado2010}

Third, conditional on values of dynamic parameters  $\bPhi_{\seq1t} =(\btheta_{\seq1t}, \V_{\seq1t}),$ the  MCMC  draws new agent states from
$ p( \X_{\seq1t} |  \bPhi_{\seq1t}, \y_{\seq1t}, \mH_{\seq1t} ).$
As with the univariate case,  the $\X_t$ are
conditionally independent over time $t$ in this conditional distribution, with time $t$
conditionals
$ p( \X_t|  \bPhi_t, \y_t, \mH_t) \propto N(\y_t|\F_t\btheta_t, \V_t) \prod_{j=\seq1J} h_{tj}(\x_{tj}).$
In cases when all of the agents' forecasts are multivariate normal, the posterior is a multivariate normal that is trivially sampled using the properties of conditional normal.
For a more central and practically important case of forecasts being multivariate T distribution, each $h_{tj}(\cdot)$ can be represented as a scale mixture of normals, and augmenting the posterior MCMC to include the implicit underlying latent scale factors generates conditional multivariate normals for each $\X_t$ coupled with conditional inverse gammas for those scales.
In other cases, augmenting the MCMC utilizing Metropolis-Hastings simulator or an augmentation can be used.
More   discussion of these algorithmic details is given in Appendix~\ref{supp:comp}.

\paragraph{Forecasting 1-Step Ahead.}
At time $t$ we forecast 1-step ahead by generating \lq\lq synthetic futures" from the BPS
model, as follows.   First, for each sampled $\bPhi_t$ from the posterior MCMC above,
draw $\V_{t+1}$ from its discount volatility evolution
model, and then  $\btheta_{t+1}$ conditional on $\btheta_t,\V_{t+1}$
from the evolution model~\eqno{DLMb}-- this  gives a draw
$\bPhi_{t+1} = \{ \btheta_{t+1}, \V_{t+1} \}$
from $p(\bPhi_{t+1} |\y_{\seq1t}, \mH_{\seq1t} )$.
Second, draw $\X_{t+1}$ via independent sampling of the $h_{t+1,j}(\x_{t+1,j}),$  $(j=\seq1J).$
Third, bring these samples together and
draw a synthetic 1-step outcome $\y_{t+1}$ from the conditional normal of~\eqno{DLMa} given these sampled
parameters and agent states.
Repeating this generates a random Monte Carlo sample from the 1-step ahead
forecast distribution for time $t+1$.

\subsection{Multi-Step Ahead Forecasting \label{sec:k-step}}
In many applications involving multivariate analysis, long term forecasting and analysis is often of equal or greater importance than the basic 1-step ahead horizon.
For example, in terms of economic policy and macroeconomic time series-- based on traditional monthly or quarterly data-- the most important horizons of interest are 1-3 years ahead.
This is especially true when dealing with monthly data, as knowing a month ahead has very little utility compared to understanding the long term dynamics and structure over multiple years.
Thus, economic policy makers advise policy decisions based on inputs from their own forecast models, judgemental inputs, views of
other economists, and forecast over the next year or 2-3 years.
However, forecast difficulty increases as the horizon increases, especially when models are calibrated on the short-term basis.
Traditional statistical evaluation of  time series models is inherently based on 1-step ahead forecasts, raising interest in
developing BPS to address longer-term forecasting goals.

BPS provides two methods for multi-step ahead forecasting, as laid out in \cite{McAlinnWest2017}.
The first method is direct sequential projection of $\bPhi_{t+1}, \bPhi_{t+2}, \ldots, \bPhi_{t+k}$, updating the parameters over time until it reaches $t+k$ ($k$ being the horizon of interest) and plugging in the relevant forecasts, $\X_{t+k}$, sampled from $h_{t,\seq1J}(\x_{t+k})$.
The second is denoted by BPS$(k)$, referring to applying the BPS model to  synthesise  $k$-step ahead forecasts directly, as expanded upon below.
BPS$(k)$ often outperforms direct projection in terms of forecast accuracy.
This is sensible, as we can expect some agents to perform differently (relative to other agents) for different forecast horizons.
In the context of macroeconomic forecasts, we might observe that economists, who rely on qualitative information and policy experience, to outperform purely quantitative models on a long term basis and thus calibrating the forecasts on their short term predictive ability-- for which quantitative models often are superior-- can be problematic.

\paragraph{BPS$(k)$ for customized multi-step forecasting.}
BPS provides $\mD$  a flexible strategy to focus on the horizon $k$ of interest as it is customizable to the forecasting goals.
This involves a trivial modification of methodology in Section~\ref{sec:dynamicBPS} in which the model at time $t-1$ for predicting $\y_t$ changes as follows.
For a {\em specific forecast horizon} $k>1$,   replace $h_{t-k,j}(\x_{tj})$ with $h_{tj}(\x_{tj})$ so BPS is calibrated using the forecasts from $t-k$.
Doing this results in dynamic model parameters $\{ \btheta_t, \V_t\}$ to be explicitly geared to the $k$-step horizon, calibrating and ``tuning" to the horizon $k$ of interest.
Forecasting, then, simply follows the model extrapolation via simulation as in Section~\ref{sec:comp}.

\section{Case Study in US Macroeconomic Forecasting \label{sec:Inf}}

\subsection{Data, Forecasting Models and Implementation \label{sec:data}}

\paragraph{Time Series Data}

\begin{figure}[htbp!]
\centering
\includegraphics[width=0.9\textwidth]{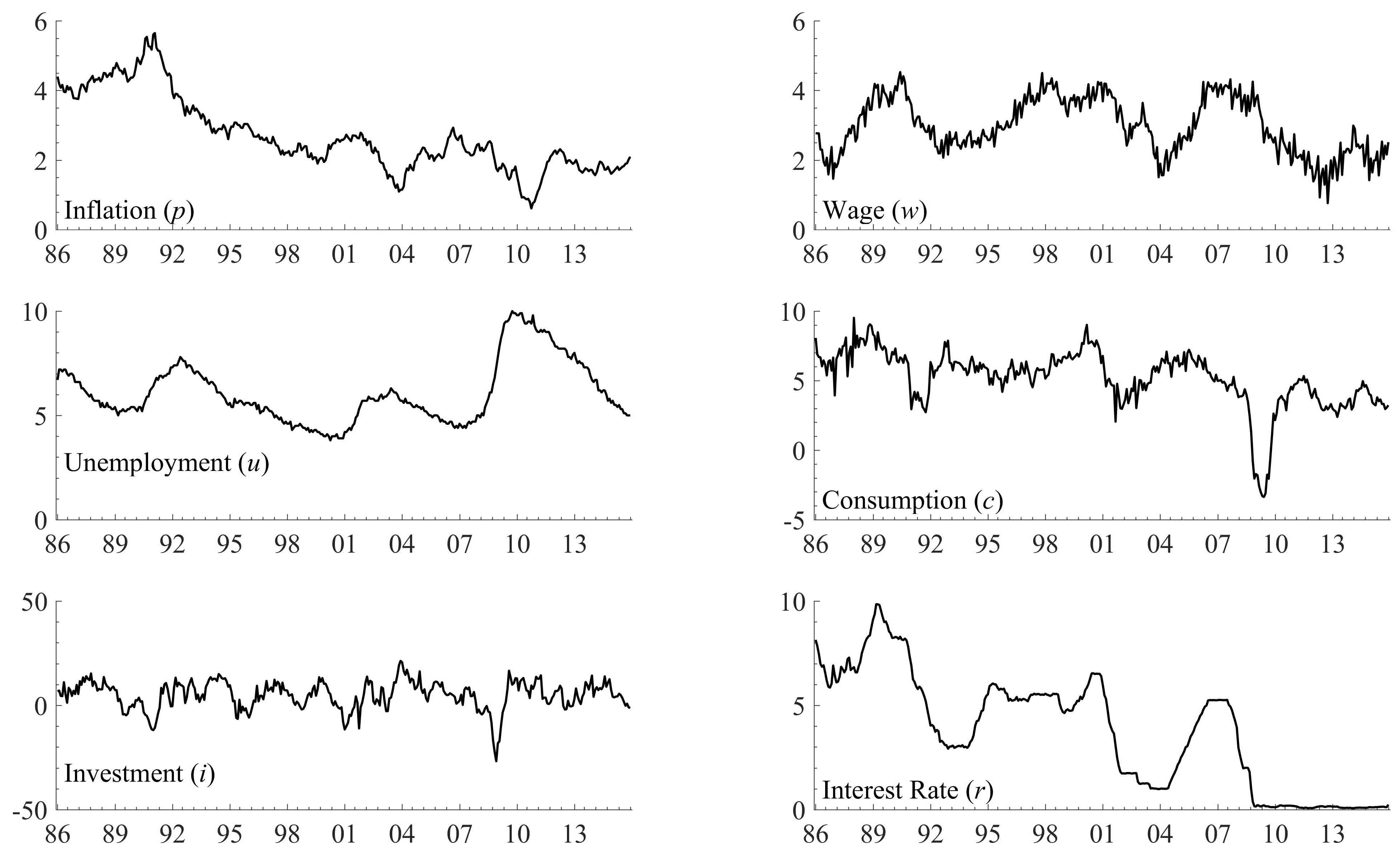}
\caption{US macroeconomic data 1986/1-2015/12: US macroeconomic time series (indices $\times$100 for $\%$ basis):
annual inflation rate $(p)$, wage $(w)$, unemployment rate $(u)$, consumption $(c)$, investment $(i)$, and short-term nominal interest rate $(r)$.
\label{data}}
\end{figure}

We analyze monthly US macroeconomic data, focusing on forecasting six macroeconomic time series with 1-, 12-, and 24-month ahead interests.
The study involves the following monthly macro series:  annual inflation rate $(p)$, wage $(w)$, unemployment rate $(u)$, consumption $(c)$, investment $(i)$, and short-term nominal interest rate $(r)$ in the US economy from
1986/1 to 2015/12, a context of topical interest~\citep{Cogley2005,Primiceri2005,Koop2009,Nakajima2010}.
The inflation rate is the consumer price index for all urban consumers: all items less food and energy, not seasonally adjusted, wage is the average hourly earnings of production and nonsupervisory employees: total private, not seasonally adjusted, the unemployment rate is the civilian unemployment rate, seasonally adjusted, consumption is the personal consumption expenditures, seasonally adjusted annual rate, investment is the ISM manufacturing: new orders index, and the interest rate is the effective federal funds rate, not seasonally adjusted; the first four being annual changes, investment being monthly changes, and monthly interest rates.
Fig.~\ref{data} shows the data for the six series over the time span considered.
We focus on forecasting
the six series, with an emphasis on inflation, using  past values of the six series as candidate predictors underlying a set of five time series models-- the $J=5$ agents--  to be evaluated, calibrated, and synthesized.

During the period of analysis, the sub-prime mortgage crisis and great recession of the late 2000s warrant special attention.
This period involved a series of significant, unique shocks to the US economy, so any analysis is challenged in terms of
predictive ability in short and longer terms.
For any combination strategy to be effective and useful, its predictive performance must be robust under these conditions.
Additionally, due to the structural changes in the overall economy~\citep[e.g.][]{Aastveit2016}, there is also interest in understanding
changes in  the inter-dependencies among series over the crisis periods. On this goal,  multivariate analysis offers opportunity for
improved understanding that simple univariate analyses just cannot.

In our BPS$(k)$ analyses, for $k=12$ we take investment as the cumulative value of the previous year of monthly differences, since investment measures monthly difference and forecasting the change from 11th to 12th month is of little relevance to the policy maker.
Similarly,  for $k=24$, investment is the cumulative value of the 24 months  of monthly differences.
Additionally, inflation, wage, unemployment, and consumption are defined as changes from the current time.
Agents will produce forecasts according to the target value by either summing the forecasts to the target horizon, or summing certain periods within the horizon.
In this way, the target forecast is directly in line with what policy makers are interested in and focus on for decision making.

\paragraph{Agent Models and BPS Specification.}
For the $J=5$ agents we use time varying parameter vector autoregressive (TVP-VAR) models that cover multiple dynamic structures utilized in the literature~\citep{Cogley2005,Primiceri2005,Koop2009,Nakajima2010} and in practice.
Labelled as  M*, the agent models are:
M1-  VAR(1);
M2- VAR(12);
M3- VAR(3);
M4- VAR(1:3:9);
M5- VAR(1:6:12).
The numbers in parentheses are the lags and the number between colons represent intervals (e.g. 1:3:9 uses lags of 1, 3, 6, and 9).
Each M* is a standard TVP-VAR model  \citep[or exchangeable time series;][Chap 10]{Prado2010} with the residual volatility following a matrix-beta/Wishart random walk so that model fitting and generation of forecasts is routine.
Though more recent variants of these core models-- such as, for example, Bayesian latent threshold TVP-VARs with stochastic volatility as
in~\citet{Nakajima2010}-- might be considered, one benefit and appeal of forecast synthesis is making improvements over a set of relatively simple models. That is, we explore BPS applied to rather standard, and currently accepted variants of models that are traditional and whose basic model forms (up to assumptions about lags and variables to include) are accepted in the applied macroeconomic forecasting community.
%

In the dynamic BPS models for forecast horizons $k=1,12,24,$ we take initial priors using independence across series $r$ and
with $\btheta_{tr}\sim N(\a_0, \R_0)$ with $\a_0=(0,\one'/J)'$ and where $\R_0$ is diagonal with diagonal elements of 1 apart from: (a) elements of 0.001
           for the intercept coefficients; and (b) elements of 0.1 for coefficients on investment.
           Coupled with this, we take $\V_0\sim IW(7, 7*0.01\I)$.  Discount factors in the BPS$(1)$ model are set at
$(\beta,\delta)=(0.99,0.99)$, and variants for
BPS$(k)$ for $k=12,24$-month ahead forecasting are as discussed in Section~\ref{sec:k-step}, though now we increase the prior variance on the intercept from $0.001$ to $0.01$ and $0.1$, for $k=12,24$, respectively, to reflect the increased uncertainty about the relevance of the agent forecasts at longer horizons.

We have explored analyses across ranges of choices of initial priors and discount factors, and chosen these values as they lead to good agent-specific and BPS forecasting accuracy; conclusions about the main questions-- how BPS can improve forecasts while
generating insights into agent characteristics and dependencies over time-- to not change materially
 with different values close to those chosen for the
summary examples.

\paragraph{Data Analysis and Forecasting.}
The 5 agent models are analyzed and synthesized as follows.
First, the agent models are analyzed in parallel over 1986/1-1993/6 as a training period to calibrate the VARs.
This continues over 1993/7-2001/12 while at each month $t$ during this period,
the MCMC-based BPS analysis is run in parallel using data from 1993/7 up to time $t$ in an ``expanding window" fashion, adding data as we move forward in time.
We do this for the traditional 1-step focused BPS model, and-- separately and in parallel-- for the $k=12,24$-step
ahead focused BPS$(k)$ model as discussed in  Section~\ref{sec:k-step}.
This continues over the third period to the end of the series,  2001/1-2015/12, generating forecasts (for the agents and BPS) for each $t$ until the end of the testing period.
This testing period spans over a decade and a half and includes 180 data points, providing a good measure on how the agents and BPS perform under different economic situations; most notably before, during, and after the sub-prime mortgage crisis.
Out-of-sample forecasting is thus conducted and evaluated in a way that mirrors the realities facing
decision and policy makers.

\paragraph{Forecast Accuracy and Comparisons.}

Following the recent literature on macroeconomic forecasting, we compare both point and density forecasts to give a broader understanding of the predictive abilities of the agents and BPS. For the point forecasts, we compute and compare mean squared forecast error (MSFE) over the forecast horizons of interest and for each series.
For density forecasts with BPS, we evaluate log predictive density ratios (LPDR); at horizon $k$ and across time indices $t$ for the joint set of series, this is
\begin{align*}
	\mathrm{LPDR}_{\seq1t}(k)=\sum_{s=\seq1t}\mathrm{log}\{p_{j}(\y_{s+k}|\y_{1{:}s})/p_{\mathrm{BPS}}(\y_{s+k}|\y_{1{:}s})\}
\end{align*}
where $p_j(\y_{s+k}|\y_{1{:}s})$ is the predictive density under each agent indexed by $j$, at each time $s$ over the next $k$ time points.  The LPDR measures are, at each $t$,  baselined against the corresponding BPS forecasts over each  horizon $k$.
LPDR provides a direct statistical assessment of the distributional accuracy of a forecast relative to, in this case, BPS for multiple horizons, extending the 1-step focused Bayes' factors.
They compare the location and dispersion of the forecasts, giving an assessment of risk, elaborating on MSFE measure, and have been increasingly used in broader model comparison and forecast accuracy studies~\citep[e.g.][]{Nakajima2010,Aastveit2015}.

\subsection{Dynamic BPS and Forecasting \label{sec:outsample}}

\paragraph{1-step ahead forecasting.} Table~\ref{table:1step} summarizes the predictive measures compared for the 1-step ahead forecasts.
Looking at point forecasts, BPS exhibits clear improvements over agent forecasts 5 out of the 6 series; the 5 are inflation, wage, consumption, investment, and interest rate.
Even for the series for which BPS does not show substantial improvement over the models, the difference between the best model is within 1\%.
On the series BPS on which makes an improvement, the gains are at least 2\%, except compared to one model for interest rate.
It is also notable that the best model differs for each series.
VAR$(1{:}3{:}9)$ is best for inflation while it is the worst for wage, for example.
Under traditional model combination strategies, such as Bayesian model averaging (BMA)  in which each model is assessed only on 1-step ahead density forecast
accuracy and for the full multivariate forecast, accuracy is inherently aggregated over series.
BPS, due to its flexible synthesis function, is able to synthesize forecasts on each series, while retaining the inter-series dependencies.
This leads to BPS improving on multiple series without trading-off one over another. In fact, the results in Table~\ref{table:1step} shows that BPS exhibits improvements for all 6 series relative to BMA.

BPS demonstrates an ability to substantially improves characterization of forecast uncertainties as well as adaptation in forecast locations, reflected in the LPDR measures. Note that the best model, in terms of LPDR, is only best for wage in terms of MSFE and performs average for the other series. This indicates how LPDR measures for multiple series favor overall performance over models that are good for some but bad for others. Model combination schemes that are dependent on likelihood measures-- such as BMA-- heavily favor the average performing model (VAR$(3)$ in the case of 1-step ahead forecasts). In contrast, BPS dynamically synthesizes forecasts for each series, while improving uncertainty assessment (per series and dependence between series), to improve in terms of overall distribution forecasts as well as point forecasts. This feature of BPS is critical, as $\mD$ typically has priorities among the series being forecasted.

\begin{figure}[htbp!]
\centering
\includegraphics[width=0.75\textwidth]{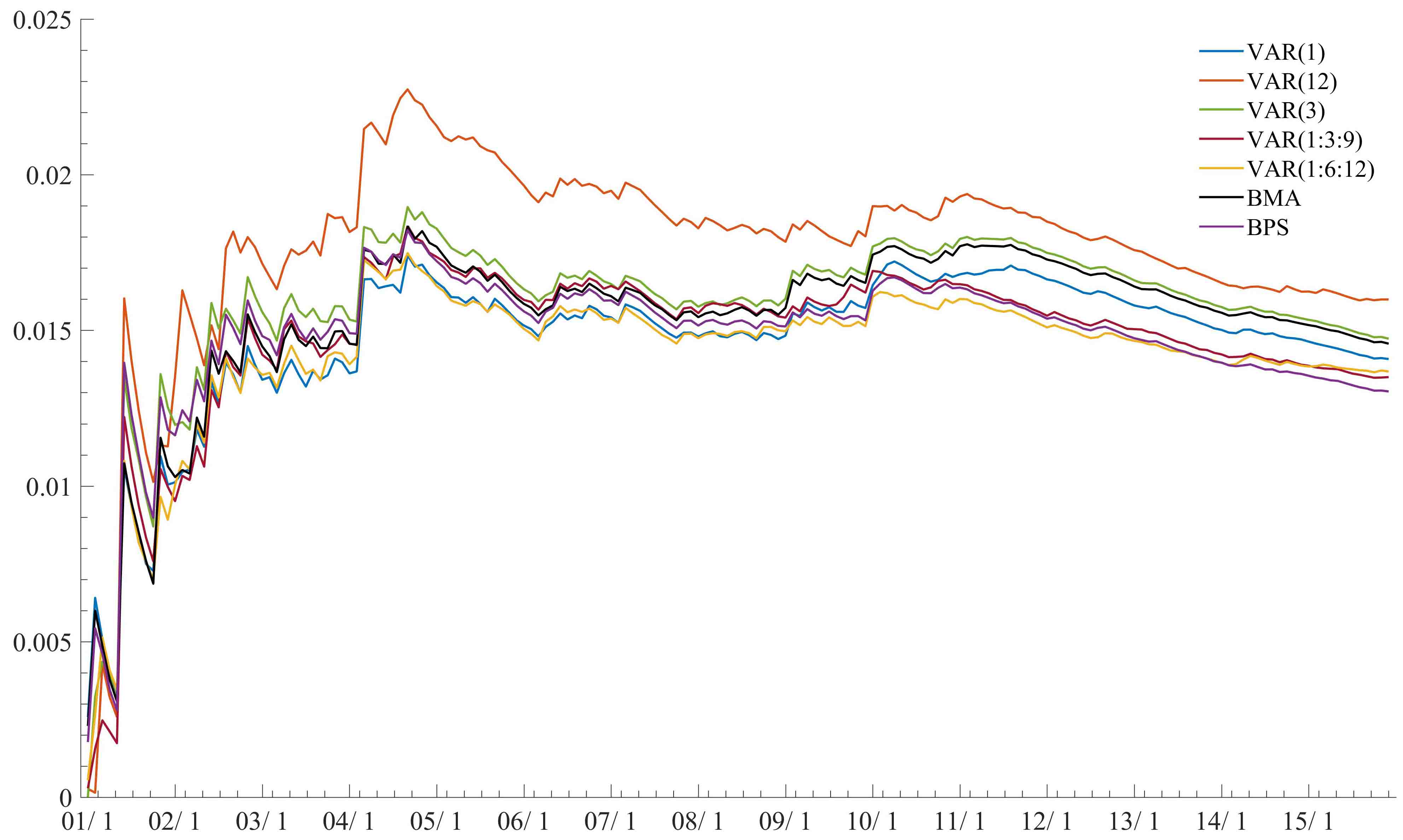}
\caption{US macroeconomic forecasting 2001/1-2015/12: Mean squared 1-step ahead forecast errors MSFE$_{1{:}t}(1)$ of inflation $(p)$ sequentially revised at each of the $t=\seq1{180}$ months.
\label{1mse1}}
\end{figure}

\begin{figure}[htbp!]
\centering
\includegraphics[width=0.75\textwidth]{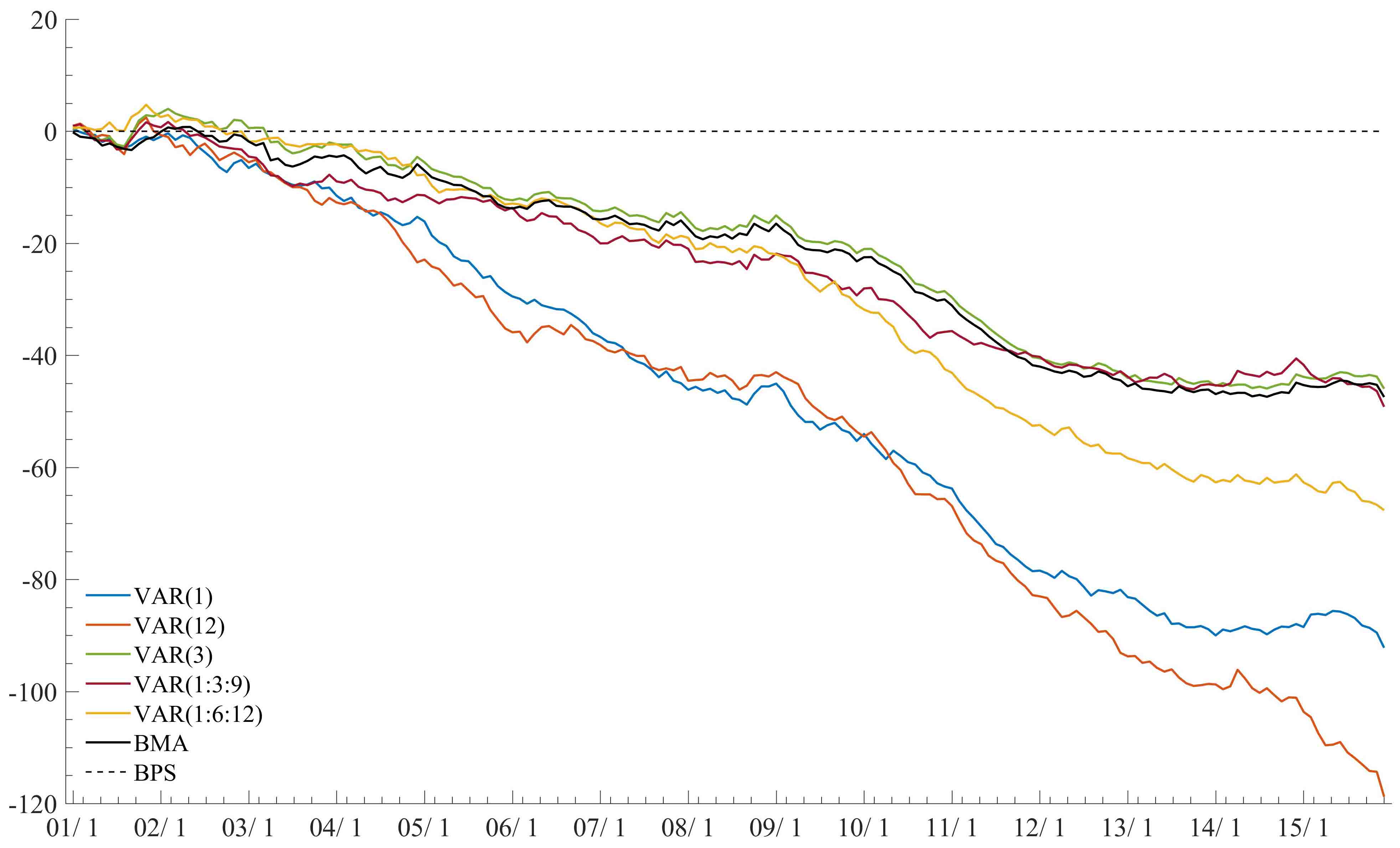}
\caption{US macroeconomic forecasting 2001/1-2015/12: 1-step ahead log predictive density ratios LPDR$_{1{:}t}(1)$  sequentially revised at each of the $t=\seq1{180}$ months. The baseline at 0 over all $t$ corresponds to the standard BPS model.
\label{1lpdr}}
\end{figure}

\begin{figure}[htbp!]
\centering
\includegraphics[width=0.75\textwidth]{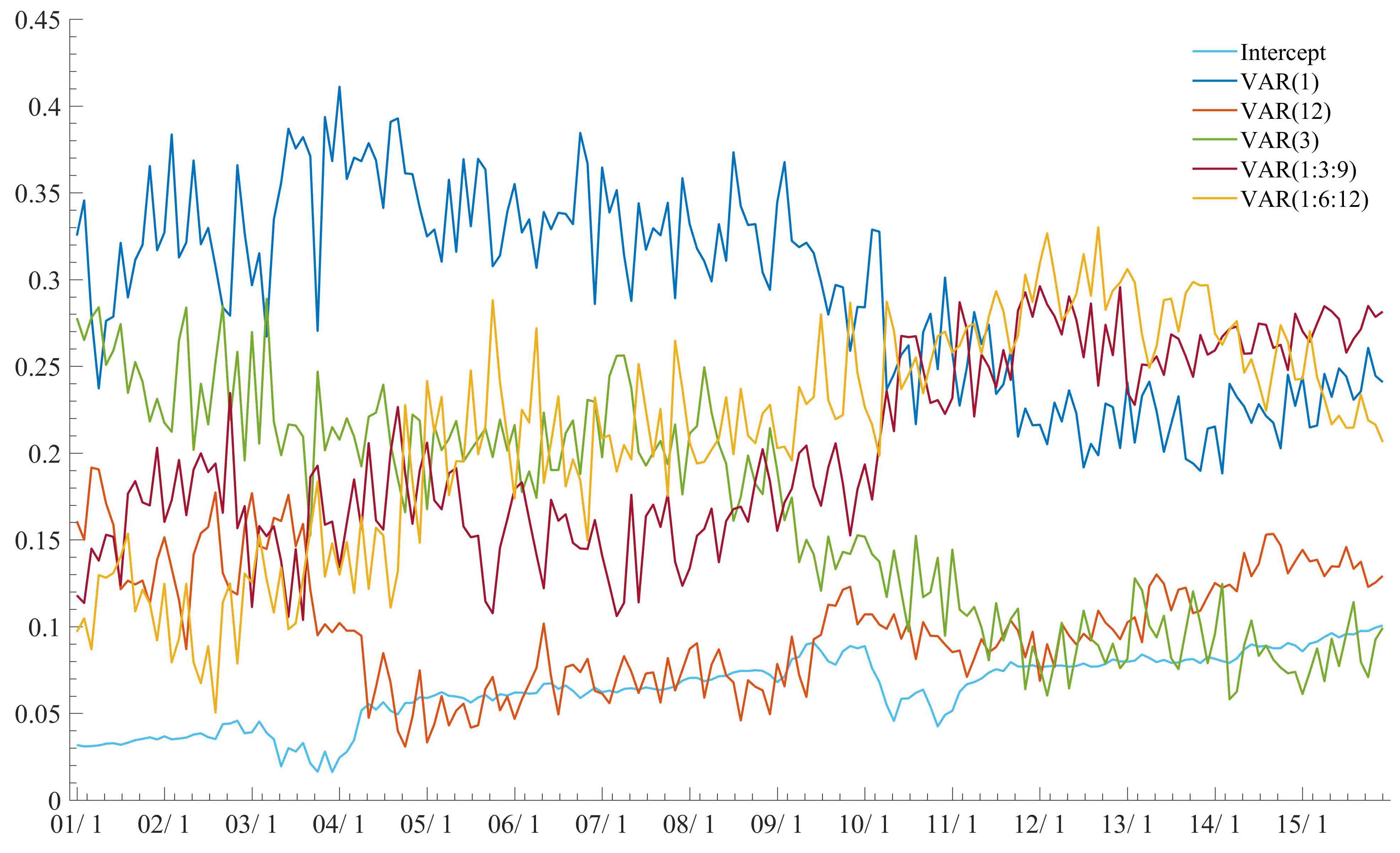}
\caption{US macroeconomic forecasting 2001/1-2015/12:  On-line posterior means of BPS model
 coefficients for inflation $(p)$ sequentially computed at each of the $t=\seq1{180}$ months.
\label{1coeff1}}
\end{figure}

\begin{figure}[htbp!]
\centering
\includegraphics[width=0.75\textwidth]{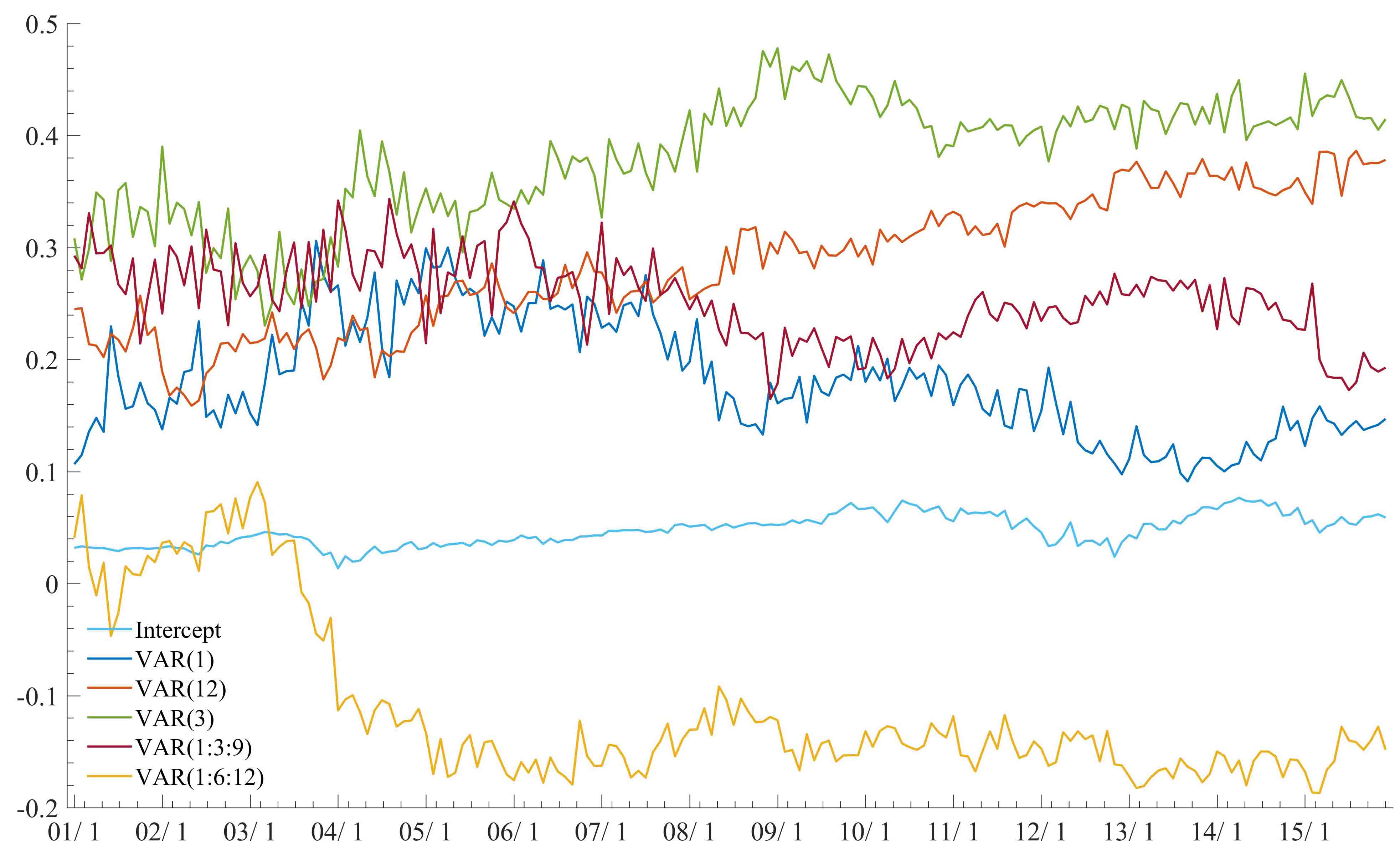}
\caption{US macroeconomic forecasting 2001/1-2015/12:  On-line posterior means of BPS model
 coefficients for wage $(w)$sequentially computed at each of the $t=\seq1{180}$ months.
\label{1coeff2}}
\end{figure}

\begin{figure}[htbp!]
\centering
\includegraphics[width=0.75\textwidth]{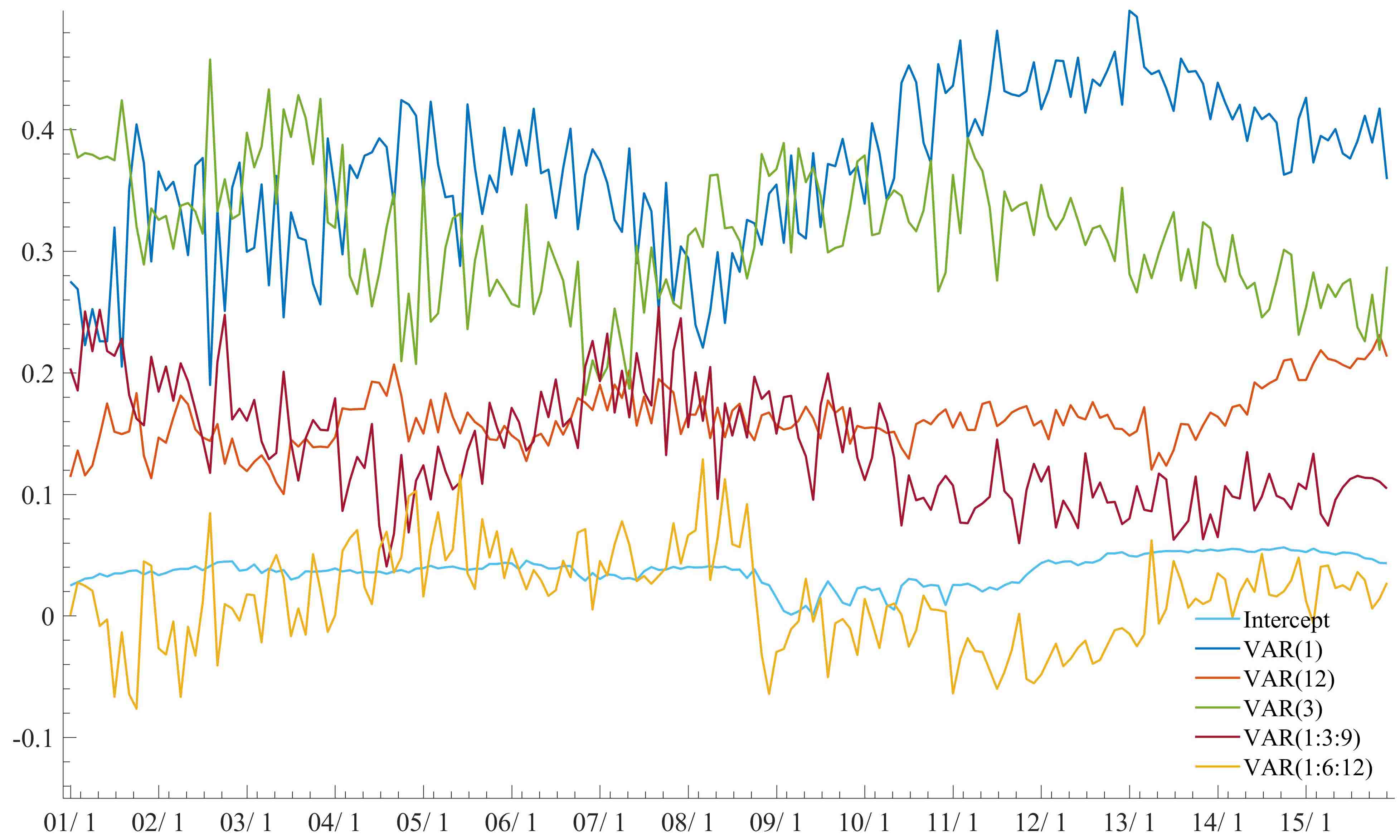}
\caption{US macroeconomic forecasting 2001/1-2015/12:  On-line posterior means of BPS model
 coefficients for unemployment $(u)$ sequentially computed at each of the $t=\seq1{180}$ months.
\label{1coeff3}}
\end{figure}

\begin{figure}[htbp!]
\centering
\includegraphics[width=0.75\textwidth]{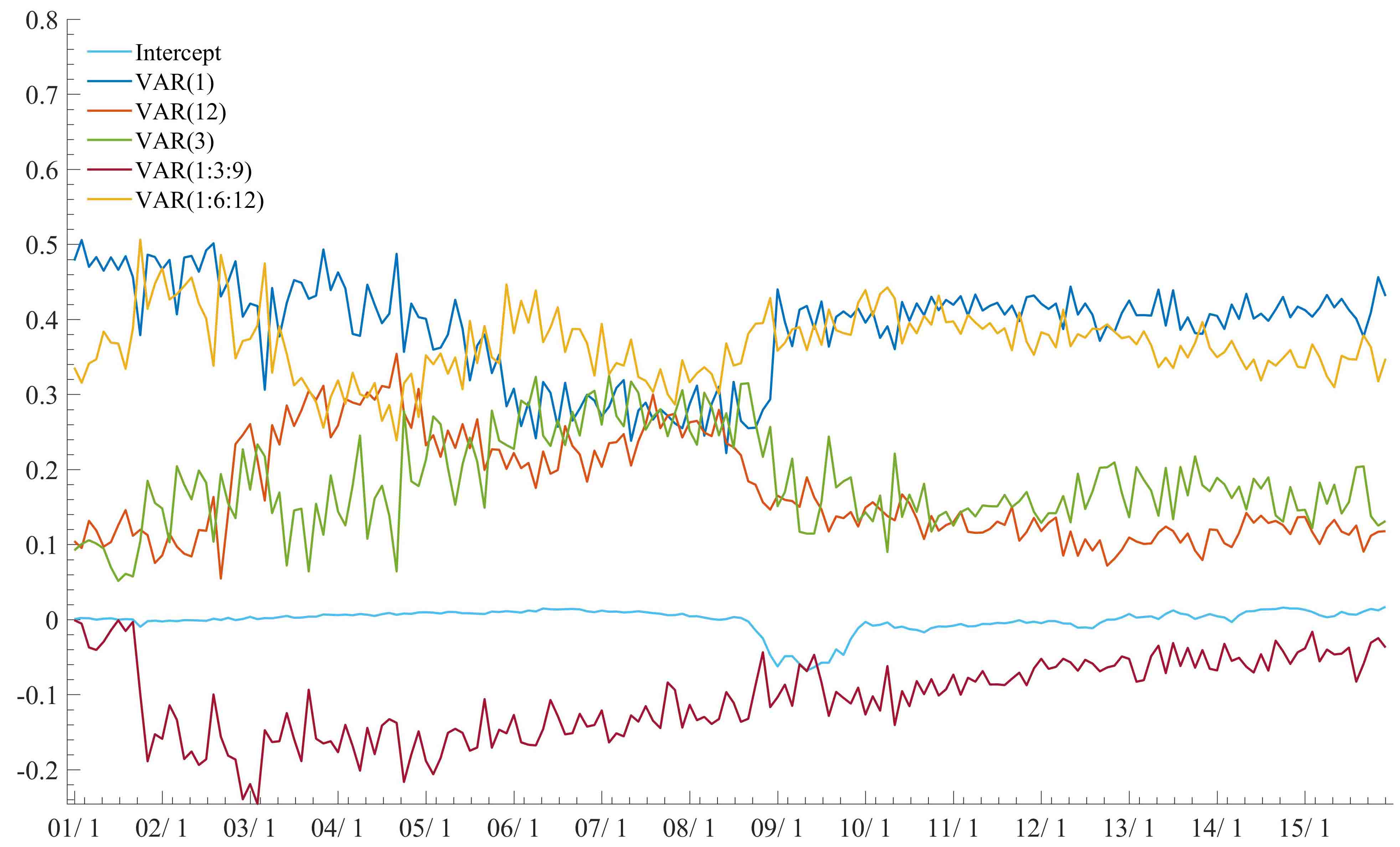}
\caption{US macroeconomic forecasting 2001/1-2015/12:  On-line posterior means of BPS model
 coefficients for consumption $(c)$ sequentially computed at each of the $t=\seq1{180}$ months.
\label{1coeff4}}
\end{figure}

\begin{figure}[htbp!]
\centering
\includegraphics[width=0.75\textwidth]{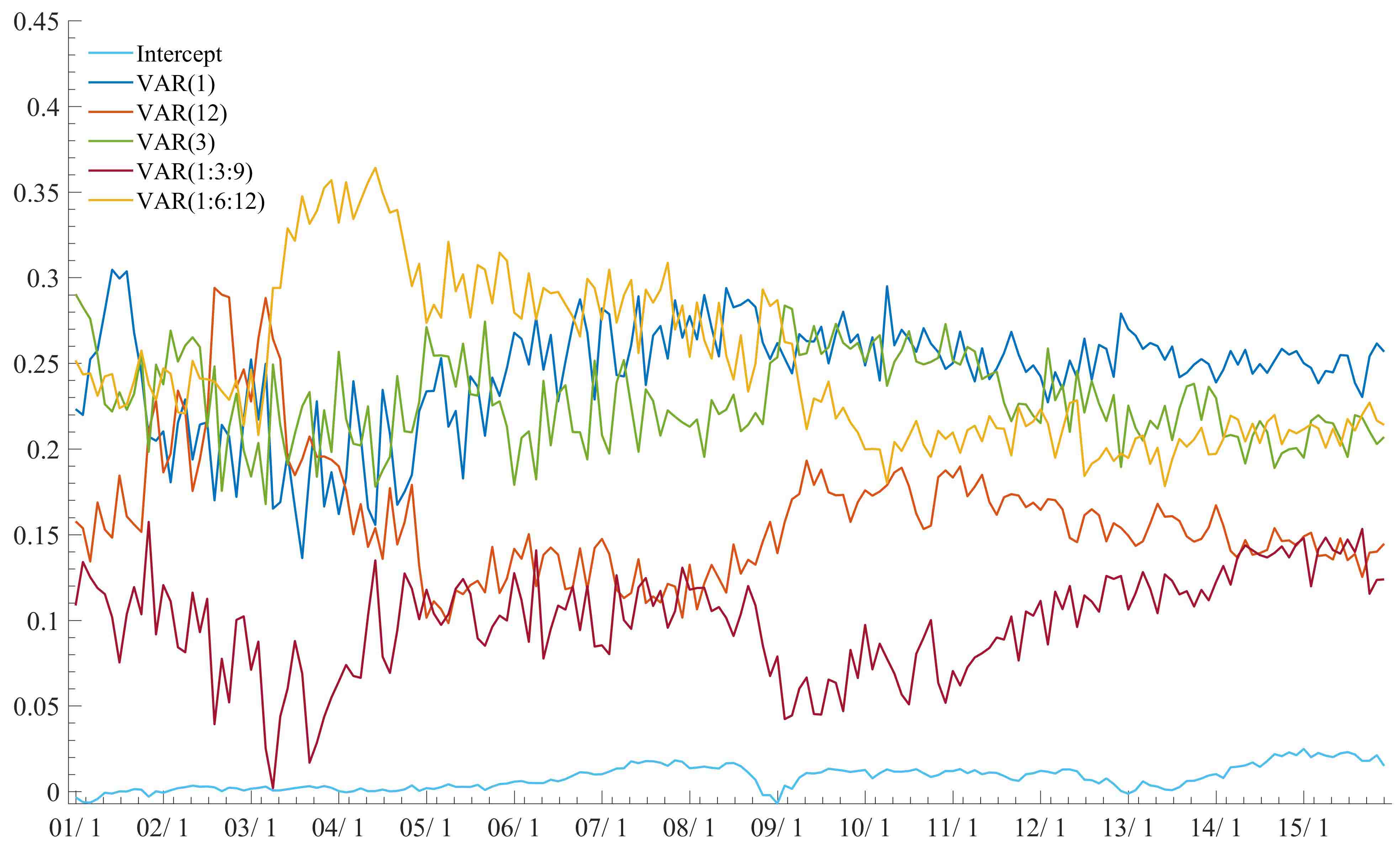}
\caption{US macroeconomic forecasting 2001/1-2015/12:  On-line posterior means of BPS model
 coefficients for investment $(i)$ sequentially computed at each of the $t=\seq1{180}$ months.
\label{1coeff5}}
\end{figure}

\begin{figure}[htbp!]
\centering
\includegraphics[width=0.75\textwidth]{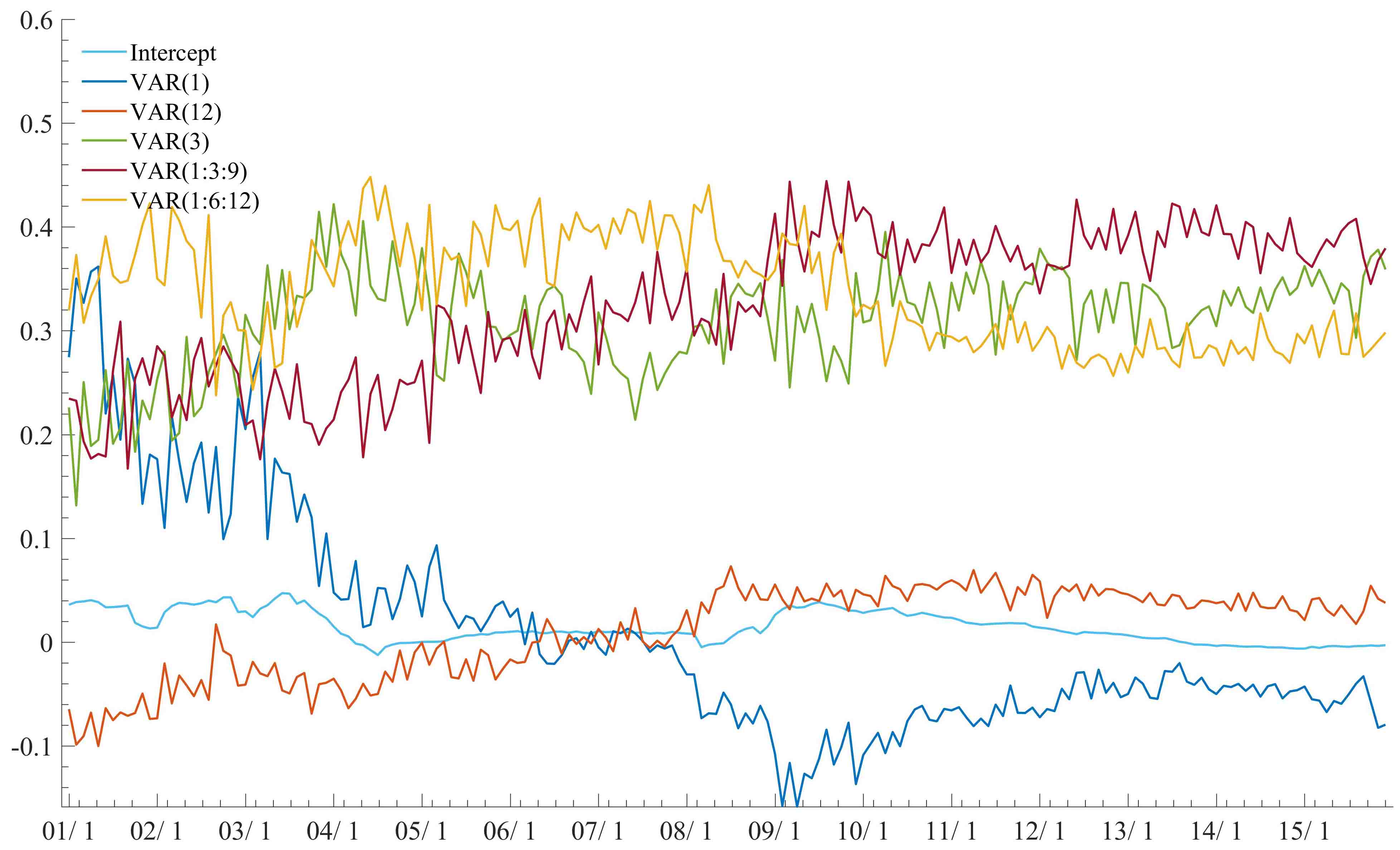}
\caption{US macroeconomic forecasting 2001/1-2015/12:  On-line posterior means of BPS model
 coefficients for interest rate $(r)$ sequentially computed at each of the $t=\seq1{180}$ months.
\label{1coeff6}}
\end{figure}

\begin{figure}[htbp]
\centering
\includegraphics[width=0.9\textwidth]{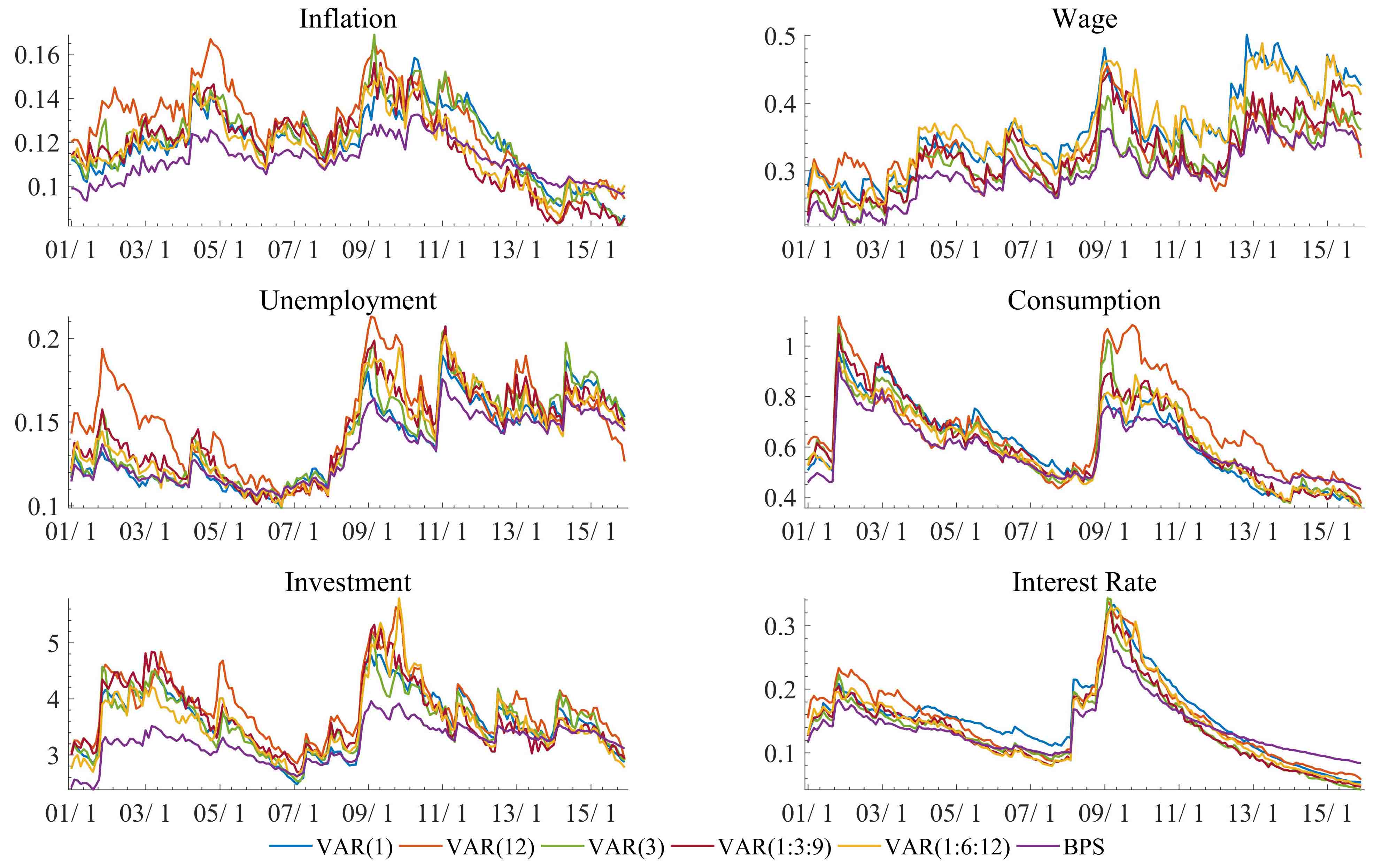}
\caption{US macroeconomic forecasting 2001/1-2015/12: 1-step ahead forecast standard deviations
 sequentially computed at each of the $t=\seq1{180}$ months.
\label{1var}}
\end{figure}

\begin{figure}[htbp!]
\centering
\includegraphics[width=0.7\textwidth]{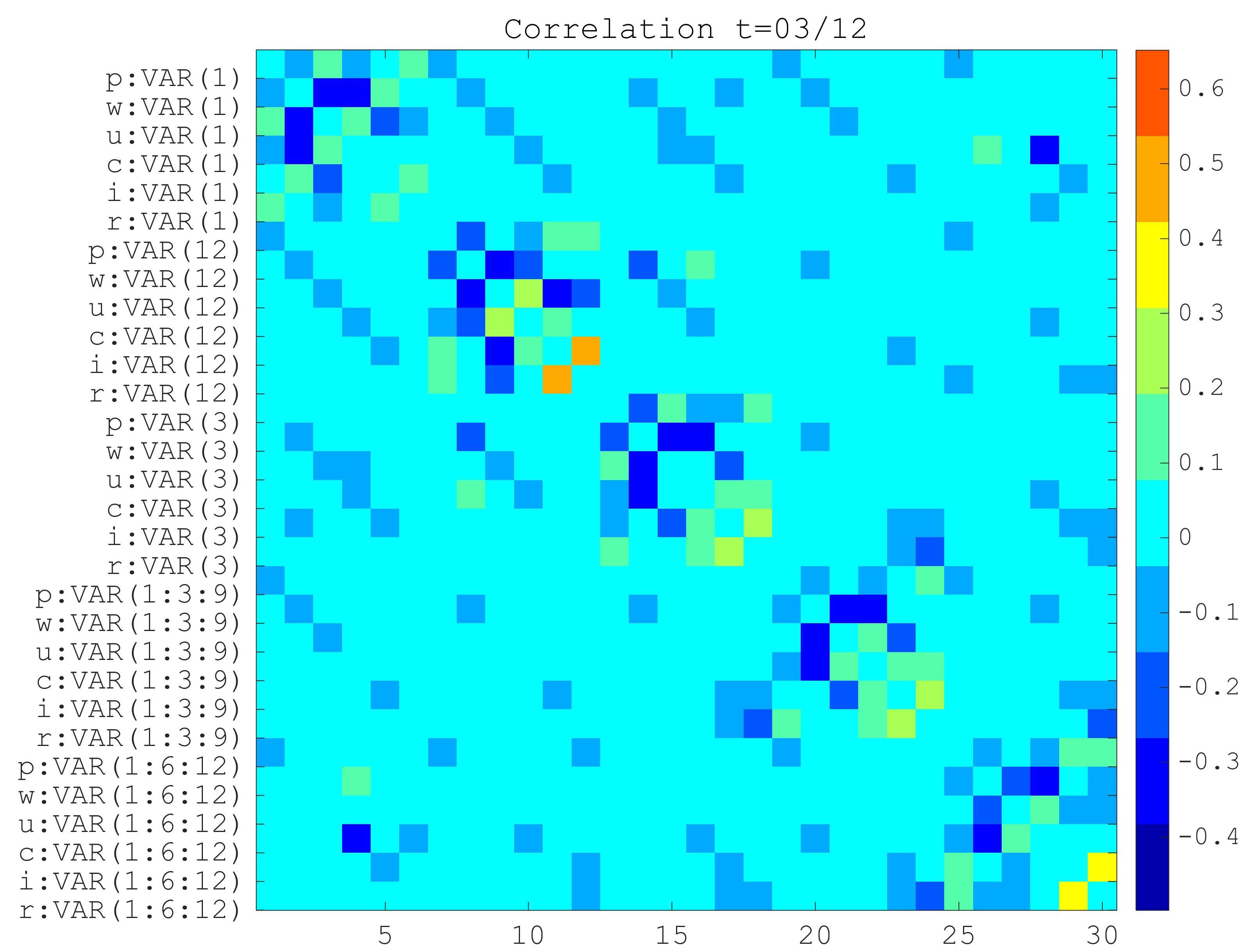}
\caption{US macroeconomic forecasting 2001/1-2015/12:  Retrospective posterior correlations of latent agent factors at 2003/12.
\label{1corr1}}
\end{figure}

\begin{figure}[htbp!]
\centering
\includegraphics[width=0.7\textwidth]{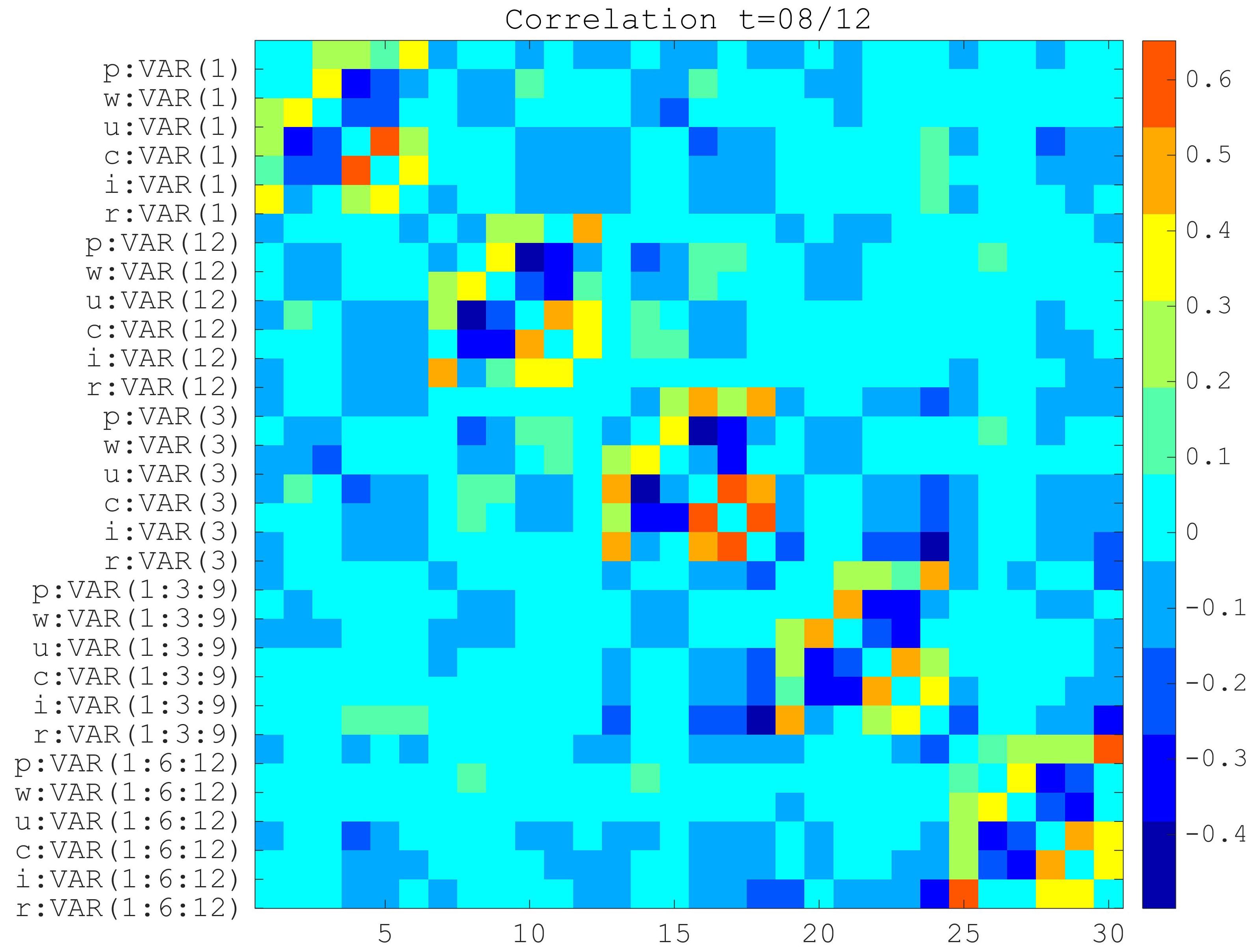}
\caption{US macroeconomic forecasting 2001/1-2015/12:  Retrospective posterior correlations of latent agent factors at 2008/12.
\label{1corr3}}
\end{figure}

\begin{figure}[htbp!]
\centering
\includegraphics[width=0.7\textwidth]{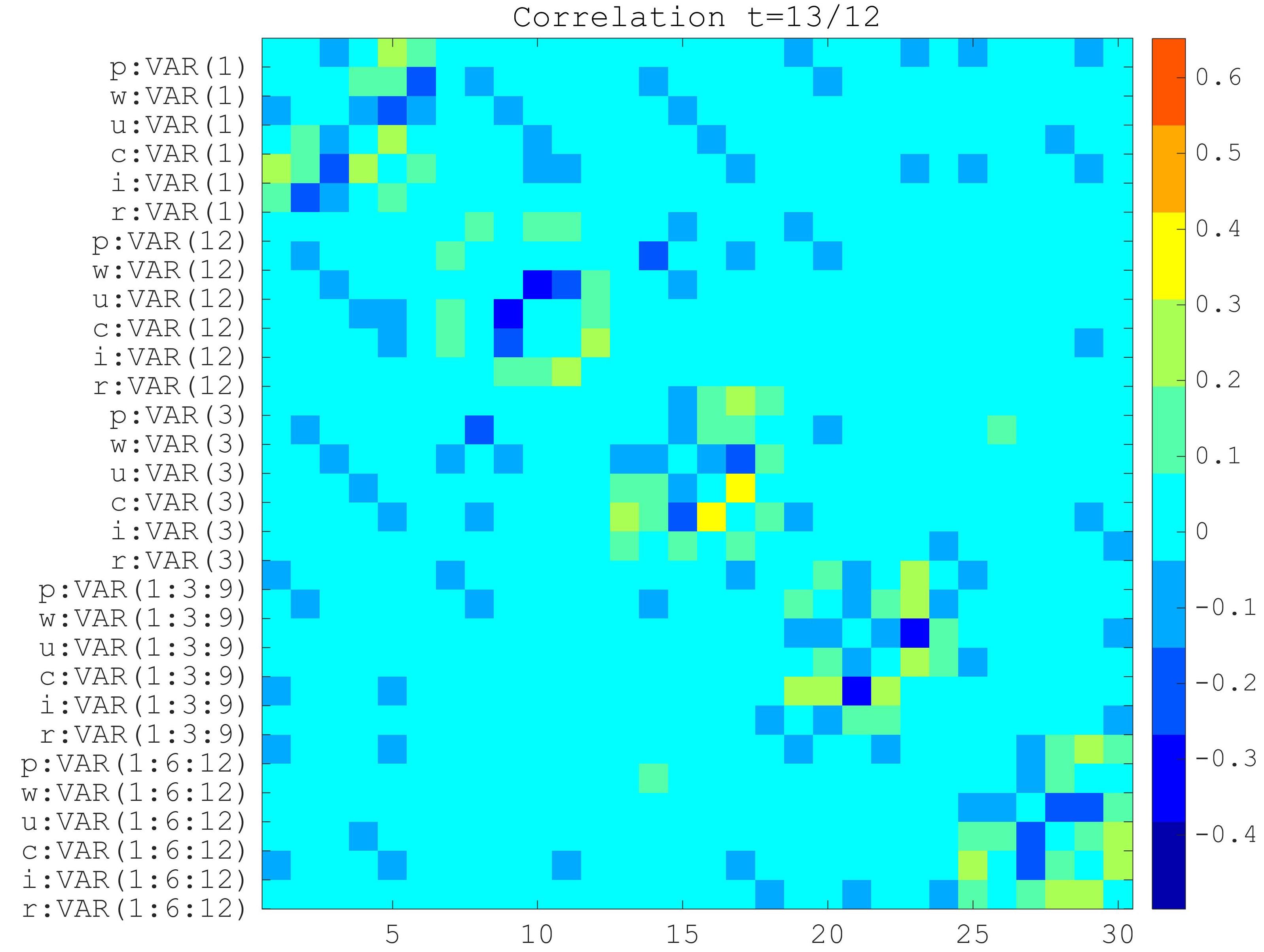}
\caption{US macroeconomic forecasting 2001/1-2015/12:  Retrospective posterior correlations of latent agent factors at 2013/12.
\label{1corr4}}
\end{figure}

\begin{figure}[htbp!]
\centering
\includegraphics[width=0.7\textwidth]{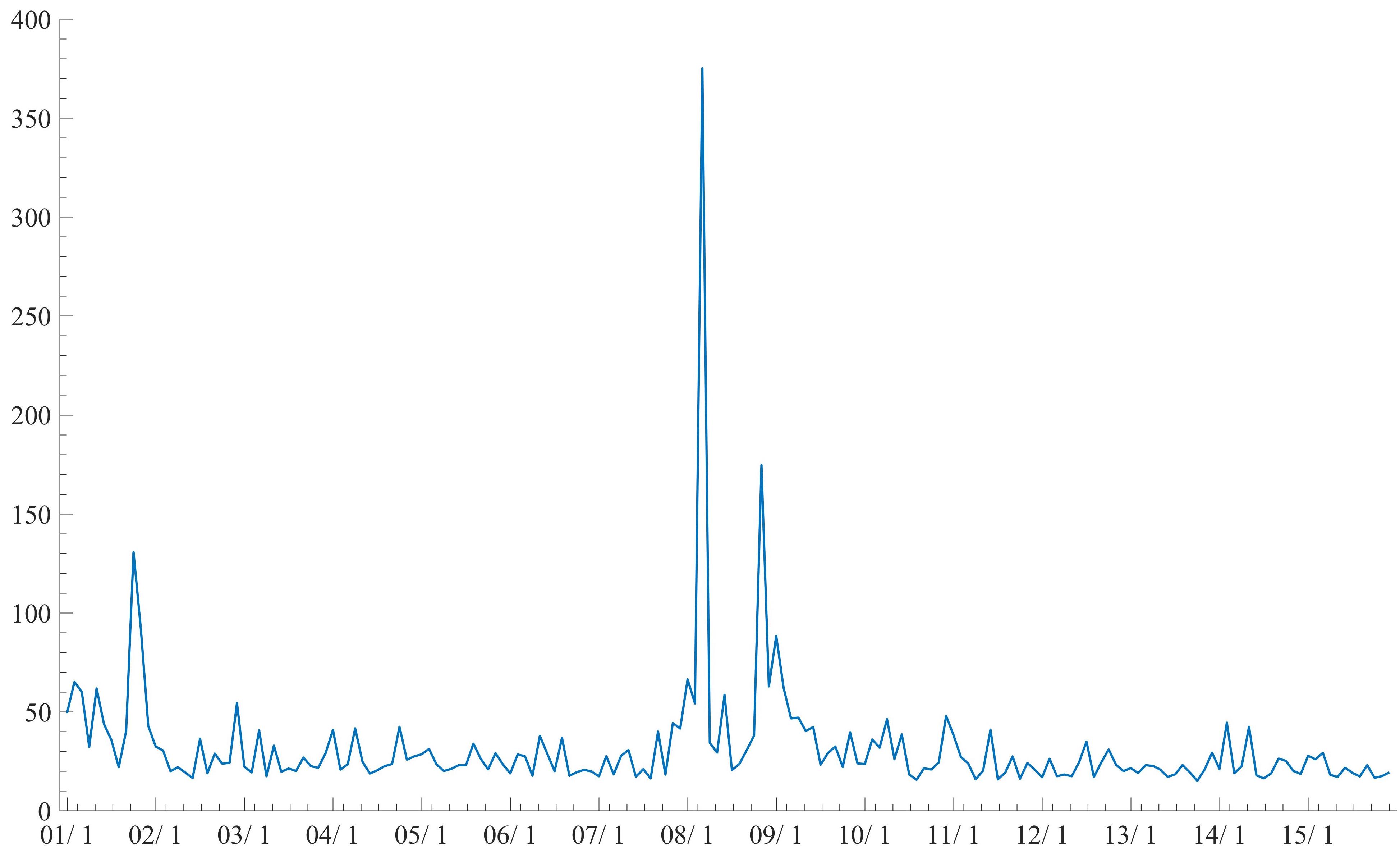}
\caption{US macroeconomic forecasting 2001/1-2015/12: Retrospective Kullback-Leibler  divergence between prior agent forecasts and posterior agent forecasts computed at each of the $t=\seq1{180}$ months.
\label{KL}}
\end{figure}

We next review summary graphs showing aspects of analysis evolving over time during the testing period, a period that includes challenging economic times that impede good predictive performance. Figs.~\ref{1mse1}--\ref{KL} summarize sequential analysis for 1-step forecasting.

Fig.~\ref{1mse1} shows the 1-step ahead measures MSFE$_{1{:}t}(1)$ over time $t$ in forecasting inflation.
The other series are omitted for the sake of brevity, but the patterns in inflation are consistent (see supplementary appendix material).
Additionally, forecasting inflation is one of the most important tasks for an economic policy maker, and therefore focusing on inflation is appropriate for this example.
While BPS does not outperform the other models over the whole testing period, we see that it is on par with the best models considered.
BPS ends up improving on the other models based on its performance during and after the sub-prime mortgage crisis, demonstrating how BPS dynamically adapts over time to produce robust forecasts over crisis periods and changing regimes.

Fig.~\ref{1lpdr} confirms that BPS performs uniformly better than the other models based on LPDR measures.
The gradual decline in LPDR and a more drastic decline after the crisis is indicative of how BPS dynamically adapts its location and uncertainty to improve its distribution forecasts.

One crucial aspect of the BPS model is that it can adapt coefficients specific to each series.
Figs.~\ref{1coeff1}-\ref{1coeff6} are the on-line posterior means of BPS model coefficients for 1-step ahead forecasts for each series.
We first note how the coefficients for each series evolve and are different across series, reflecting how different  models are better at forecasting different series and how the relative accuracy differs in time.
Additionally, we note that the dynamic coefficients can appear to be somewhat erratic, which is reflective of the level of uncertainty in the latent agent states.

The somewhat erratic coefficient trajectories for the 1-step ahead forecasts--particularly  compared to those from the
12- and 24-step analyses (Figs.~\ref{12coeff1}-\ref{24coeff6})-- arises due to a number of factors.
It is partly a result of the constraining initial prior on the intercept,
heavily favoring small values, so the on-line posteriors for the BPS coefficients adapt more dramatically than were the intercept to be \lq\lq freer'' to move around.  This is coupled with the generally strong positive inter-dependencies among agent forecasts that
lead to high collinearity.
Coefficients trajectories are most erratic during the stable pre-crisis period, where agent forecasts are particularly similar, and less erratic when agent forecasts diverge more substantially post-crisis.
%
%
For 12- and 24-step ahead forecasts, while the agent forecasts are generally poorer, their inter-dependencies are
much weaker yielding more stable coefficient trajectories based on sustained differences in relative forecasting accuracy across agents
even in the context of high uncertainties.

For inflation (Fig.~\ref{1coeff1}), the coefficients clearly exhibit a structural change after the sub-prime mortgage crisis.
VAR$(1)$ and VAR$(3)$, which are relatively simple models with short lags, have the highest coefficients up until the crisis, but quickly drop off, replaced by VAR$(1{:}3{:}9)$ and VAR$(1{:}6{:}12)$, which are more complex models with longer lags.
This can be viewed as a structural change where simpler dynamics are being replaced by longer, more complex, dynamics after the crisis.

Coefficients of wage (Fig.~\ref{1coeff2}), on the other hand, are relatively stable over time.
The VAR$(3)$ model, using a quarters worth of lags, has the highest coefficient and that stays the highest throughout.
In comparison to inflation, VAR$(1)$, the simplest model has limited impact, while the most complex VAR$(12)$
model has a persistent negative coefficient which, to some degree, balances the impact of the simpler, short-lag models.

In Fig.~\ref{1coeff3}, the estimated trajectory of coefficients for unemployment exhibits an increase for the
VAR$(1)$ model and a gradual decrease for the more complex VAR$(3)$ model after the sub-prime mortgage crisis, as well as overall small effects from the other more complex models.
Due to unemployment being heavily impacted by the crisis, this characteristic is understandable.
Long term unemployment trends become irrelevant in light of the recent shock to the economy, and the coefficients reflect that shift.

For   consumption Fig.~\ref{1coeff4}, we see a clear grouping of agents: that of VAR$(1)$-VAR$(1{:}6{:}12)$ and that of VAR$(3)$-VAR$(12)$.
Before the crisis, we see that these two groups converging to almost equal weight (except for VAR$(1{:}3{:}9)$, which is almost always negative), then quickly re-separating again after the crisis.
This suggests  that consumption is mainly driven by biannual lags.

The coefficients for investment (Fig.~\ref{1coeff5}) uses a more restrictive initial prior due to experience in the
training data period of extremely high uncertainty in the agent forecasts linked to the highly volatile nature of this series.
This results in relatively more stable coefficients trajectories, with all  being above zero and around equal weight.
However, there are still clear patterns that emerge; notably, we see an upward spike in VAR$(1{:}6{:}12)$ at around 2003 and some fluctuations during the subprime mortgage crisis.

For interest rate (Fig.~\ref{1coeff6}) the coefficients favor more complex models with longer lags.
Interestingly, we see a gradual decrease in coefficients on the VAR$(1)$ model up until the sub-prime mortgage crisis, at which point it stays level.
Long term dynamics, we can infer, were taking over short term dynamics leading up to the crisis, bringing up interesting questions about lending and credit characteristics pre-crisis.
We also note that the introduction of zero interest rates after the crisis does not seem to effect the coefficients at all.


Figs.~\ref{1var}-\ref{1corr4} exhibit selected aspects of inferences on the trends in uncertainty and dependence between and within agents over time.
The posterior latent agent variances (the diagonal elements of the covariance matrix) displays how the uncertainty measures of the forecasts change over time; see Fig.~\ref{1var}.
Complex models for multiple series-- that require estimation methods that are also complex-- often produce large forecast
standard deviations coming from the model, data, estimation method, or all of the above.
Large VAR models are popular in practice, due to  modeling flexibility and interpretability, but naturally lead to inflated uncertainty measures due to large numbers of parameters, colinearities and resulting estimation uncertainties.
BPS, on the other hand, has smaller uncertainty in synthesized forecasts, resulting in decreased forecast uncertainty relative to each of the agents.  This is a critical benefit of BPS.
Though underestimating real risk is as dangerous as overestimating it, the LPDR results indicate that the BPS uncertainty estimates are valid-- point forecasts are generally improved and lower predictive uncertainties couple with that to lead to a win-win analysis.

We move attention to posterior distributions on the posterior latent agent factors over time to explore
inter-dependencies among agent forecasts and their temporal evolutions.
One set of numerical summary of dependencies among agent forecasts is given by the retrospective posterior correlation of $\X_t$.
The posterior dependencies among agents are measured through the off block diagonal elements of the correlation, which is zero when the forecasters provide their forecasts.
For instance, high positive dependencies among
agent forecasts will generate high negative correlations among the corresponding dynamic regression coefficients on the
agent latent factors, and vice-versa; discovering the underlying temporal dependencies.  We look at these measures of dependence at 3 specific time points that represent   different regimes in the testing period; pre-sub-prime mortgage crisis (2003/12), immediately after the crisis (2008/12), and post-crisis (2014/12).  Correlations are arranged in order agents.
See  Figs.~\ref{1corr1}-\ref{1corr4}.

At the beginning of the testing period,  where the series are relatively stable (Fig.~\ref{1corr1}), we note an overall strong negative and positive dependence  within agents, with some notable positive dependence between 12-month (long term) investment and interest rate and strong negative dependence for unemployment and consumption across all agents.
The negative correlation within series is expected, as these are modeled by the agents and updated through BPS.
Across agents, we note some diagonal patterns that are present, albeit weak.
For example, looking at the off diagonal block between VAR$(1)$ and VAR$(12)$, we can see negative dependencies between all series except for interest rates.
This indicates some long terms dependencies, although we do expect them to exist since VAR$(1)$ is nested within VAR$(12)$.

Immediately after the crisis (Fig.~\ref{1corr3}), we note a drastic change in correlations across and within agents appearing, as seen in the positive correlations (yellow/red) radiating from and within series and negative correlations across agents.
Focusing on VAR$(1{:}3{:}9)$, there are strong negative correlations between VAR$(3)$, but comparatively very little between the other agents.
Other patterns, notably block patterns appear between VAR$(1)$, VAR$(12)$, and VAR$(3)$ that were not present pre-crisis.
These emergent patterns suggest a large shift in dependence during the crisis that would have been overlooked without BPS.
Post crisis (Fig.~\ref{1corr4}), we see a new dependence structure appearing, with almost no dependence across--  and even within-- series, except for mild dependencies within series and some diagonal structure seen in Fig.~\ref{1corr1} reappearing.
In particular, the correlation between VAR$(1{:}3{:}9)$ and VAR$(12)$ and VAR$(3)$ has seem to have completely vanished. 

The patterns of changes in dependencies among agents and across series over time provide insights into how the economy changes with respect to economic  shocks defining different regimes.
Figs.~\ref{1corr1} and~\ref{1corr4} are both snapshots of relatively stable periods, yet the characteristics exhibited through the
estimated  dependencies are  different.
These differences in economic structure are not unexpected, though graphically visualizing the differences through the lens of agents
and BPS-defined inferences on inter-dependencies provides new insight and perceptions into the overall change in economy.

Finally, Fig.~\ref{KL} measures the Kullback-Leibler divergence from the prior (forecast densities from the agents with block diagonal covariance structure) to the posterior (learned through BPS) agent forecasts, measured at each time $t$.
The Kullback-Leibler divergence for each time $t$ is defined as follows:
\begin{align*}
D_{KL}(t)=\mathbb{E}_{p(\X_{t})}[log(p(\X_{t})/h(\X_{t}))]
\end{align*}
where $h(\X_{t})$ is the prior agent forecast density and $p(\X_{t})=p( \X_{t} |  \bPhi_{t}, \y_{t}, \mH_{t} )$ is the posterior agent forecast density.
Note that, while the Kullback-Leibler divergence is analytically estimable for two multivariate Gaussian densities, for other distributions, or samples from posterior densities, the Kullback-Leibler divergence must be approximated by a Gaussian density (if suitable) or by computing the following Monte Carlo approximation
\begin{align*}
D^{approx}_{KL}(t)=\frac{1}{n}\sum_{i=1{:}n} [log(p(\X_{ti})/h(\X_{ti}))].
\end{align*}
Since the Kullback-Leibler divergence measures the information gained from $h(\X_{t})$ to $p(\X_{t})$, it is a direct measurement/summary of the amount of biases and inter-dependencies that are learned through BPS using the data and past agent forecast information.
For example, if an agent produces significantly biased forecasts during economic distress, BPS will learn and calibrate that bias, and the difference will be captured in the Kullback-Leibler divergence.
Likewise, we expect and observe inter-dependencies across agents to increase during crises and shocks, as seen in Fig.~\ref{1corr3}.
Under these conditions, the Kullback-Leibler divergence will naturally increase, since the prior agent inter-dependencies are zero.
Looking at Fig.~\ref{KL}, we see three major spikes that are notable; the first during the dot-com bubble, and two before and after the great financial crisis.
The first and third spike is understandable, as the results in Fig.~\ref{1corr3} indicate a large amount of inter-dependencies across agents to appear during a crisis.
The second spike is perhaps more interesting, as early 2008 was when-- in hindsight-- early signs of the crisis were emerging.
Because the subtle shifts in agent forecasts (increased biases and inter-dependencies not reported in the prior agent forecasts) are captured using BPS, the Kullback-Leibler divergence is able to pick up the early signals of a recession.
Although the results presented here are retrospective (i.e. the posterior is in regard to $t=1{:}T$), it is indicative of the strengths of BPS in capturing crises and shocks measured through how much information is gained via Bayesian learning.

\paragraph{k-step ahead forecasting.} Long term forecasting for economic policy   is far more important than 1-step ahead forecasting.
For this study, we forecast 12- (one year) and 24- (two years) step ahead to demonstrate the effectiveness of BPS over the set of agents at practically important horizons.

Tables~\ref{table:12step} and \ref{table:24step} summarizes the predictive measures compared for the two forecast horizons.
For point forecasts, BPS$(k)$ outperforms all other models for all series with the exception of wage growth.
The improvement hold  for all $k$-step ahead forecasts considered and the improvements of BPS$(k)$  significantly increase with $k$.
The improvements come from BPS$(k)$ directly synthesizing the $k$-step ahead forecasts from the agents, calibrating, adapting, and learning the latent dependencies and biases over the $k$-step ahead quantity of interest.
For 24-step ahead forecasts of inflation,     one of the most important series for a central bank when setting their key policy rate, BPS$(k)$ greatly improves on the agent models, with massive gains over    the best agent model.
As with 1-step ahead forecasting, it is also notable that agent model performances   vary significantly across series.
In contrast to BPS, traditional model combination schemes, such as BMA,  fail to improve over all series by sacrificing improved accuracy for one series over others; in fact,  for both 12- and 24-step ahead forecasts, BMA-based analysis degenerates to the VAR(1) model. In addition, BPS$(k)$ significantly improves in quantifying uncertainty in forecasts, as evident in the comparison of LPDR.
Creating long-term   forecasts for multiple time series is a very difficult problem due to the nature of these models being built and
trained on  1-step ahead forecasting metrics (likelihood)  and failing to propagate forward accurately.
In contrast, BPS$(k)$ synthesizes the $k$-step ahead forecasts directly, adjusting and calibrating uncertainty according to the actual quantity of interest.
Thus, no matter how the agent uncertainty forecasts are over- or under-estimating, BPS$(k)$ can re-adjust accordingly by learning how the agents over- or under-estimate.
The consistency of the LPDR improvements over multiple $k$-steps demonstrate this key feature of BPS$(k)$.

\begin{figure}[htbp!]
\centering
\includegraphics[width=0.75\textwidth]{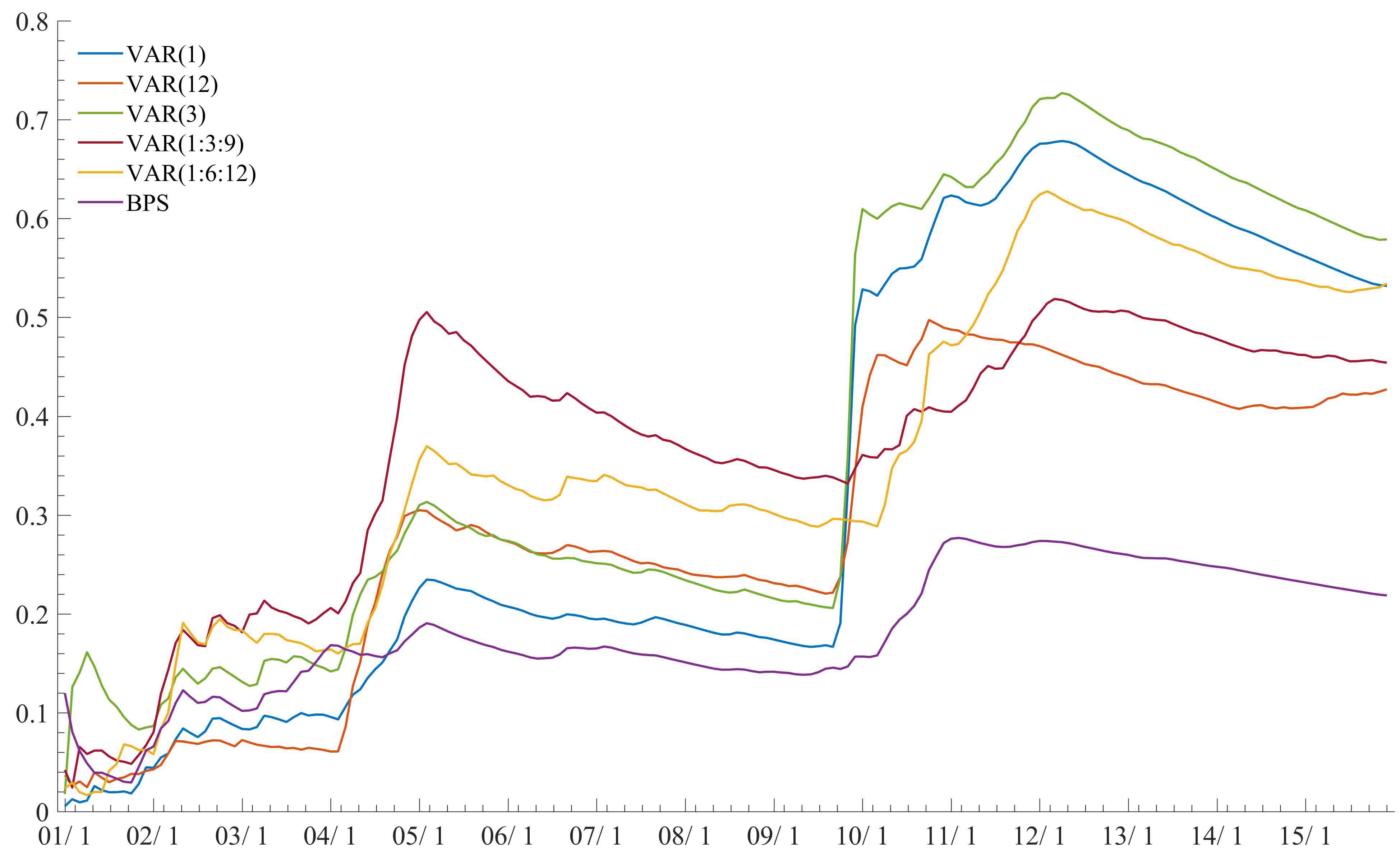}
\caption{US macroeconomic forecasting 2001/1-2015/12: Mean squared 12-step ahead forecast errors MSFE$_{1{:}t}(12)$ of inflation $(p)$ sequentially revised at each of the $t=\seq1{180}$ months.
\label{12mse1}}
\end{figure}

\begin{figure}[htbp!]
\centering
\includegraphics[width=0.75\textwidth]{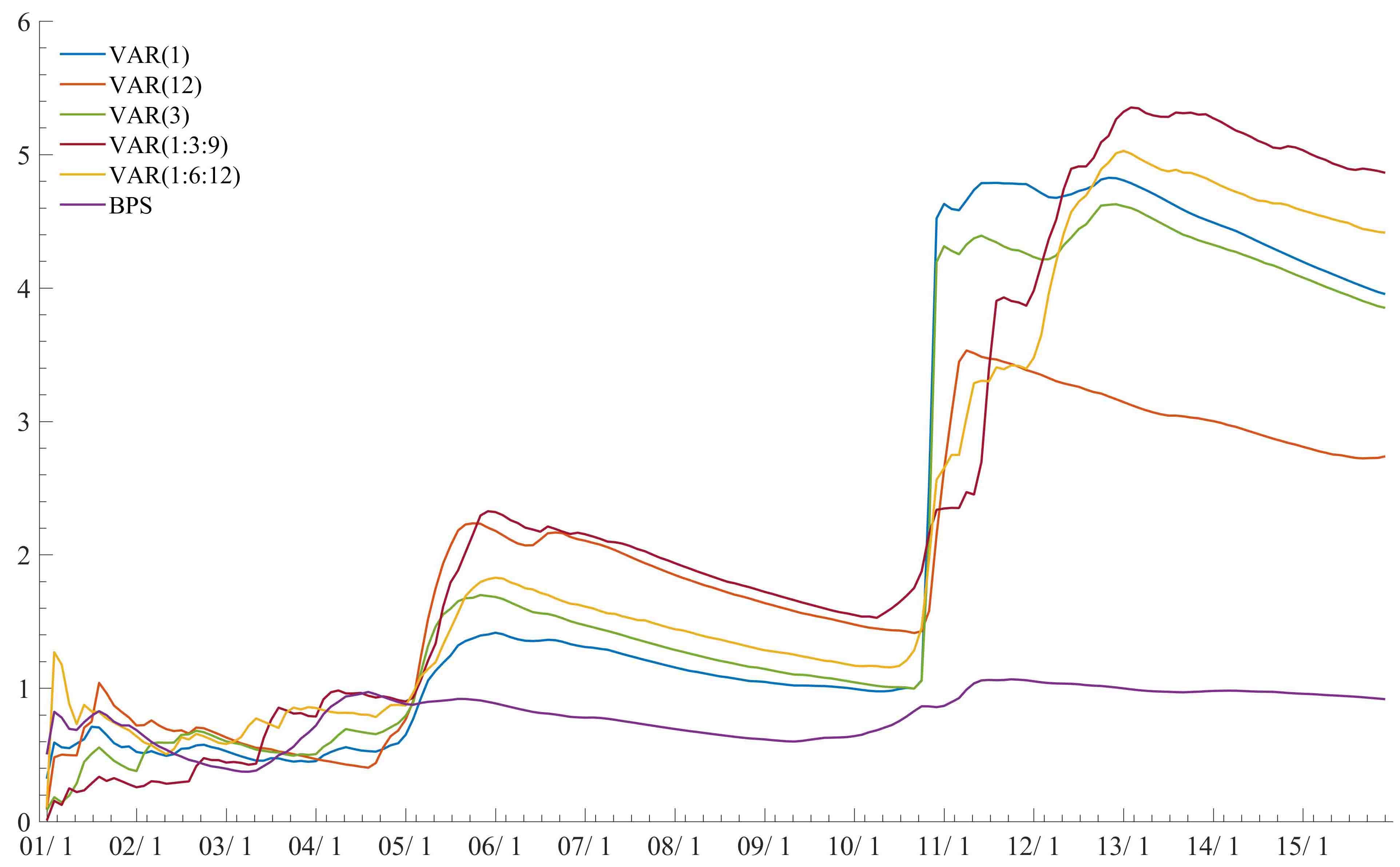}
\caption{US macroeconomic forecasting 2001/1-2015/12: Mean squared 24-step ahead forecast errors MSFE$_{1{:}t}(24)$ of inflation $(p)$ sequentially revised at each of the $t=\seq1{180}$ months.
\label{24mse1}}
\end{figure}

As with the sequential MSFE results for 1-step ahead forecasts, we focus now on multi-step MSFE results for inflation.
The characteristics of the results for inflation are similar to those of the other series and are omitted for the sake of brevity.
Figs.~\ref{12mse1} and \ref{24mse1} exhibit MSFE comparisons for inflation over the testing period for $k=12,24$-step ahead forecasts.
Although the scale is different for each $k$, there are notable common characteristics that characterize the BPS$(k)$ results.
For example, agent models experience several large shocks in precision over the testing period; this occurs,  in particular,
around the time of the advent of  the sub-prime mortgage crisis in the late 2000s.
These shocks particularly effect the precision of the agent forecasts, especially the  24-step ahead forecasts.
In comparison, BPS$(k)$ stays relatively robust throughout multiple shocks and structural  breaks.

\begin{figure}[htbp!]
\centering
\includegraphics[width=0.75\textwidth]{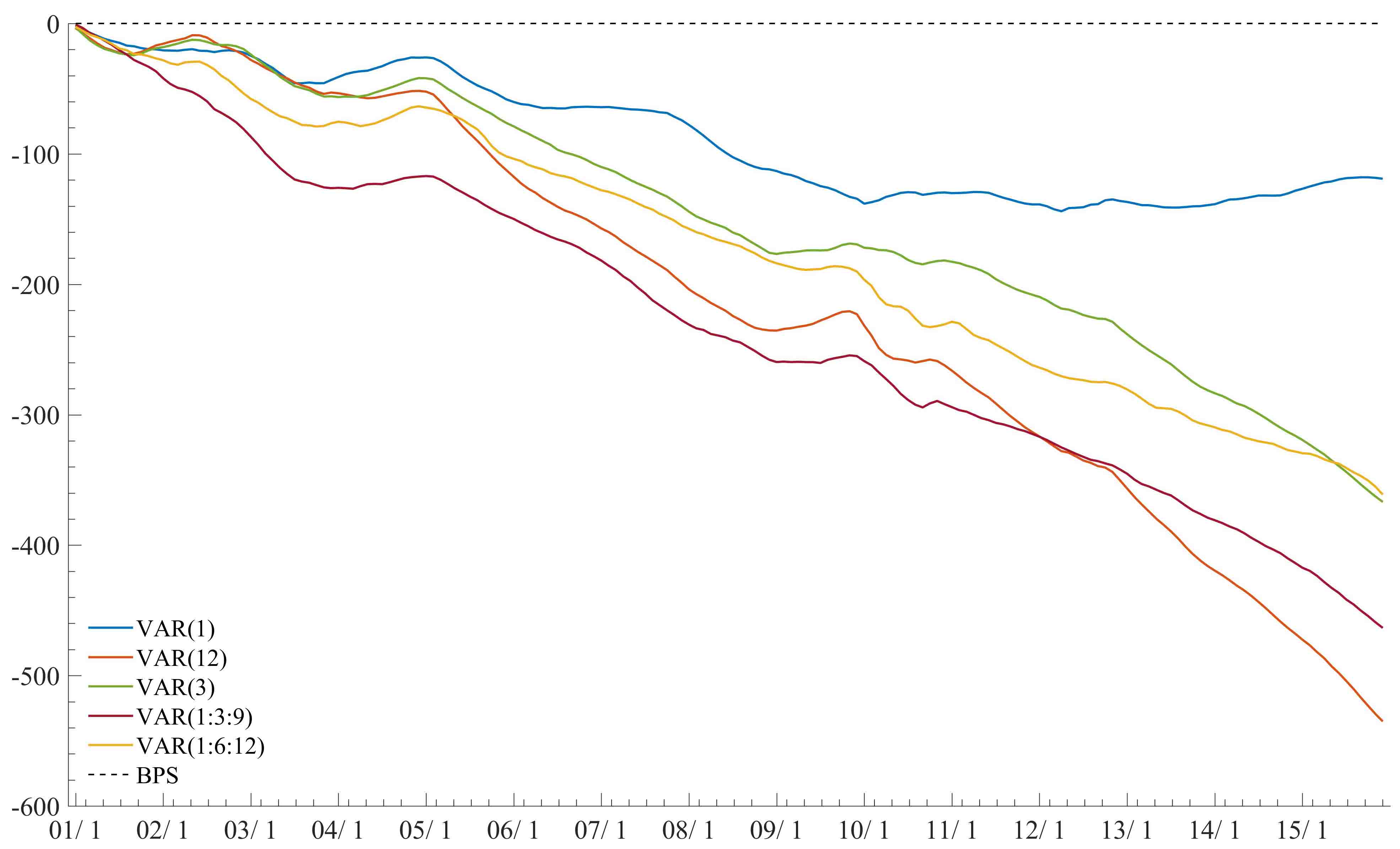}
\caption{US macroeconomic forecasting 2001/1-2015/12: 12-step ahead log predictive density ratios LPDR$_{1{:}t}(12)$  sequentially revised at each of the $t=\seq1{180}$ months. The baseline at 0 over all $t$ corresponds to BPS(12).
\label{12lpdr}}
\end{figure}

\begin{figure}[htbp!]
\centering
\includegraphics[width=0.75\textwidth]{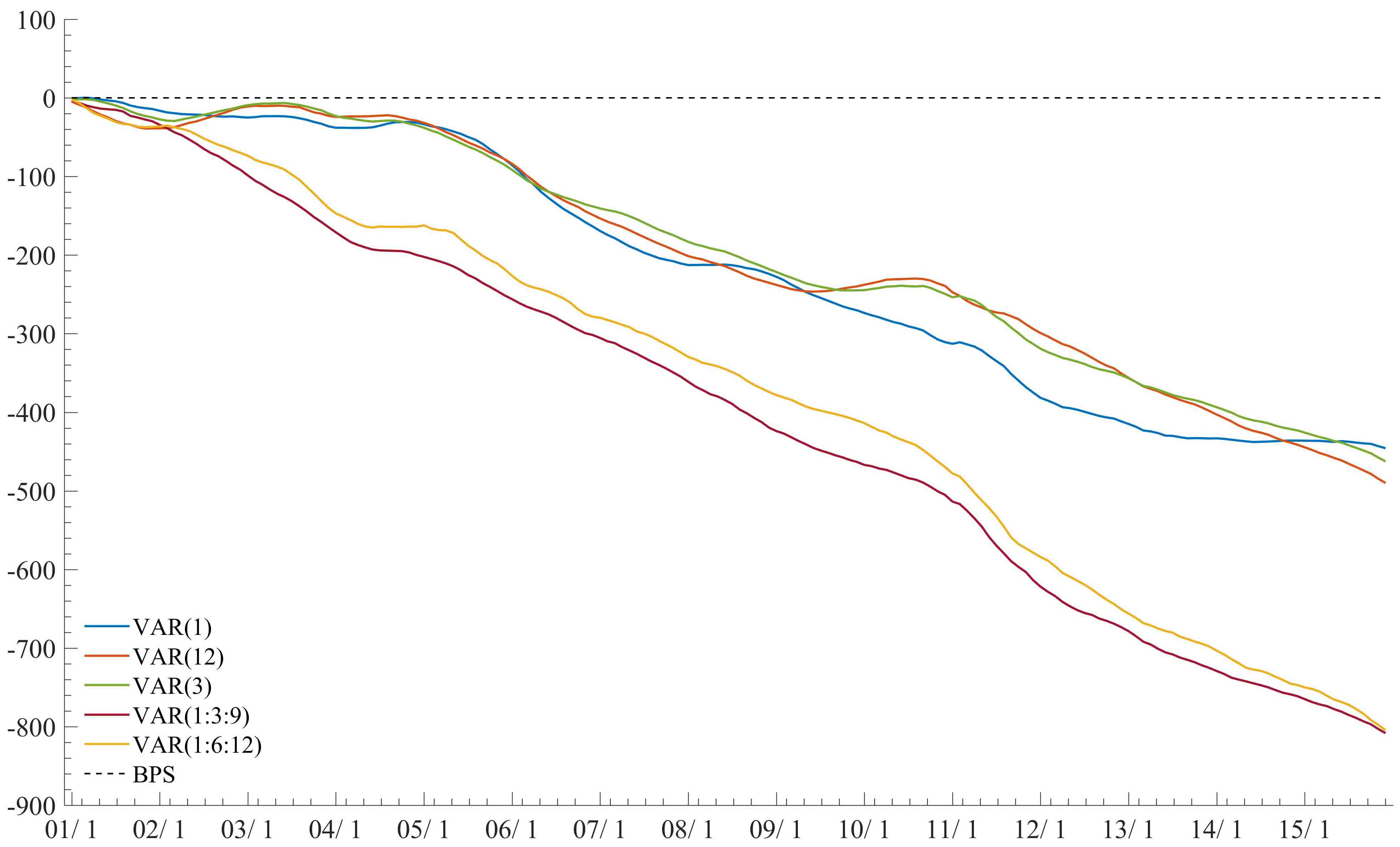}
\caption{US macroeconomic forecasting 2001/1-2015/12: 24-step ahead log predictive density ratios LPDR$_{1{:}t}(24)$  sequentially revised at each of the $t=\seq1{180}$ months. The baseline at 0 over all $t$ corresponds to BPS(24).
\label{24lpdr}}
\end{figure}

Looking at LPDR evolutions over time (Figs.~\ref{12lpdr} and \ref{24lpdr}), BPS$(k)$ improves over the agent models over all of the time period considered, except for slight increases in favoring the simpler VAR$(1)$ model immediately post crisis.
BPS$(k)$ is able to adapt to maintain improved forecasting performance both in terms of location and uncertainty assessment, a key positive feature for decision makers tasked with forecasting risk and quantiles for long horizons under possible shocks and regime change.

Figs.~\ref{24coeff1}-\ref{24coeff6} exhibit the on-line posterior means of BPS model coefficients for the 24-step ahead forecasts.
The coefficients for 6- and 12-step ahead forecasts are omitted-- similar conclusions arise in those analyses.
Overall, the BPS$(k)$ coefficients are relatively stable compared to the 1-step ahead results,   due somewhat to the lack of signal from the agent forecasts.
The agent forecasts' ability for 24-steps are considerably worse than from their 1-step ahead counterparts, leading to less useful information to be synthesized by BPS$(k)$.
The lack of signal from all of the agent models leads to less movement in the coefficients, and in turn, an increase in adaptability in the intercept.



\section{Additional Comments \label{sec:summary}}

Our extensions and development of multivariate BPS define a theoretically and conceptually sound framework to compare and synthesize multivariate density forecasts in a dynamic context.   The approach will enable decision makers to dynamically calibrate, learn, and update predictions based on ranges of forecasts  from sets of models, as well as from more subjective sources such as individual forecasters or agencies. While it will be of interest to develop future studies in which agents are represented by sets of more elaborate macroeconomic models-- such as dynamic threshold models and dynamic stochastic general equilibrium (DSGE) models-- and to integrate forecasts coming from professional forecasters and economists,
the current case study already demonstrates the real practical potential. In our sequential BPS analysis of
multiple US macroeconomic   series, we have highlighted questions of forecast synthesis methodology with respect to forecasting goals: interest in 12 or 24 month-ahead forecasting demands-- from a formal Bayesian perspective-- analysis customized to the horizon,
and the results bear out the practical relevance of that perspective.  The studies show that the flexible and  interpretable DFSUR models
can (i) adapt to time-varying biases and miscalibration of multiple models or forecasters, (ii) adaptively and practically account for-- while generating useful insights into-- patterns of time-varying relationships and dependencies among sets of models or forecasters, and (iii) improving forecast accuracy-- in some cases, most substantially-- for each of several multiple macroeconomic series together, at multiple horizons. The predictive performance of BPS is robust in times of severe economic distress, which is important for practical applications.
Additionally, inference on the inter-dependencies among forecasting models-- linked to the BPS foundational latent factor structure and aspects of inference on time-varying parameters characterizing that structure-- provides both illumination of the inter-dependencies, and
how they may vary across subsets of the multivariate series. This also provides the decision maker with the opportunity to respond
and change or intervene in the  BPS modeling for continued forecast synthesis into the future.

Multivariate BPS has further potential in applications to other fields and data where inter-dependencies between series have impact on the decision making, and where multiple forecasts, whether from forecasters or models, are available.
Such applications include financial data, such as stocks, indexes, and bonds with portfolio decisions in mind.
Further methodological extensions that warrant investigation include non-normal forecasts and discrete data,
and missing or incomplete/partial forecasts.   A second, major area for extension  arises from the observation
that our  BPS setting is not, in fact, as general as it could be in terms of
building on the historical Bayesian agent analysis literature.  Referring back to Section~\ref{sec:BPSbackground}, each agent
$\mA_j$ could provides a forecast density $h_j(\z_j)$ for a set of outcomes $\z_j$ that is not, in fact, exactly the
outcome $\y$ of interest to $\mD.$ For example,  $\z_j$ may be a subset of the macroeconomic variables in $\y$, but not all of them;
and/or it may include additional, relevant variables not included in $\y.$   The same foundational theory applies, and this provides
potential to explore BPS when different models or forecasters have differing areas of expertise as well as different strategies in
forecasting collections of related series.

Questions about  the computational aspects of fitting and forecasting with BPS are also relevant.
The current analysis, as developed and exemplified in this paper, relies on repeat reanalysis using MCMC for each time $t$.
This is a common strategy in the application of Bayesian dynamic latent factor models of other forms in a sequential forecasting context,
and state-of-the-art as an approach if full and accurate numerical Bayesian analysis is to be achieved.
In the case study developed, the computational burdens are not at all detrimental,  and
the potential improvements in forecasting accuracy and insights  that our example illustrates outweighs the computational cost.
In other contexts requiring fast data processing in sequential   analyses, it may be that some form of sequential Monte Carlo (SMC, e.g.~\citealp{LopesTsay2011}) may prove useful, especially if enabled by decoupling of multivariate series into small or univariate
but linked subsets, within each of which customized SMC methods might be more effective. This idea would aim to
exploit and extend to a BPS framework the concept of decouple/recouple in modeling increasingly high-dimensional
time series in other contexts~\citep[e.g.][]{GruberWest2016BA,ChenETALdynets2016JASA,GruberWest2017ECOSTA}.
Further, beyond our development of DFSUR models in the current study, some of these referenced modeling approaches may be
of interest in themselves as candidates for defining relationships among outcome time series and the inherent latent factors
in dynamic multivariate BPS models.




\bibliographystyle{asa} 
\bibliography{mBPS}

\begin{thebibliography}{54}
\newcommand{\enquote}[1]{``#1''}
\expandafter\ifx\csname natexlab\endcsname\relax\def\natexlab#1{#1}\fi

\bibitem[{Aastveit et~al.(2017{\natexlab{a}})Aastveit, Carriero, Clark, and
  Marcellino}]{Aastveit2016}
Aastveit, K.~A., Carriero, A., Clark, T.~E., and Marcellino, M.
  (2017{\natexlab{a}}), \enquote{{Have standard {VARs} remained stable since
  the crisis?}} \textit{Journal of Applied Econometrics}, 32, 931--951.

\bibitem[{Aastveit et~al.(2014)Aastveit, Gerdrup, Jore, and
  Thorsrud}]{Aastveit2014}
Aastveit, K.~A., Gerdrup, K.~R., Jore, A.~S., and Thorsrud, L.~A. (2014),
  \enquote{Nowcasting {GDP} in real time: {A} density combination approach,}
  \textit{Journal of Business \& Economic Statistics}, 32, 48--68.

\bibitem[{Aastveit et~al.(2017{\natexlab{b}})Aastveit, Ravazzolo, and van
  Dijk}]{Aastveit2015}
Aastveit, K.~A., Ravazzolo, F., and van Dijk, H.~K. (2017{\natexlab{b}}),
  \enquote{Combined density {N}owcasting in an uncertain economic environment,}
  \textit{Journal of Business \& Economic Statistics}, in press- published on
  line April 27, 2017.

\bibitem[{Amendola and Storti(2015)}]{amendola2015model}
Amendola, A. and Storti, G. (2015), \enquote{Model uncertainty and forecast
  combination in high-dimensional multivariate volatility prediction,}
  \textit{Journal of Forecasting}, 34, 83--91.

\bibitem[{Amisano and Geweke(2017)}]{AmisanoGeweke2017}
Amisano, G. and Geweke, J. (2017), \enquote{Prediction using several
  macroeconomic models,} \textit{Review of Economics and Statistics}, 99,
  912--925.

\bibitem[{Amisano and Giacomini(2007)}]{Amisano2007}
Amisano, G.~G. and Giacomini, R. (2007), \enquote{Comparing density forecasts
  via weighted likelihood ratio tests,} \textit{Journal of Business \& Economic
  Statistics}, 25, 177--190.

\bibitem[{Andersson and Karlsson(2008)}]{andersson2008bayesian}
Andersson, M.~K. and Karlsson, S. (2008), \enquote{Bayesian forecast
  combination for {VAR} models,} in \textit{Bayesian Econometrics (Advances in
  Econometrics, Volume 23)}, eds. S.~Chib, W., Griffiths, Koop, G., and
  Terrell, D., Emerald Group Publishing Limited, pp. 501--524.

\bibitem[{Bates and Granger(1969)}]{BatesGranger1969}
Bates, J.~M. and Granger, C. W.~J. (1969), \enquote{The combination of
  forecasts,} \textit{Operational Research Quarterly}, 20, 451--468.

\bibitem[{Benati and Surico(2008)}]{benati2008evolving}
Benati, L. and Surico, P. (2008), \enquote{Evolving {US} monetary policy and
  the decline of inflation predictability,} \textit{Journal of the European
  Economic Association}, 6, 634--646.

\bibitem[{Billio et~al.(2013)Billio, Casarin, Ravazzolo, and van
  Dijk}]{Billio2013}
Billio, M., Casarin, R., Ravazzolo, F., and van Dijk, H.~K. (2013),
  \enquote{Time-varying combinations of predictive densities using nonlinear
  filtering,} \textit{Journal of Econometrics}, 177, 213--232.

\bibitem[{Casarin et~al.(2015)Casarin, Grassi, Ravazzolo, and van
  Dijk}]{Casarin2015}
Casarin, R., Grassi, S., Ravazzolo, F., and van Dijk, H.~K. (2015),
  \enquote{Parallel sequential {M}onte {C}arlo for efficient density
  combination: {T}he {DeCo} {MATLAB} Toolbox,} \textit{Journal of Statistical
  Software, Articles}, 68, 1--30.

\bibitem[{Chan et~al.(2012)Chan, Koop, Leon-Gonzalez, and
  Strachan}]{chan2012time}
Chan, J.~C., Koop, G., Leon-Gonzalez, R., and Strachan, R.~W. (2012),
  \enquote{Time varying dimension models,} \textit{Journal of Business \&
  Economic Statistics}, 30, 358--367.

\bibitem[{Chen et~al.(2017)Chen, Irie, Banks, Haslinger, Thomas, and
  West}]{ChenETALdynets2016JASA}
Chen, X., Irie, K., Banks, D., Haslinger, R., Thomas, J., and West, M. (2017),
  \enquote{Scalable Bayesian modeling, monitoring and analysis of dynamic
  network flow data,} \textit{Journal of the American Statistical Association},
  in press- published online July 10, 2016, arXiv:1607.02655.

\bibitem[{Clemen(1989)}]{Clemen1989}
Clemen, R.~T. (1989), \enquote{Combining forecasts: A review and annotated
  bibliography,} \textit{International Journal of Forecasting}, 5, 559--583.

\bibitem[{Clemen and Winkler(1999)}]{ClemenWinkler1999}
Clemen, R.~T. and Winkler, R.~L. (1999), \enquote{Combining probability
  distributions from experts in risk analysis,} \textit{Risk Analysis}, 19,
  187--203.

\bibitem[{Clemen and Winkler(2007)}]{ClemenWinkler2007}
--- (2007), \enquote{Aggregating probability distributions,} in
  \textit{Advances in Decision Analysis: From Foundations to Applications},
  eds. W.~Edwards, R.~M. and von Winterfeldt, D., Cambridge University Press,
  chap.~9, pp. 154--176.

\bibitem[{Cogley and Sargent(2005)}]{Cogley2005}
Cogley, T. and Sargent, T.~J. (2005), \enquote{Drifts and volatilities:
  {M}onetary policies and outcomes in the post {WWII U.S.}} \textit{Review of
  Economic Dynamics}, 8, 262--302.

\bibitem[{Dawid et~al.(1995)Dawid, DeGroot, Mortera, Cooke, French, Genest,
  Schervish, Lindley, McConway, and Winkler}]{APDetal1995}
Dawid, A.~P., DeGroot, M.~H., Mortera, J., Cooke, R., French, S., Genest, C.,
  Schervish, M.~J., Lindley, D.~V., McConway, K.~J., and Winkler, R.~L. (1995),
  \enquote{Coherent combination of experts' opinions,} \textit{Test}, 4,
  263--313.

\bibitem[{Del~Negro et~al.(2016)Del~Negro, Hasegawa, and
  Schorfheide}]{Negro2016}
Del~Negro, M., Hasegawa, R.~B., and Schorfheide, F. (2016), \enquote{{Dynamic
  prediction pools: An investigation of financial frictions and forecasting
  performance},} \textit{Journal of Econometrics}, 192, 391--405.

\bibitem[{French(2011)}]{French2011}
French, S. (2011), \enquote{Aggregating expert judgement,} \textit{Revista de
  la Real Academia de Ciencias Exactas, Fisicas y Naturales (Serie A:
  Matematicas)}, 105, 181--206.

\bibitem[{Fr{\"u}hwirth-Schnatter(1994)}]{Schnatter1994}
Fr{\"u}hwirth-Schnatter, S. (1994), \enquote{Data augmentation and dynamic
  linear models,} \textit{Journal of Time Series Analysis}, 15, 183--202.

\bibitem[{Genest and Schervish(1985)}]{GenestSchervish1985}
Genest, C. and Schervish, M.~J. (1985), \enquote{Modelling expert judgements
  for {B}ayesian updating,} \textit{Annals of Statistics}, 13, 1198--1212.

\bibitem[{Geweke and Amisano(2012)}]{Geweke2012}
Geweke, J. and Amisano, G.~G. (2012), \enquote{Prediction with misspecified
  models,} \textit{The American Economic Review}, 102, 482--486.

\bibitem[{Geweke and Amisano(2011)}]{Geweke2011}
Geweke, J.~F. and Amisano, G.~G. (2011), \enquote{Optimal prediction pools,}
  \textit{Journal of Econometrics}, 164, 130--141.

\bibitem[{Gruber and West(2016)}]{GruberWest2016BA}
Gruber, L.~F. and West, M. (2016), \enquote{GPU-accelerated Bayesian learning
  in simultaneous graphical dynamic linear models,} \textit{Bayesian Analysis},
  11, 125--149.

\bibitem[{Gruber and West(2017)}]{GruberWest2017ECOSTA}
--- (2017), \enquote{Bayesian forecasting and scalable multivariate volatility
  analysis using simultaneous graphical dynamic linear models,}
  \textit{Econometrics and Statistics}, 3, 3--22, arXiv:1606.08291.

\bibitem[{Hall and Mitchell(2007)}]{HallMitchell2007}
Hall, S.~G. and Mitchell, J. (2007), \enquote{Combining density forecasts,}
  \textit{International Journal of Forecasting}, 23, 1--13.

\bibitem[{Hoogerheide et~al.(2010)Hoogerheide, Kleijn, Ravazzolo, Van~Dijk, and
  Verbeek}]{Hooger2010}
Hoogerheide, L., Kleijn, R., Ravazzolo, F., Van~Dijk, H.~K., and Verbeek, M.
  (2010), \enquote{Forecast accuracy and economic gains from {B}ayesian model
  averaging using time-varying weights,} \textit{Journal of Forecasting}, 29,
  251--269.

\bibitem[{Jore et~al.(2010)Jore, Mitchell, and Vahey}]{JoreMitchellVahey2010}
Jore, A.~S., Mitchell, J., and Vahey, S.~P. (2010), \enquote{Combining forecast
  densities from VARs with uncertain instabilities,} \textit{Journal of Applied
  Econometrics}, 25, 621--634.

\bibitem[{Kapetanios et~al.(2015)Kapetanios, Mitchell, Price, and
  Fawcett}]{Fawcett2014}
Kapetanios, G., Mitchell, J., Price, S., and Fawcett, N. (2015),
  \enquote{Generalised density forecast combinations,} \textit{Journal of
  Econometrics}, 188, 150--165.

\bibitem[{Kascha and Ravazzolo(2010)}]{Kascha2010}
Kascha, C. and Ravazzolo, F. (2010), \enquote{Combining inflation density
  forecasts,} \textit{Journal of Forecasting}, 29, 231--250.

\bibitem[{Koop et~al.(2010)Koop, Korobilis, et~al.}]{koop2010bayesian}
Koop, G., Korobilis, D., et~al. (2010), \enquote{Bayesian multivariate time
  series methods for empirical macroeconomics,} \textit{Foundations and Trends
  in Econometrics}, 3, 267--358.

\bibitem[{Koop et~al.(2009)Koop, Leon-Gonzalez, and Strachan}]{Koop2009}
Koop, G., Leon-Gonzalez, R., and Strachan, R.~W. (2009), \enquote{On the
  evolution of the monetary policy transmission mechanism,} \textit{Journal of
  Economic Dynamics and Control}, 33, 997--1017.

\bibitem[{Korobilis(2013)}]{korobilis2013var}
Korobilis, D. (2013), \enquote{{VAR} forecasting using {B}ayesian variable
  selection,} \textit{Journal of Applied Econometrics}, 28, 204--230.

\bibitem[{Lindley et~al.(1979)Lindley, Tversky, and Brown}]{DVL1979}
Lindley, D.~V., Tversky, A., and Brown, R.~V. (1979), \enquote{On the
  reconciliation of probability assessments,} \textit{Journal of the Royal
  Statistical Society (Series A: General)}, 142, 146--180.

\bibitem[{Lopes and Tsay(2011)}]{LopesTsay2011}
Lopes, H.~F. and Tsay, R.~S. (2011), \enquote{Particle filters and {B}ayesian
  inference in financial econometrics,} \textit{Journal of Forecasting}, 30,
  168--209.

\bibitem[{McAlinn and West(2017)}]{McAlinnWest2017}
McAlinn, K. and West, M. (2017), \enquote{Dynamic Bayesian predictive synthesis
  in time series forecasting,} \textit{Journal of Econometrics}, forthcoming,
  arXiv:1601.07463.

\bibitem[{Nakajima(2011)}]{nakajima2011time}
Nakajima, J. (2011), \enquote{Time-varying parameter VAR model with stochastic
  volatility: An overview of methodology and empirical applications,}
  \textit{Monetary and Economic Studies}, 29, 107--142.

\bibitem[{Nakajima and West(2013{\natexlab{a}})}]{Nakajima2010}
Nakajima, J. and West, M. (2013{\natexlab{a}}), \enquote{Bayesian analysis of
  latent threshold dynamic models,} \textit{Journal of Business \& Economic
  Statistics}, 31, 151--164.

\bibitem[{Nakajima and West(2013{\natexlab{b}})}]{Nakajima2011a}
--- (2013{\natexlab{b}}), \enquote{Bayesian dynamic factor models: {L}atent
  threshold approach,} \textit{Journal of Financial Econometrics}, 11,
  116--153.

\bibitem[{Pettenuzzo and Ravazzolo(2016)}]{Pettenuzzo2015}
Pettenuzzo, D. and Ravazzolo, F. (2016), \enquote{Optimal portfolio choice
  under decision-based model combinations,} \textit{Journal of Applied
  Econometrics}, 31, 1312--1332.

\bibitem[{Prado and West(2010)}]{Prado2010}
Prado, R. and West, M. (2010), \textit{Time Series: Modelling, Computation \&
  Inference}, Chapman \& Hall/CRC Press.

\bibitem[{Primiceri(2005)}]{Primiceri2005}
Primiceri, G.~E. (2005), \enquote{Time varying structural vector
  autoregressions and monetary policy,} \textit{Review of Economic Studies},
  72, 821--852.

\bibitem[{Sims(1993)}]{sims1993nine}
Sims, C.~A. (1993), \enquote{A nine-variable probabilistic macroeconomic
  forecasting model,} in \textit{Business Cycles, Indicators and Forecasting},
  eds. Stock, J.~H. and Watson, M.~W., University of Chicago Press, pp.
  179--212.

\bibitem[{Sims and Zha(1998)}]{sims1998bayesian}
Sims, C.~A. and Zha, T. (1998), \enquote{Bayesian methods for dynamic
  multivariate models,} \textit{International Economic Review}, 96, 949--968.

\bibitem[{Stock and Watson(1996)}]{stock1996evidence}
Stock, J.~H. and Watson, M.~W. (1996), \enquote{Evidence on structural
  instability in macroeconomic time series relations,} \textit{Journal of
  Business \& Economic Statistics}, 14, 11--30.

\bibitem[{Timmermann(2004)}]{Timmermann2004}
Timmermann, A. (2004), \enquote{Forecast combinations,} in \textit{Handbook of
  Economic Forecasting}, eds. Elliott, G., Granger, C. W.~J., and Timmermann,
  A., North Holland, vol.~1, chap.~4, pp. 135--196.

\bibitem[{West(1984)}]{West1984}
West, M. (1984), \enquote{Bayesian aggregation,} \textit{Journal of the Royal
  Statistical Society (Series A: General)}, 147, 600--607.

\bibitem[{West(1988)}]{West1988}
--- (1988), \enquote{Modelling expert opinion (with discussion),} in
  \textit{Bayesian Statistics 3}, eds. Bernardo, J.~M., DeGroot, M.~H.,
  Lindley, D.~V., and Smith, A. F.~M., Oxford University Press, pp. 493--508.

\bibitem[{West(1992)}]{West1992d}
--- (1992), \enquote{Modelling agent forecast distributions,} \textit{Journal
  of the Royal Statistical Society (Series B: Methodological)}, 54, 553--567.

\bibitem[{West and Crosse(1992)}]{West1992c}
West, M. and Crosse, J. (1992), \enquote{Modelling of probabilistic agent
  opinion,} \textit{Journal of the Royal Statistical Society (Series B:
  Methodological)}, 54, 285--299.

\bibitem[{West and Harrison(1997)}]{WestHarrison1997book2}
West, M. and Harrison, P.~J. (1997), \textit{Bayesian Forecasting \& Dynamic
  Models}, Springer Verlag, 2nd ed.

\bibitem[{Zellner(1962)}]{Zellner1962}
Zellner, A. (1962), \enquote{An efficient method of estimating seemingly
  unrelated regressions and tests for aggregation bias,} \textit{Journal of the
  American Statistical Association}, 57, 348--368.

\bibitem[{Zhou et~al.(2014)Zhou, Nakajima, and West}]{Zhou2012}
Zhou, X., Nakajima, J., and West, M. (2014), \enquote{Bayesian forecasting and
  portfolio decisions using dynamic dependent sparse factor models,}
  \textit{International Journal of Forecasting}, 30, 963--980.

\end{thebibliography}

\newpage

\begin{center}
{\Large Multivariate Bayesian Predictive Synthesis \\in Macroeconomic Forecasting}

\bigskip
{\large Kenichiro McAlinn, Knut Are Aastveit, Jouchi Nakajima \& Mike West}

\bigskip
{\Large  Tables and Figures}

\bigskip\bigskip
\end{center}

\begin{table}[htbp!]
\centering
\caption{US macroeconomic forecasting 2001/1-2015/12: 1-step ahead forecast evaluations for monthly US macroeconomic series over the 15 years 2001/1-2015/12,
 comparing mean squared forecast errors   and log predictive density ratios  for this $T=180$ months. The column $\%$ denotes improvements over the standard BPS model. Note: LPDR$_{1:T}$ is relative to the standard BPS model.}
\label{table:1step}
\begin{tabular}{lrrrrrr}
                                 & \multicolumn{6}{c}{MSFE$_{1:T}$}                                                                                                                    \\
\multicolumn{1}{l|}{1-step}      & Infl                             & \%                   & Wage                 & \%                   & Unemp                & \%                   \\ \hline
\multicolumn{1}{l|}{VAR(1)}      & 0.0141                           & $-$8.22                & 0.1444               & $-$35.91               & 0.0206               & 0.74                \\
\multicolumn{1}{l|}{VAR(12)}     & 0.0160                           & $-$22.93               & 0.1110               & $-$4.44                & 0.0230               & $-$10.73               \\
\multicolumn{1}{l|}{VAR(3)}      & 0.0147                           & $-$13.24               & 0.1105               & $-$3.96                & 0.0219               & $-$5.67                \\
\multicolumn{1}{l|}{VAR(1:3:9)}  & 0.0135                           & $-$3.76                & 0.1198               & $-$12.77               & 0.0222               & $-$7.18                \\
\multicolumn{1}{l|}{VAR(1:6:12)} & 0.0137                           & $-$5.14                & 0.1449               & $-$36.40               & 0.0215               & $-$3.70                \\
\multicolumn{1}{l|}{BMA} & 0.0146                           & $-$12.20                & 0.1111               & $-$4.53               & 0.0218               & $-$5.26                \\
\multicolumn{1}{l|}{BPS}         & 0.0130                           & $-$                    & 0.1063               & $-$                    & 0.0207               & $-$                    \\
                                 & \multicolumn{1}{l}{}             & \multicolumn{1}{l}{} & \multicolumn{1}{l}{} & \multicolumn{1}{l}{} & \multicolumn{1}{l}{} & \multicolumn{1}{l}{} \\
\multicolumn{1}{}{}            & \multicolumn{6}{c}{MSFE$_{1:T}$}                                                                                                                   \\
\multicolumn{1}{l|}{1-step}      & Cons                             & \%                   & Invest               & \%                   & Interest             & \%                   \\ \hline
\multicolumn{1}{l|}{VAR(1)}      & 0.3908                           & $-$3.50                & 13.2183              & $-$2.99                & 0.0275               & $-$35.07               \\
\multicolumn{1}{l|}{VAR(12)}     & 0.4697                           & $-$24.41               & 15.3571              & $-$19.65               & 0.0246               & $-$20.92               \\
\multicolumn{1}{l|}{VAR(3)}      & 0.3982                           & $-$5.48                & 13.3210              & $-$3.79                & 0.0211               & $-$3.74                \\
\multicolumn{1}{l|}{VAR(1:3:9)}  & 0.4049                           & $-$7.25                & 13.8918              & $-$8.24                & 0.0204               & $-$0.55                 \\
\multicolumn{1}{l|}{VAR(1:6:12)} & 0.3889                           & $-$3.02                & 13.4301              & $-$4.64                & 0.0228               & $-$12.02                \\
\multicolumn{1}{l|}{BMA} & 0.3971                           & $-$5.18                & 13.2145              & $-$2.96                & 0.0215               & $-$5.80                \\
\multicolumn{1}{l|}{BPS}         & 0.3775                           & $-$                    & 12.8346              & $-$                    & 0.0203               & $-$                    \\
                                 & \multicolumn{1}{l}{}             & \multicolumn{1}{l}{} & \multicolumn{1}{l}{} & \multicolumn{1}{l}{} & \multicolumn{1}{l}{} & \multicolumn{1}{l}{} \\
\multicolumn{1}{l|}{1-step}     & \multicolumn{1}{c}{LPDR$_{1:T}$} & \multicolumn{1}{l}{} & \multicolumn{1}{l}{} & \multicolumn{1}{l}{} & \multicolumn{1}{l}{} & \multicolumn{1}{l}{} \\ \cline{1-2}
\multicolumn{1}{l|}{VAR(1)}      & $-$77.25                         & \multicolumn{1}{l}{} & \multicolumn{1}{l}{} & \multicolumn{1}{l}{} & \multicolumn{1}{l}{} & \multicolumn{1}{l}{} \\
\multicolumn{1}{l|}{VAR(12)}     & $-$103.82                        & \multicolumn{1}{l}{} & \multicolumn{1}{l}{} & \multicolumn{1}{l}{} & \multicolumn{1}{l}{} & \multicolumn{1}{l}{} \\
\multicolumn{1}{l|}{VAR(3)}      & $-$31.00                         & \multicolumn{1}{l}{} & \multicolumn{1}{l}{} & \multicolumn{1}{l}{} & \multicolumn{1}{l}{} & \multicolumn{1}{l}{} \\
\multicolumn{1}{l|}{VAR(1:3:9)}  & $-$34.22                         & \multicolumn{1}{l}{} & \multicolumn{1}{l}{} & \multicolumn{1}{l}{} & \multicolumn{1}{l}{} & \multicolumn{1}{l}{} \\
\multicolumn{1}{l|}{VAR(1:6:12)} & $-$52.69                         & \multicolumn{1}{l}{} & \multicolumn{1}{l}{} & \multicolumn{1}{l}{} & \multicolumn{1}{l}{} & \multicolumn{1}{l}{} \\
\multicolumn{1}{l|}{BMA} & $-$32.48                         & \multicolumn{1}{l}{} & \multicolumn{1}{l}{} & \multicolumn{1}{l}{} & \multicolumn{1}{l}{} & \multicolumn{1}{l}{} \\
\multicolumn{1}{l|}{BPS}         & $-$                                & \multicolumn{1}{l}{} & \multicolumn{1}{l}{} & \multicolumn{1}{l}{} & \multicolumn{1}{l}{} & \multicolumn{1}{l}{}
\end{tabular}
\end{table}

\begin{table}[htbp!]
\centering
\caption{US macroeconomic forecasting 2001/1-2015/12: 12-step ahead forecast evaluations for monthly US macroeconomic series over the 15 years 2001/1-2015/12,
 comparing mean squared forecast errors   and log predictive density ratios  for this $T=180$ months. The column $\%$ denotes improvements over BPS(12). Note: LPDR$_{1:T}$ is relative to BPS(12).}
\label{table:12step}
\begin{tabular}{lrrrrrr}
\multicolumn{1}{}{}            & \multicolumn{6}{c}{MSFE$_{1:T}$}                                                                                                                   \\
\multicolumn{1}{l|}{12-step}     & Infl                             & \%                   & Wage                 & \%                   & Unemp                & \%                   \\ \hline
\multicolumn{1}{l|}{VAR(1)}      & 0.5317                           & $-$143.15                     & 0.4453                        & 19.50                       & 1.2028                       & $-$10.66                     \\
\multicolumn{1}{l|}{VAR(12)}     & 0.4272                           & $-$95.35                      & 0.7750                        & $-$40.12                      & 1.6918                       & $-$55.65                     \\
\multicolumn{1}{l|}{VAR(3)}      & 0.5789                           & $-$164.74                     & 0.5215                        & 5.72                        & 1.1788                       & $-$8.45                      \\
\multicolumn{1}{l|}{VAR(1:3:9)}  & 0.4541                           & $-$107.69                     & 1.1207                        & $-$102.62                     & 1.6353                       & $-$50.46                     \\
\multicolumn{1}{l|}{VAR(1:6:12)} & 0.5342                           & $-$144.30                     & 0.8934                        & $-$61.52                      & 1.3585                       & $-$24.99                     \\
\multicolumn{1}{l|}{BPS(12)}     & 0.2187                           & $-$                           & 0.5531                        & $-$                           & 1.0869                       & $-$                          \\
                                 & \multicolumn{1}{l}{}             & \multicolumn{1}{l}{} & \multicolumn{1}{l}{} & \multicolumn{1}{l}{} & \multicolumn{1}{l}{} & \multicolumn{1}{l}{} \\
\multicolumn{1}{}{}            & \multicolumn{6}{c}{MSFE$_{1:T}$}                                                                                                                   \\
\multicolumn{1}{l|}{12-step}     & Cons                             & \%                   & Invest               & \%                   & Interest             & \%                   \\ \hline
\multicolumn{1}{l|}{VAR(1)}                          & 7.2471                           & $-$23.21                 & 7067.67                  & $-$65.55                 & 5.5916                       & $-$68.74                 \\
\multicolumn{1}{l|}{VAR(12)}                        & 18.4145                          & $-$213.07                & 8824.02                  & $-$106.68                & 6.1707                       & $-$86.22                 \\
\multicolumn{1}{l|}{VAR(3)}                          & 7.3142                           & $-$24.35                 & 6378.42                 & $-$49.40                 & 4.8222                       & $-$45.52                 \\
\multicolumn{1}{l|}{VAR(1:3:9)}                    & 10.3823                          & $-$76.51                 & 9111.99                  & $-$113.43                & 4.6622                       & $-$40.69                 \\
\multicolumn{1}{l|}{VAR(1:6:12)}                       & 10.1116                          & $-$71.91                 & 10013.45                 & $-$134.54                & 7.4612                       & $-$125.16                \\
\multicolumn{1}{l|}{BPS(12)}                          & 5.8818                           & $-$                      & 4269.33                  & $-$                      & 3.3137                       & $-$                      \\
                                 & \multicolumn{1}{l}{}             & \multicolumn{1}{l}{} & \multicolumn{1}{l}{} & \multicolumn{1}{l}{} & \multicolumn{1}{l}{} & \multicolumn{1}{l}{} \\
\multicolumn{1}{l|}{12-step}     & \multicolumn{1}{c}{LPDR$_{1:T}$} & \multicolumn{1}{l}{} & \multicolumn{1}{l}{} & \multicolumn{1}{l}{} & \multicolumn{1}{l}{} & \multicolumn{1}{l}{} \\ \cline{1-2}
\multicolumn{1}{l|}{VAR(1)}                           & $-$119.05                        &                        &                            &                        &                              &                        \\
\multicolumn{1}{l|}{VAR(12)}                         & $-$535.09                        &                        &                            &                        &                              &                        \\
\multicolumn{1}{l|}{VAR(3)}                           & $-$366.85                        &                        &                            &                        &                              &                        \\
\multicolumn{1}{l|}{VAR(1:3:9)}                      & $-$463.46                        &                        &                            &                        &                              &                        \\
\multicolumn{1}{l|}{VAR(1:6:12)}                       & $-$361.20                        &                        &                            &                        &                              &                        \\
\multicolumn{1}{l|}{BPS(12)}    & $-$                                &                        &                            &                        &                              &
\end{tabular}
\end{table}

\begin{table}[]
\centering
\caption{US macroeconomic forecasting 2001/1-2015/12: 24-step ahead forecast evaluations for monthly US macroeconomic series over the 15 years 2001/1-2015/12,
 comparing mean squared forecast errors   and log predictive density ratios  for this $T=180$ months. The column $\%$ denotes improvements over BPS(24). Note: LPDR$_{1:T}$ is relative to BPS(24).}
\label{table:24step}
\begin{tabular}{lrrrrrr}
                                 & \multicolumn{6}{c}{MSFE$_{1:T}$}                                                                                                                    \\
\multicolumn{1}{l|}{24-step}     & Infl                             & \%                   & Wage                 & \%                   & Unemp                & \%                   \\ \hline
\multicolumn{1}{l|}{VAR(1)}      & 3.9536       & $-$331.10 & 2.4117     & 7.71     & 16.46  & $-$55.68   \\
\multicolumn{1}{l|}{VAR(12)}     & 2.7373       & $-$198.47 & 4.5054     & $-$72.41   & 18.32  & $-$73.28   \\
\multicolumn{1}{l|}{VAR(3)}      & 3.8504       & $-$319.85 & 3.1877     & $-$21.98   & 13.78  & $-$30.35   \\
\multicolumn{1}{l|}{VAR(1:3:9)}  & 4.8627       & $-$430.23 & 8.8723     & $-$239.52  & 21.06  & $-$99.17   \\
\multicolumn{1}{l|}{VAR(1:6:12)} & 4.4141       & $-$381.32 & 8.4162     & $-$222.06  & 16.99  & $-$60.65   \\
\multicolumn{1}{l|}{BPS(24)}     & 0.9171       & $-$       & 2.6132     & $-$        & 10.58  & $-$        \\
                                 & \multicolumn{1}{l}{}             & \multicolumn{1}{l}{} & \multicolumn{1}{l}{} & \multicolumn{1}{l}{} & \multicolumn{1}{l}{} & \multicolumn{1}{l}{} \\
                                 & \multicolumn{6}{c}{MSFE$_{1:T}$}                                                                                                                    \\
\multicolumn{1}{l|}{24-step}     & Cons                             & \%                   & Invest               & \%                   & Interest             & \%                   \\ \hline
\multicolumn{1}{l|}{VAR(1)}      & 56.27      & $-$104.54 & 51937 & $-$776.23  & 31.68  & $-$480.56  \\
\multicolumn{1}{l|}{VAR(12)}     & 118.09     & $-$329.23 & 38151 & $-$543.65  & 25.89  & $-$374.58  \\
\multicolumn{1}{l|}{VAR(3)}      & 46.80      & $-$70.09  & 39671 & $-$569.30  & 21.84  & $-$300.31  \\
\multicolumn{1}{l|}{VAR(1:3:9)}  & 78.73      & $-$186.15 & 80278 & $-$1254.37 & 25.41  & $-$365.71  \\
\multicolumn{1}{l|}{VAR(1:6:12)} & 72.54      & $-$163.67 & 86671 & $-$1362.23 & 62.16  & $-$1039.32 \\
\multicolumn{1}{l|}{BPS(24)}     & 27.51      & $-$       & 5927  & $-$        & 5.46   & $-$        \\
                                 & \multicolumn{1}{l}{}             & \multicolumn{1}{l}{} & \multicolumn{1}{l}{} & \multicolumn{1}{l}{} & \multicolumn{1}{l}{} & \multicolumn{1}{l}{} \\
\multicolumn{1}{l|}{24-step}     & \multicolumn{1}{c}{LPDR$_{1:T}$} & \multicolumn{1}{l}{} & \multicolumn{1}{l}{} & \multicolumn{1}{l}{} & \multicolumn{1}{l}{} & \multicolumn{1}{l}{} \\ \cline{1-2}
\multicolumn{1}{l|}{VAR(1)}      & $-$445.81                        & \multicolumn{1}{l}{} & \multicolumn{1}{l}{} & \multicolumn{1}{l}{} & \multicolumn{1}{l}{} & \multicolumn{1}{l}{} \\
\multicolumn{1}{l|}{VAR(12)}     & $-$489.98                        & \multicolumn{1}{l}{} & \multicolumn{1}{l}{} & \multicolumn{1}{l}{} & \multicolumn{1}{l}{} & \multicolumn{1}{l}{} \\
\multicolumn{1}{l|}{VAR(3)}      & $-$462.48                        & \multicolumn{1}{l}{} & \multicolumn{1}{l}{} & \multicolumn{1}{l}{} & \multicolumn{1}{l}{} & \multicolumn{1}{l}{} \\
\multicolumn{1}{l|}{VAR(1:3:9)}  & $-$808.31                        & \multicolumn{1}{l}{} & \multicolumn{1}{l}{} & \multicolumn{1}{l}{} & \multicolumn{1}{l}{} & \multicolumn{1}{l}{} \\
\multicolumn{1}{l|}{VAR(1:6:12)} & $-$804.49                       & \multicolumn{1}{l}{} & \multicolumn{1}{l}{} & \multicolumn{1}{l}{} & \multicolumn{1}{l}{} & \multicolumn{1}{l}{} \\
\multicolumn{1}{l|}{BPS(24)}     & $-$                                & \multicolumn{1}{l}{} & \multicolumn{1}{l}{} & \multicolumn{1}{l}{} & \multicolumn{1}{l}{} & \multicolumn{1}{l}{}
\end{tabular}
\end{table}

\clearpage


 \clearpage


\clearpage

\clearpage

\clearpage

\begin{figure}[htbp!]
\centering
\includegraphics[width=0.75\textwidth]{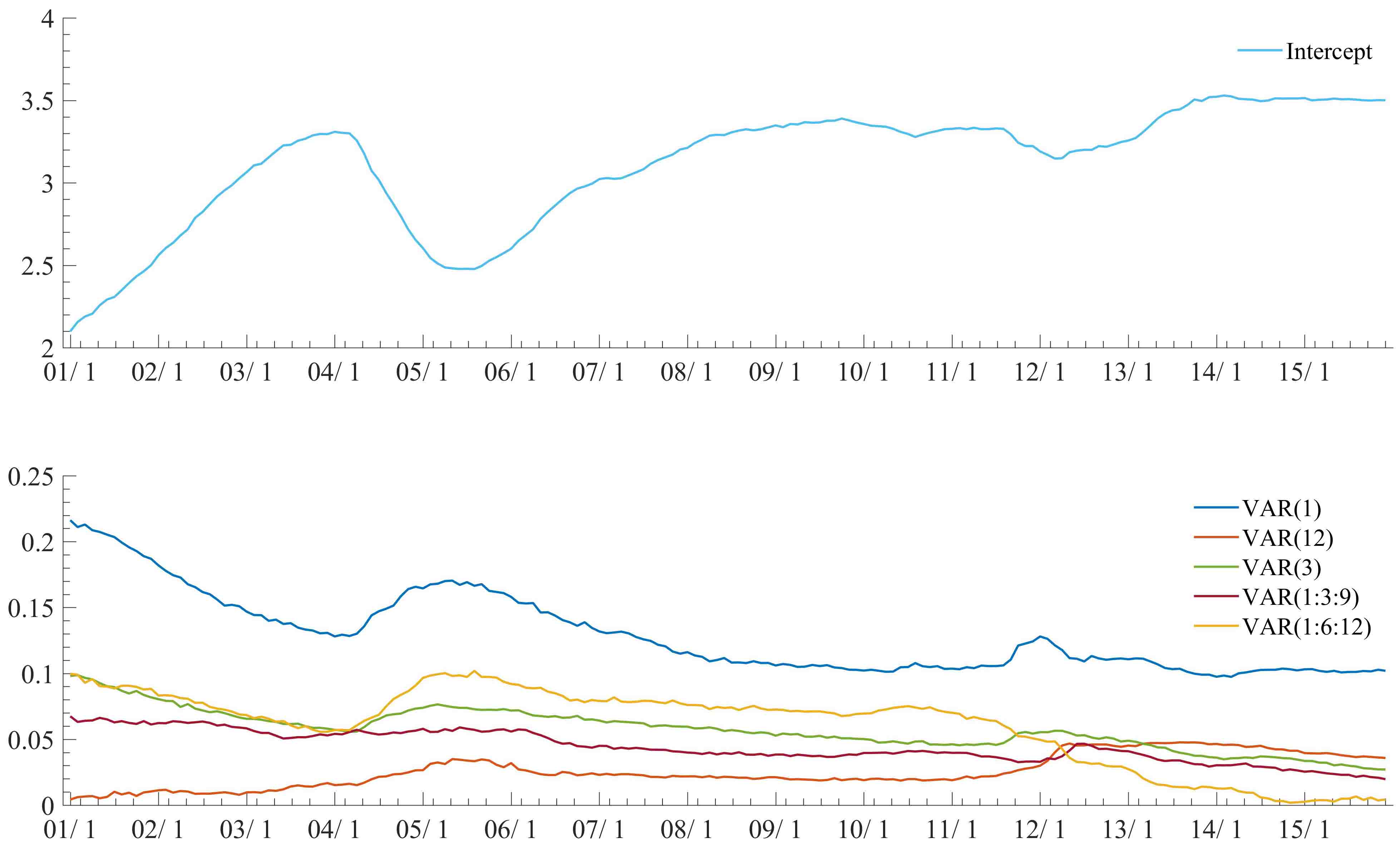}
\caption{US macroeconomic forecasting 2001/1-2015/12:  On-line posterior means of BPS(24) model
  coefficients for inflation $(p)$ sequentially computed at each of the $t=\seq1{180}$ months.
\label{24coeff1}}
\end{figure}

\begin{figure}[htbp!]
\centering
\includegraphics[width=0.75\textwidth]{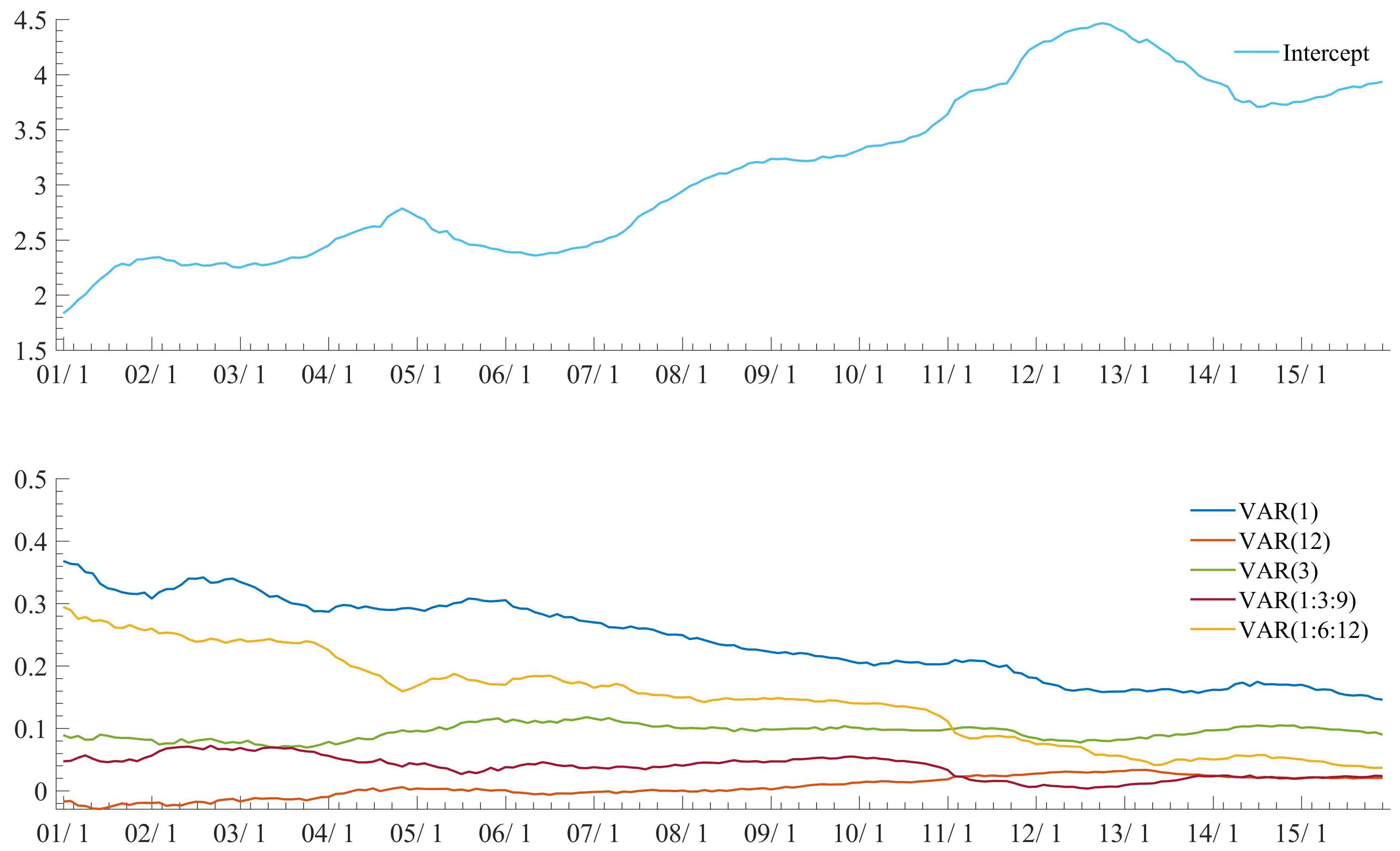}
\caption{US macroeconomic forecasting 2001/1-2015/12:  On-line posterior means of BPS(24)  model
  coefficients for wage $(w)$sequentially computed at each of the $t=\seq1{180}$ months.
\label{24coeff2}}
\end{figure}

\begin{figure}[htbp!]
\centering
\includegraphics[width=0.75\textwidth]{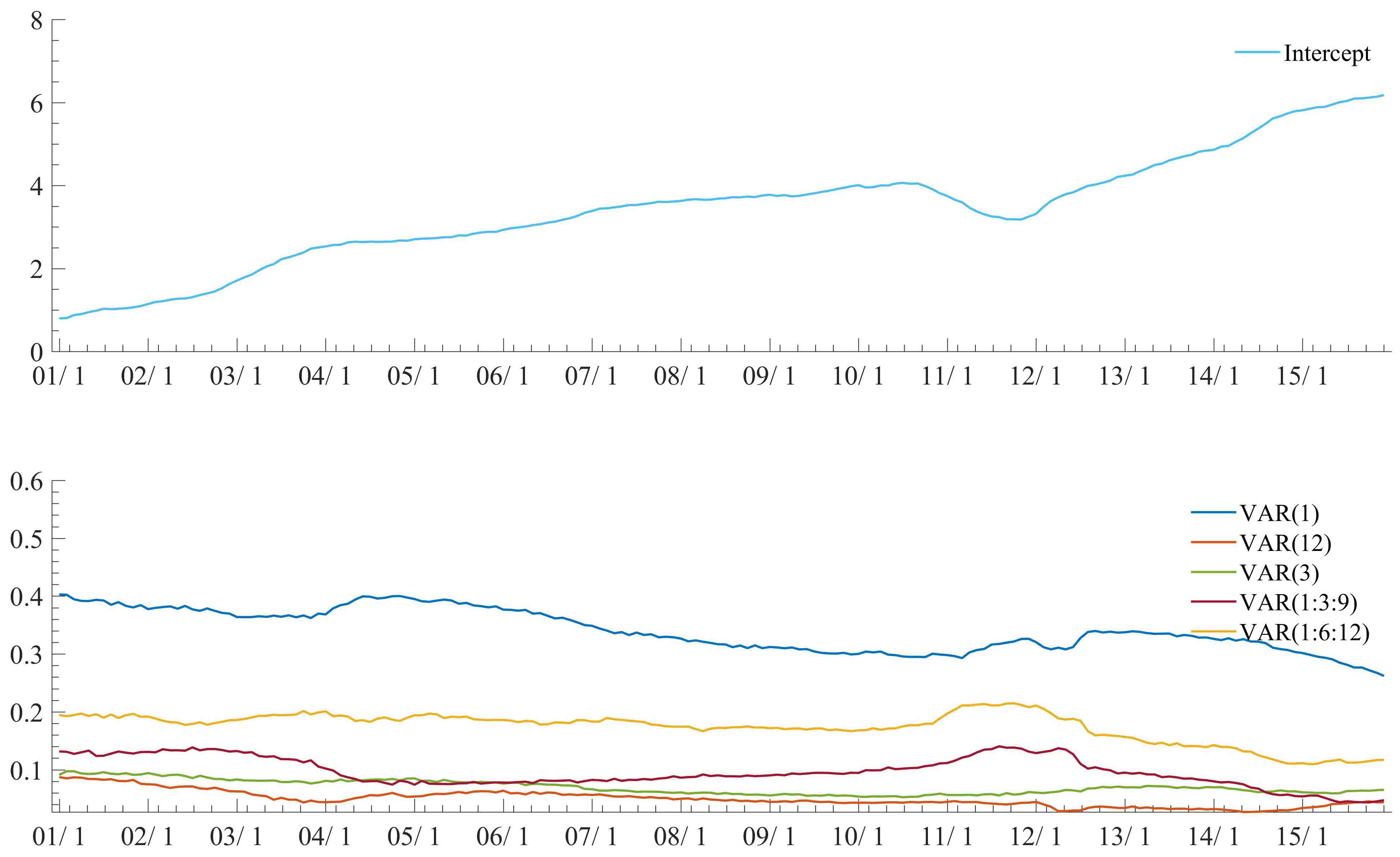}
\caption{US macroeconomic forecasting 2001/1-2015/12:  On-line posterior means of BPS(24)  model
  coefficients for unemployment $(u)$ sequentially computed at each of the $t=\seq1{180}$ months.
\label{24coeff3}}
\end{figure}

\begin{figure}[htbp!]
\centering
\includegraphics[width=0.75\textwidth]{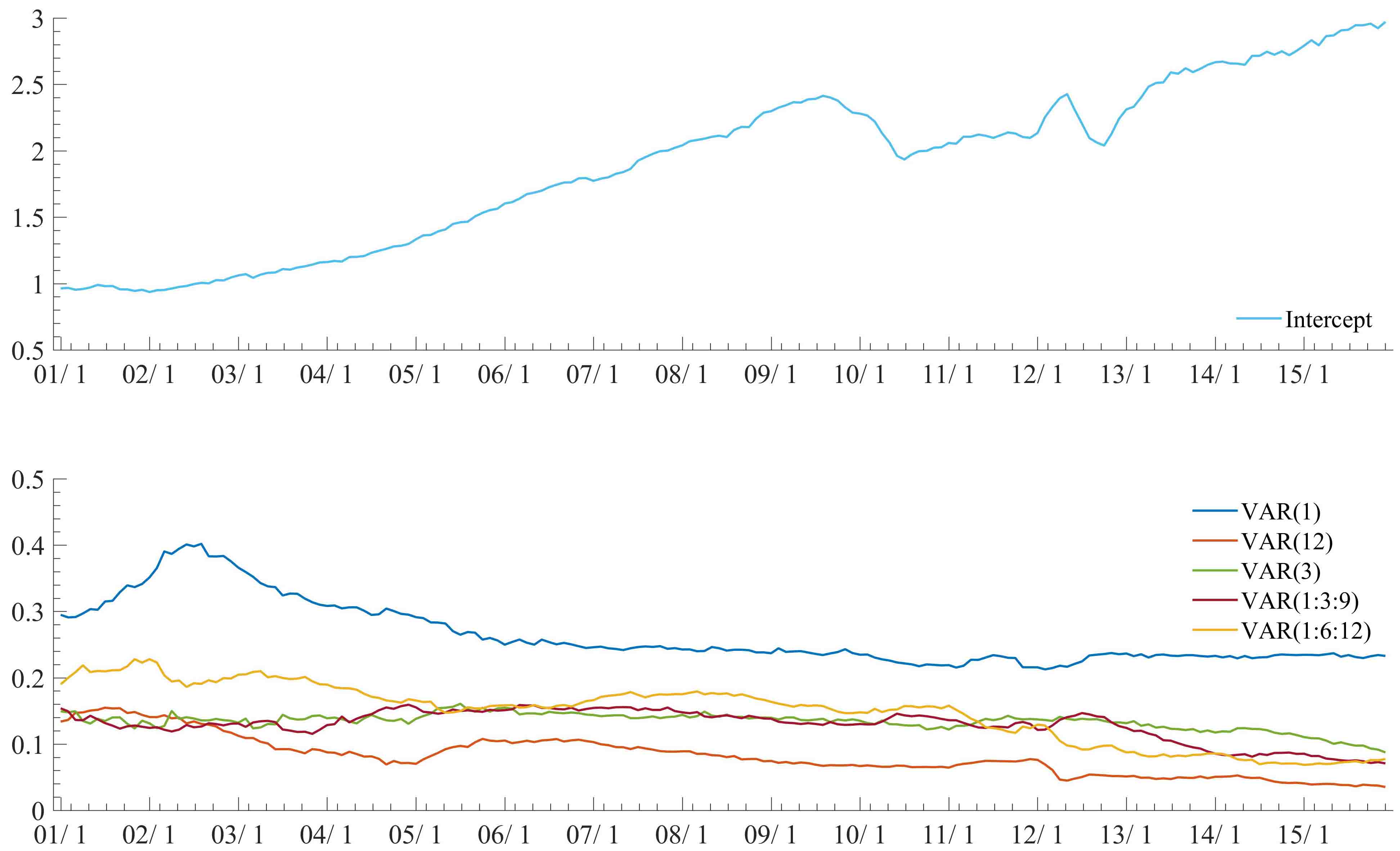}
\caption{US macroeconomic forecasting 2001/1-2015/12:  On-line posterior means of BPS(24)  model
  coefficients for consumption $(c)$ sequentially computed at each of the $t=\seq1{180}$ months.
\label{24coeff4}}
\end{figure}

\begin{figure}[htbp!]
\centering
\includegraphics[width=0.75\textwidth]{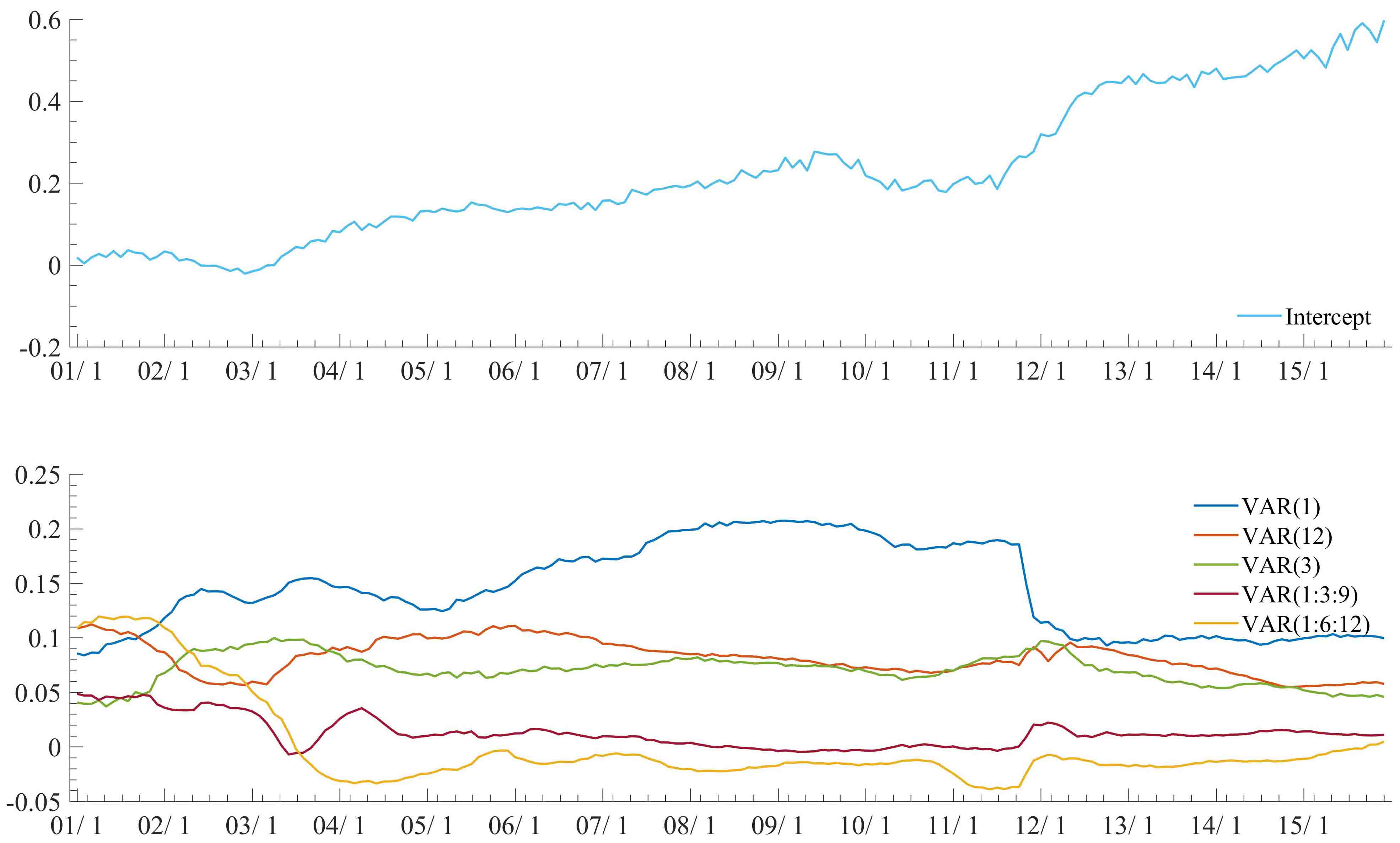}
\caption{US macroeconomic forecasting 2001/1-2015/12:  On-line posterior means of BPS(24)  model
  coefficients for investment $(i)$ sequentially computed at each of the $t=\seq1{180}$ months.
\label{24coeff5}}
\end{figure}

\begin{figure}[htbp!]
\centering
\includegraphics[width=0.75\textwidth]{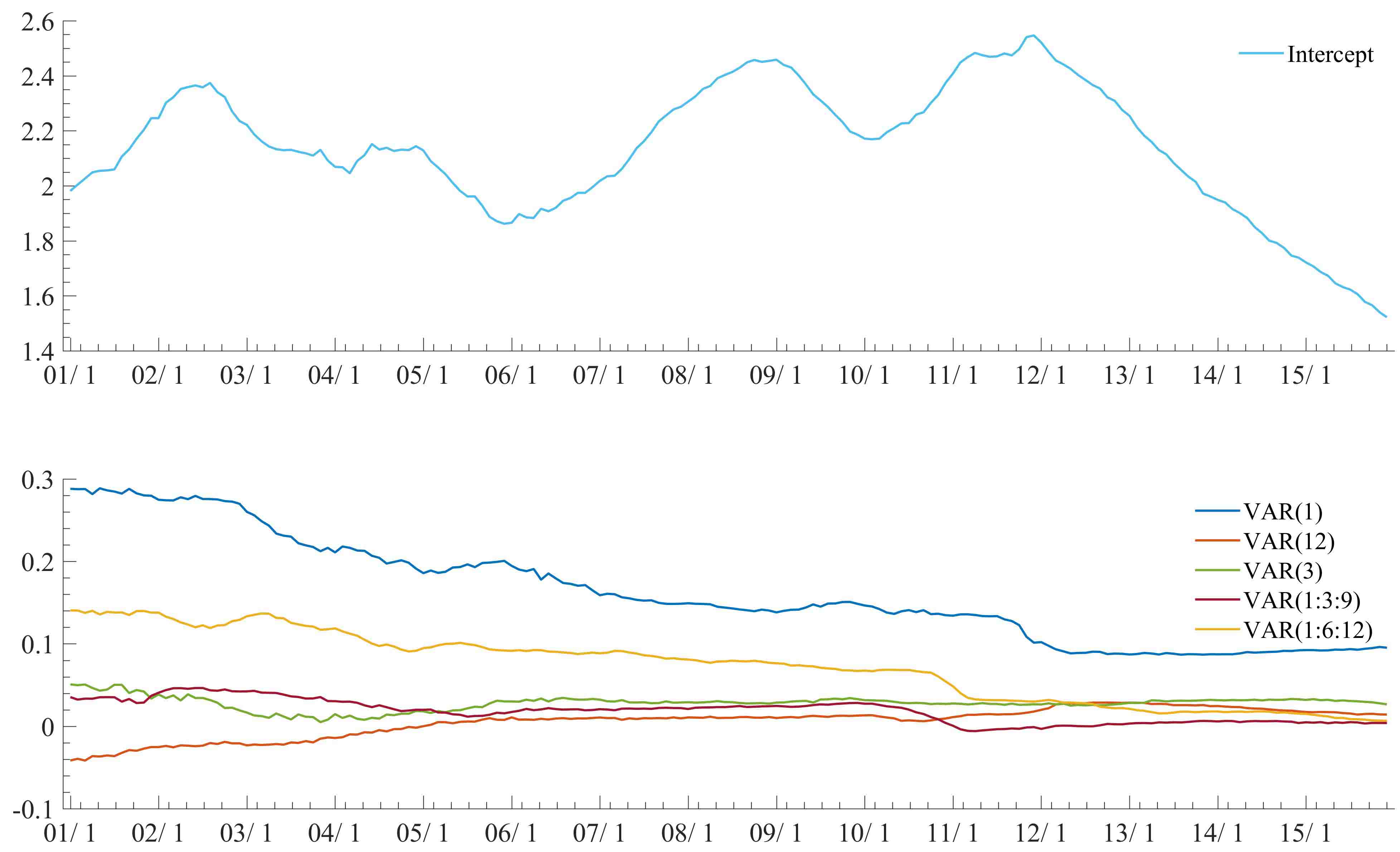}
\caption{US macroeconomic forecasting 2001/1-2015/12:  On-line posterior means of BPS(24)  model
  coefficients for interest rate $(r)$ sequentially computed at each of the $t=\seq1{180}$ months.
\label{24coeff6}}
\end{figure}


\clearpage
\setcounter{page}{1}
\begin{center}
{\Large Multivariate Bayesian Predictive Synthesis\\ in Macroeconomic Forecasting}

\bigskip
{\large Kenichiro McAlinn, Knut Are Aastveit, Jouchi Nakajima, \& Mike West}

\bigskip
{\Large  Supplementary Material}

\bigskip\bigskip
\end{center}

\appendix

 \section{Appendix: Summary of MCMC for Dynamic BPS\label{supp:comp}}

\subsection{Overview and Initialization}
This appendix summarizes algorithmic details of implementation of the MCMC computations for dynamic BPS model fitting
of Section~\ref{sec:comp}.  This involves a   standard set of steps in a customized
three-component block Gibbs sampler: the first component samples the latent agent states, the second,
the second samples the dynamic BPS model states/parameters, and the third component samples the observation variance.
The latter two involves a modified FFBS algorithm central to MCMC in all conditionally normal
DLMs~(\citealt{Schnatter1994}; \citealt[][Sect 15.2]{WestHarrison1997book2}; \citealt[][Sect 4.5]{Prado2010}).

In our sequential learning and forecasting context, the full MCMC analysis is
performed anew at each time point as time evolves and new data are observed.
We detail MCMC steps for analysis based on data over times $t=\seq1T$ for any chosen $T$.

Standing at time $t=0$, the decision maker has initial information summarized via
in terms of $\btheta_0\sim N(\m_0, \C_0)$ and $\V_0\sim IW(n_0, \D_0),$ independently.
Here the $q\times q$ variance matrix $\V_0$ has the inverse Wishart distribution
with $n>0$ degrees of freedom and  prior \lq\lq sum-of-squares'' matrix $\D_0.$
Equivalently, the precision matrix $\V_0^{-1}\sim W(h_0,\D_0^{-1})$, the
Wishart distribution with $h_0=n_0+q-1$  and mean $h_0 \D_0^{-1}$ so that the
initial estimate  $\D_0/h_0$ is the prior harmonic mean of $\V_0.$
Model specification is completed with two chosen discount factors:
$\beta$, defining the extent of time variation in the evolution of the states $\btheta_t$, and
$\delta$ defining levels of variation in the evolution of the volatility matrices $\V_t$.

At time $T,$ the decision maker has accrued information $\{ y_{\seq1T}, \mH_{\seq1T}\}$.
The MCMC analysis is then run iteratively as follows.

\paragraph{Initialization:} First, initialize by setting
\begin{equation}\label{eq:mBPSalphadynamicFtheta}
\F_t=\left(
    \begin{array}{cccccccc}
      1 &  \f_{t1}'  & 0 & \zero' & \cdots & \cdots  & 0 & \zero'  \\
      0 & \zero' &  1& \f_{t2}' &  &  & &  \vdots \\
      \vdots &  &  &  & \ddots &  &  & \vdots  \\
      0 & \cdots & \cdots & \cdots & \cdots  & \cdots & 1& \f_{tq}'
    \end{array}
  \right),
\end{equation}
 for each $t=\seq1T$, with elements set  at some chosen initial values of the
latent agent states.
Initial values can be chosen arbitrarily. One obvious and appropriate choice-- our recommended default choice-- is to simply generate agent states from their priors, i.e.,
from the agent forecast distributions,  $\x_{tj} \sim h_{tj}(\x_{tj})$ independently for all $t=\seq1T$ and $j=\seq1J$.
This is easily implemented in cases when the agent forecasts are T or normal distributions, or can be otherwise directly
sampled; we use this in our analyses reported in the paper, and recommend it as standard.
An obvious alternative initialization is to simply set $\x_{tj}=\y_t$ for each $t,j$, though we prefer to initialize with
some inherent dispersion in starting values.  Ultimately, since the MCMC is rapidly convergent, choice of initial values is not critical.
Given initial values of agent factor vector $\x_{tj} = (x_{t1j},x_{t2j},...,x_{tqj})'$ for each agent $j=1{:}J$ and each time $t,$  the
$\F_t$ matrices are initialized with series-specific row entries from
$\f_{tr}=(x_{tr1},x_{tr2},...,x_{trJ})'$ for each $r=1{:}q.$

\subsection{Three Sampling Steps in Each MCMC Iterate}

Following initialization,  the MCMC  iterates repeatedly to resample three sets of conditional posteriors to generate the
MCMC samples from the target posterior $p(\X_{\seq1T},\btheta_{\seq1T},\V_{\seq1T}|\y_{\seq1T}, \mH_{\seq1T}).$   These   conditional posteriors
and algorithmic details of their simulation are as follows.

\subsubsection{Per MCMC Iterate Step 1: Sampling BPS DLM parameters $\btheta_{\seq1T}$ }
Conditional on any values of
the latent agent states and observation error,  we are in the setting of a conditionally normal, multivariate DLM with the agent states as
known predictors based on their specific values.  The BPS  DLM form,
\begin{align*}
		\y_t&=\F_t\btheta_t+\bnu_t, \quad \nu_t\sim N(0,\V_t), \label{eq:DLMa} \\
	\btheta_t&=\btheta_{t-1}+\bomega_t, \quad \bomega_t\sim N(0, \W_t),
\end{align*}
has known elements $\F_t,\W_t$ and specified initial prior at $t=0.$ The implied conditional posterior
for  $\btheta_{\seq1T}$  then does not depend on  $\mH_{\seq1T}$, reducing to
$p(\btheta_{\seq1T}|\X_{\seq1T},\V_{\seq1T},\y_{\seq1T}).$  This is simulated using the efficient and standard FFBS
algorithm 
(e.g.~\citealt{Schnatter1994}; \citealt[][Sect 15.2]{WestHarrison1997book2}; \citealt[][Sect 4.5]{Prado2010}).
In detail, this proceeds as follows.

\begin{itemize}
\item[]{\em\bf Forward filtering:} For each $t=\seq1T$ in sequence, perform the standard one-step
filtering updates to compute and save the sequence of sufficient statistics for the on-line posteriors
$p(\btheta_t|\X_{\seq1t},\V_{\seq1t},\y_{\seq1t})$ at each $t.$ The summary technical details are as follows:
\begin{itemize}
	\item[1.]{\em Time $t-1$ posterior:}
		\begin{align*}
		\btheta_{t-1}|\X_{\seq1{t-1}},\V_{\seq1{t-1}},\y_{\seq1{t-1}}&\sim N(\m_{t-1}, \C_{t-1}),
		\end{align*}
		with point estimate $\m_{t-1}$ of $\btheta_{t-1}$.
	\item[2.]{\em Update to time $t$ prior:}
		\begin{align*}
		\btheta_{t}|\X_{\seq1{t-1}},\V_{\seq1{t-1}},\y_{\seq1{t-1}}&\sim N(\m_{t-1}, \R_t)
		\quad\textrm{with}\quad \R_{t}=\C_{t-1}/\delta,
		\end{align*}
		with (unchanged) point estimates $\m_{t-1}$ of $\btheta_{t}$,  but with
		increased uncertainty relative to the time $t-1$ posteriors, the level of increased uncertainty
			being defined by the discount factors.
	\item[3.]{\em  1-step predictive distribution:}
		$\y_t |\X_{\seq1t},\V_{\seq1t},\y_{\seq1{t-1}} \sim T_{\beta n_{t-1}}(\f_t,\Q_t)$ where
		$$\f_t=\F_t\m_{t-1}\quad \textrm{and}\quad \Q_t=\F_t\R_t\F_t'+\V_t.$$
	\item[4.]{\em  Filtering update to time $t$ posterior:}
		\begin{align*}
		\btheta_{t}|\V_{\seq1t},\X_{\seq1{t}},\y_{\seq1{t}}&\sim N(\m_{t}, \C_{t}),
		\end{align*}
		 with defining parameters  $\m_t=\m_{t-1}+\A_t \e_t$ and $ 	\C_{t}=\R_t-\A_t\Q_t\A_t',$
		based on  1-step forecast error  $ \e_t=\y_t-\f_t$ and the state adaptive coefficient vector (a.k.a. \lq\lq Kalman gain'')
		$\A_t=\R_t\F_t'\Q_t^{-1}$.
\end{itemize}
	\item[]{\em\bf Backward sampling:}  Having run the forward filtering analysis up to time $T,$ the
	 backward sampling proceeds as follows.
	 \begin{itemize}
	 \item[a.]{\em At time $T$:} Simulate $\btheta_T$ from the final multivariate normal posterior
	  $$p(\btheta_T| \X_{\seq1{T}},\V_{\seq1{T}},\y_{\seq1{T}}) = N(\m_T,\C_T).$$
	 \item[b.]{\em Recurse back over times $t=T-1, T-2, \ldots, 0:$}  At each time $t,$  simulate the state $\btheta_t$ from the conditional posterior
	 			$p(\btheta_{t}|\btheta_{t+1},\X_{\seq1{t}},\V_{\seq1t},\y_{\seq1t})$; this is multivariate normal with mean
	 			vector $\m_{t}+\delta(\btheta_{t+1}-\m_{t})$ and variance matrix
	 			$ \C_{t} (1-\delta).$ 	 			
	 \end{itemize}
\end{itemize}

\subsubsection{Per MCMC Iterate Step 2:   Sampling BPS DLM parameters $\V_{\seq1T}$}

Conditional on the sampled values of the BPS DLM parameters $\btheta_{\seq1T}$ and  latent agent states $\X_{\seq1T}$, the next step in the MCMC iterate samples the full conditional posterior of the sequence of volatility matrices, generating a draw from
$ p( \V_{\seq1T}|  \X_{\seq1T},\btheta_{\seq1T},  \y_{\seq1T})$.

\begin{itemize}
\item[]{\em\bf Forward filtering:} For each $t=\seq1T$ in sequence,  update and save the forward filtering
summaries $(n_t,\D_t)$  of on-line posteriors
\begin{equation*}	
 \V_t|\X_{\seq1t},\btheta_{\seq1t},\y_{\seq1t} \sim IW(n_t,\D_t),
\end{equation*}
given by $n_t = h_t-q+1$ where $h_t=\beta h_{t-1} + 1,$ and
 $\D_t = \beta\D_{t-1} + (\y_t-\F_t^{\prime}\btheta_t)(\y_t-\F_t^{\prime}\btheta_t)'$.
 \item[]{\em\bf Backwards sampling:}
 Having run the forward filtering analysis up to time $T,$ the
	 backward sampling proceeds as follows.
	 \begin{itemize}
	 \item[a.]{\em At time $T$:} Simulate $\V_T$ from the final inverse Wishart posterior
	  $IW(n_T,\D_T).$
	  	 \item[b.]{\em Recurse back over times $t=T-1, T-2, \ldots, 0:$}  At time $t,$ sample
	 	$\V_t$ from the conditional posterior $p(\V_t|\V_{t+1},\X_{\seq1{t}},\btheta_{\seq1t},\y_{\seq1t})$.
	 		Algorithmically, this is achieved via
	 			\begin{equation*}	
 					\V^{-1}_t = \beta\V^{-1}_{t+1}+\bUpsilon_t \quad\textrm{where}\quad \bUpsilon_t \sim W((1-\beta)h_t,\D_t^{-1}),
				\end{equation*}	 and where the $\Upsilon_t$ are independent over $t.$
	 \end{itemize}
 \end{itemize}

\subsubsection{Per MCMC Iterate Step 3:   Sampling the latent agent states $\X_{\seq1T}$}

  Conditional on most recently
sampled values of
the BPS DLM parameters $\bPhi_{\seq1T},$   the MCMC iterate completes with resampling of the
latent agent states from their full conditional posterior
$ p( \X_{\seq1t} |  \bPhi_{\seq1t}, \y_{\seq1t}, \mH_{\seq1t} ).$    It is immediate that the $\X_t$ are
conditionally independent over time $t$ in this conditional distribution, with time $t$
conditionals
\begin{equation}\label{app:ccforx}
p( \X_t|  \bPhi_t, \y_t, \mH_t) \propto N(\y_t|\F_t\btheta_t, \V_t) \prod_{j=\seq1J} h_{tj}(\x_{tj}).
 \end{equation}
Several comments are relevant to studies with different forms of the agent forecast densities.
\begin{itemize}
\item[1.] {\em Multivariate normal agent forecast densities:}
In cases when each of the agent forecast densities is normal,  the posterior in eqn.~(\ref{app:ccforx})
yields a multivariate normal distribution for vectorized $\X_t.$  Computation of its defining parameters and then
drawing a new sample vector $\X_t$  are trivial.

\item[2.] In some cases, as in our study in this paper, the agent forecast densities will be those of Student T
distributions. In our case study the five agents represent conjugate exchangeable dynamic linear models in which
all forecast densities are multivariate T, with parameters varying over time and with step-ahead forecast horizon.
In such cases,   standard Bayesian augmentation methods apply to enable simulation.    Each multivariate T distribution
is expressed as a scale mixture of multivariate normals,   with the mixing scale parameters introduced as inherent
latent variables with inverse gamma distributions.
This expansion of the parameter space makes the multivariate T distributions conditional multivariate normals, and the
mixing scales are resampled (from implied conditional posterior inverse gamma distributions)
each MCMC iterate along with the agent states. This is again a standard MCMC approach
and much used in Bayesian time series, as in other areas (e.g.~\citealt{Schnatter1994}; \citealt[][Chap 15]{WestHarrison1997book2}).
Then, conditional on the current values of these latent scales,  sampling the $\X_t$ reduces technically to that
conditional normals above.

Specifically, suppose that  $h_{tj}(\x_{tj})$ is density of the normal $ T_{n_{tj}}(\h_{tj},\H_{tj})$;
the notation means that $ (\x_{tj}-\h_{tj})/\sqrt{\H_{tj}}$ has a standard
multivariate Student T distribution with $n_{tj}$ degrees of freedom.   Then latent scale factors $\phi_{tj}$ exist such that:
{\em (i)} conditional on $\phi_{tj},$  latent agent factor $\x_{tj}$ has a conditional multivariate normal density
$\x_{tj}|\phi_{tj} \sim N(\h_{tj},\H_{tj}/\phi_{tj})$ independently over $t,j;$ {\em (ii)}  the $\phi_{tj}$ are
independent over $t,j$ with gamma distributions, $\phi_{tj} \sim G(n_{tj}/2,n_{tj}/2).$
Then, at each MCMC step, the above normal update for latent agent states is replaced by normal
simulations conditional on the $\phi_{tj}.$   Following this, we resample values of the
$\phi_{tj}$ from their trivially  implied conditional gamma posteriors.

%
\item[3.] In some cases, agent densities may be more elaborate mixtures of normals, such as (discrete or
continuous) location and/or scale mixtures that represent asymmetric distributions.  The same augmentation
strategy can be applied in such cases, with augmented parameters including location shifts in place of, or
in addition to, scale shifts.

\item[4.]  In other cases,  we may be able to directly simulate the agent forecast distributions and evaluate
forecast density functions at any point,   but do not have access to analytic forms.  One class of examples
is when the agents are simulation models, e.g., DSGE models. Another involves forecasts in terms of
histograms.  In such cases,   MCMC will proceed using some form of 	
  Metropolis-Hastings algorithm, or accept/reject methods, or importance sampling for the
  latent agent states.

  For example,  suppose we only have access to simulations from the agent forecast distributions,
   in terms of $I$ independent draws from each collated in the simulated matrix $\X_t^{(i)}$ for
    $i=\seq1I.$  We can apply importance sampling as follows:
  {\em (a)} compute the marginal likelihood
  values  $p( \y_t|  \bPhi_t, \X_t^{(i)}, \mH_t)$ for each $i=\seq1I$; {\em (b)} compute and normalize the
   implied importance sampling weights $w_{ti} \propto N( \y_t|  \bPhi_t, \X_t^{(i)}, \mH_t), $
  and then {\em (c)} resample latent agent states for this MCMC stage according to the probabilities these
   weights define.
\end{itemize}

 \newpage

\section{Appendix: Additional  Graphical Summaries from Macroeconomic Analysis  \label{supp:macroadd}}
\renewcommand{\thefigure}{C\arabic{figure}}\setcounter{figure}{0}  

This appendix lays out additional graphical summaries of results from the macroeconomic forecasting analysis in the paper, providing material supplementary to that discussed in Section~\ref{sec:Inf}.

\begin{figure}[htbp!]
\centering
\includegraphics[width=0.75\textwidth]{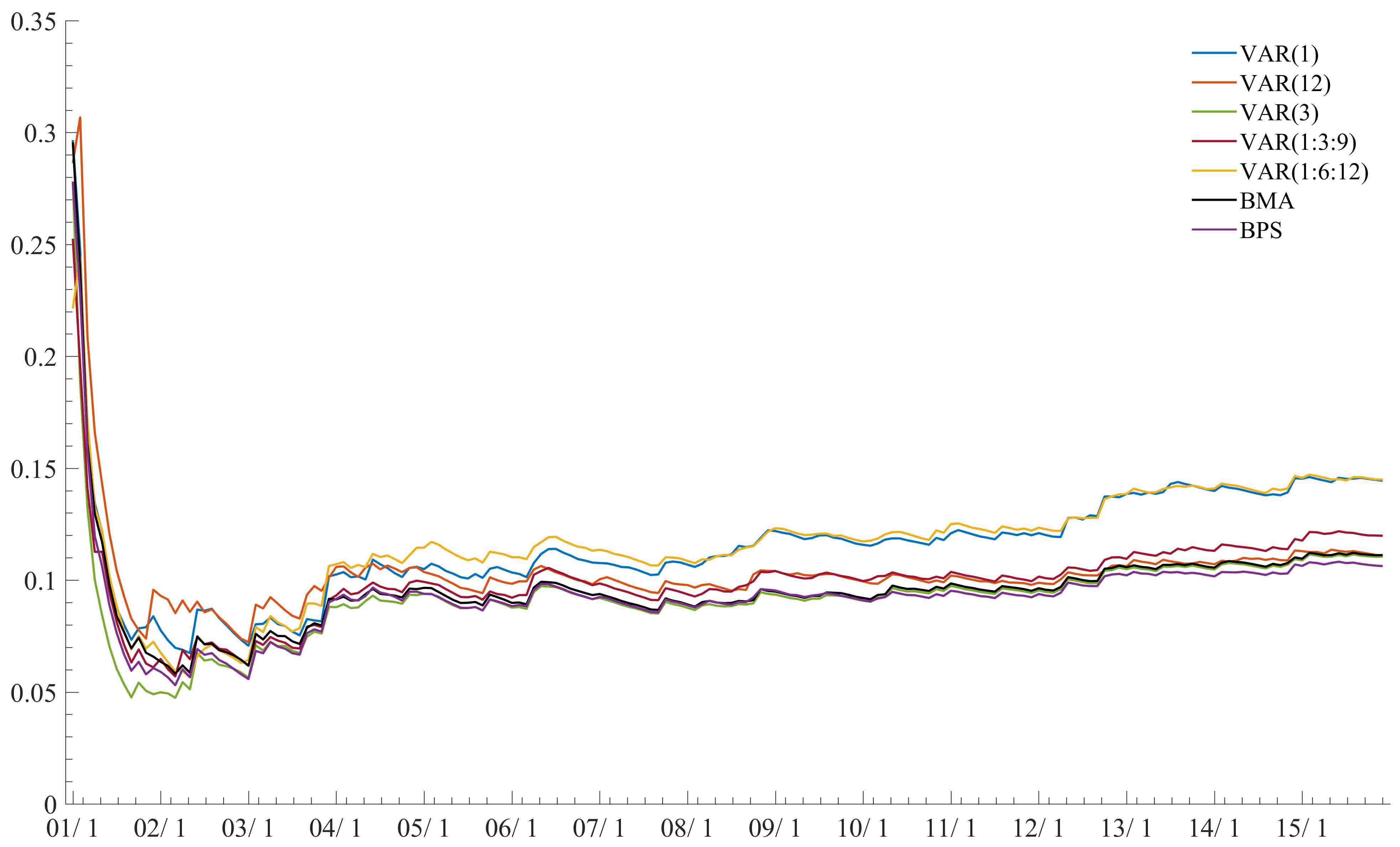}
\caption{US macroeconomic forecasting 2001/1-2015/12: Mean squared 1-step ahead forecast errors MSFE$_{1{:}t}(1)$ of wage $(w)$ sequentially revised at each of the $t=\seq1{180}$ months.
\label{1mse2}}
\end{figure}

\begin{figure}[htbp!]
\centering
\includegraphics[width=0.75\textwidth]{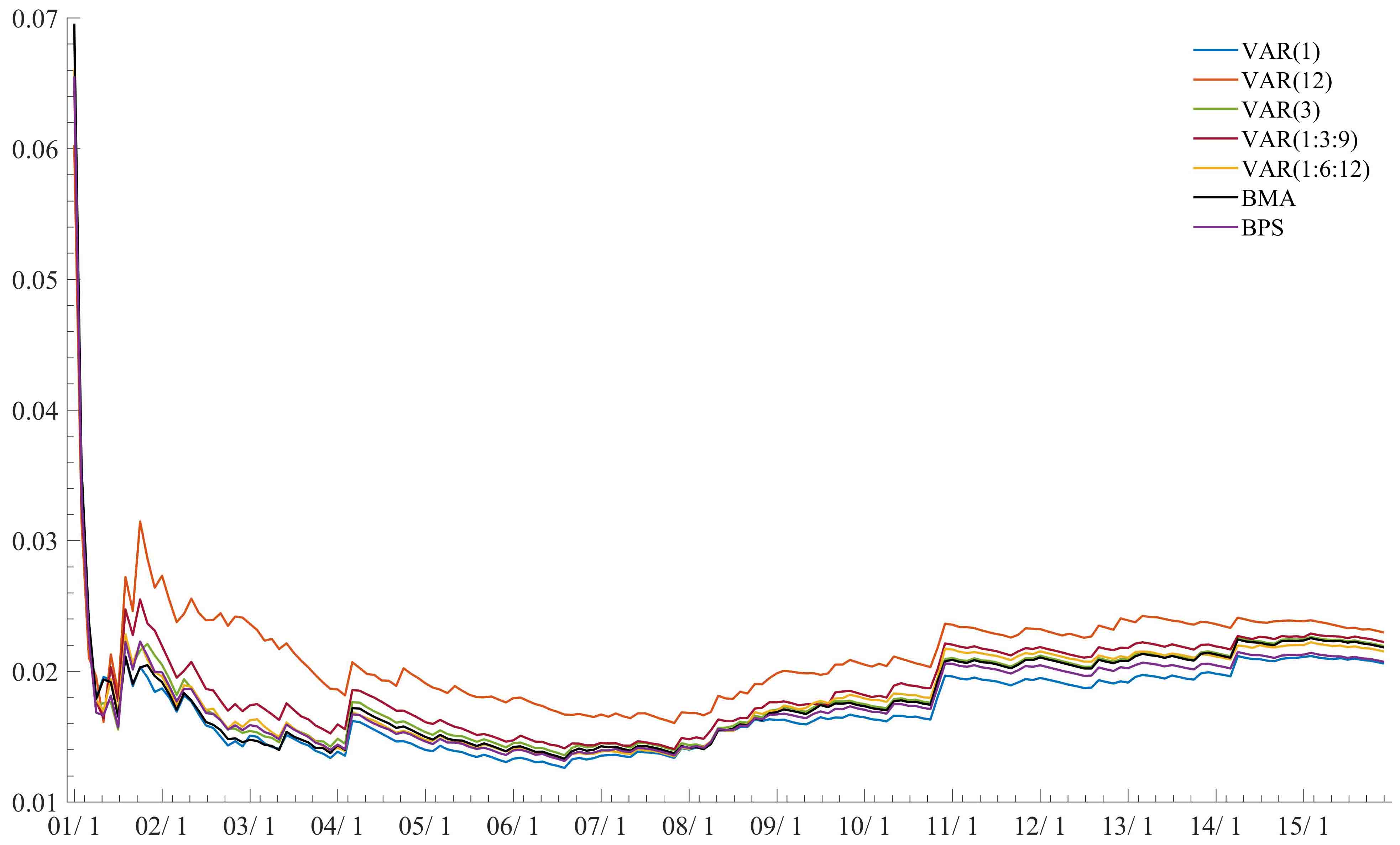}
\caption{US macroeconomic forecasting 2001/1-2015/12: Mean squared 1-step ahead forecast errors MSFE$_{1{:}t}(1)$ of unemployment rate $(u)$ sequentially revised at each of the $t=\seq1{180}$ months.
\label{1mse3}}
\end{figure}

\begin{figure}[htbp!]
\centering
\includegraphics[width=0.75\textwidth]{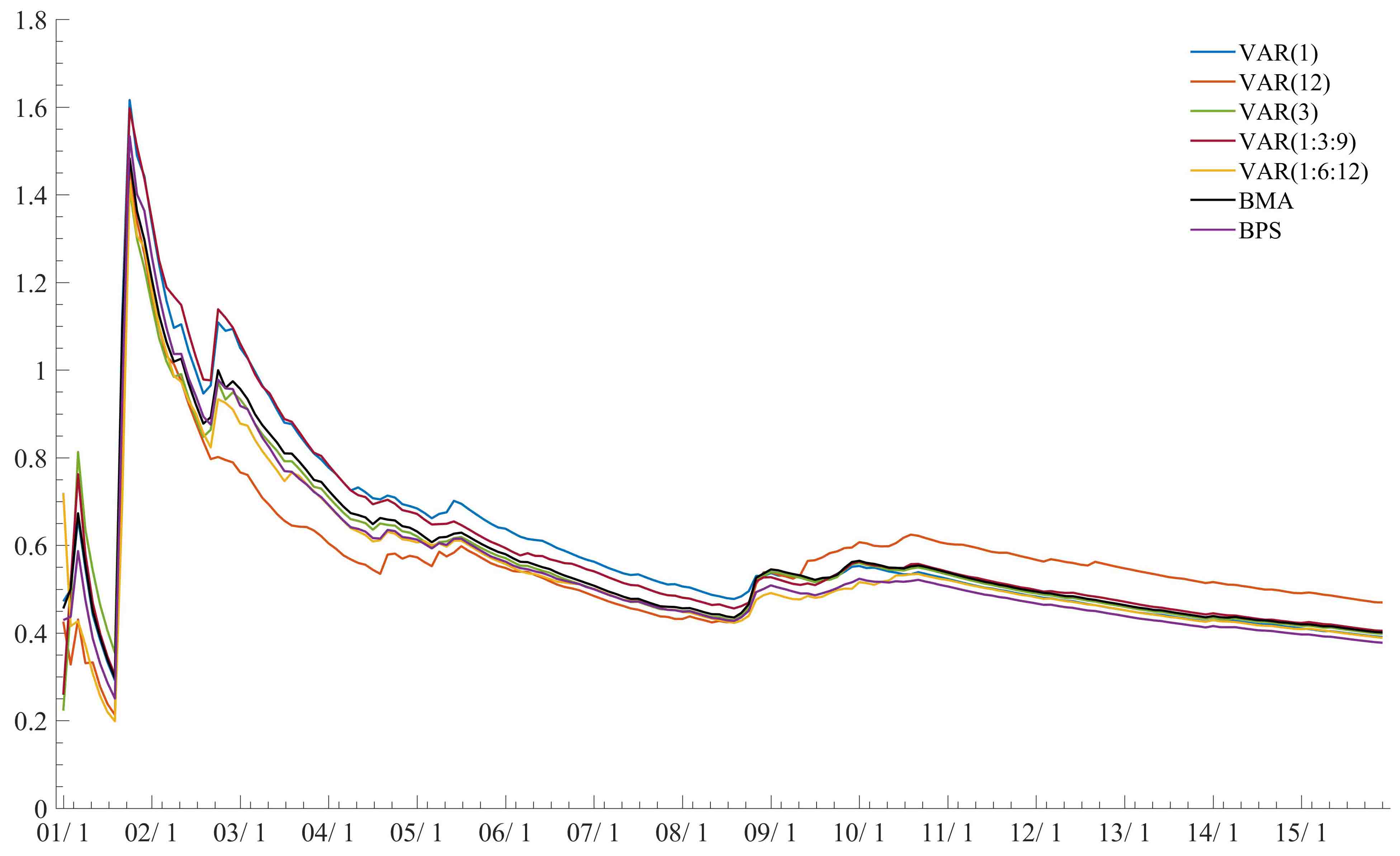}
\caption{US macroeconomic forecasting 2001/1-2015/12: Mean squared 1-step ahead forecast errors MSFE$_{1{:}t}(1)$ of consumption $(c)$ sequentially revised at each of the $t=\seq1{180}$ months.
\label{1mse4}}
\end{figure}

\begin{figure}[htbp!]
\centering
\includegraphics[width=0.75\textwidth]{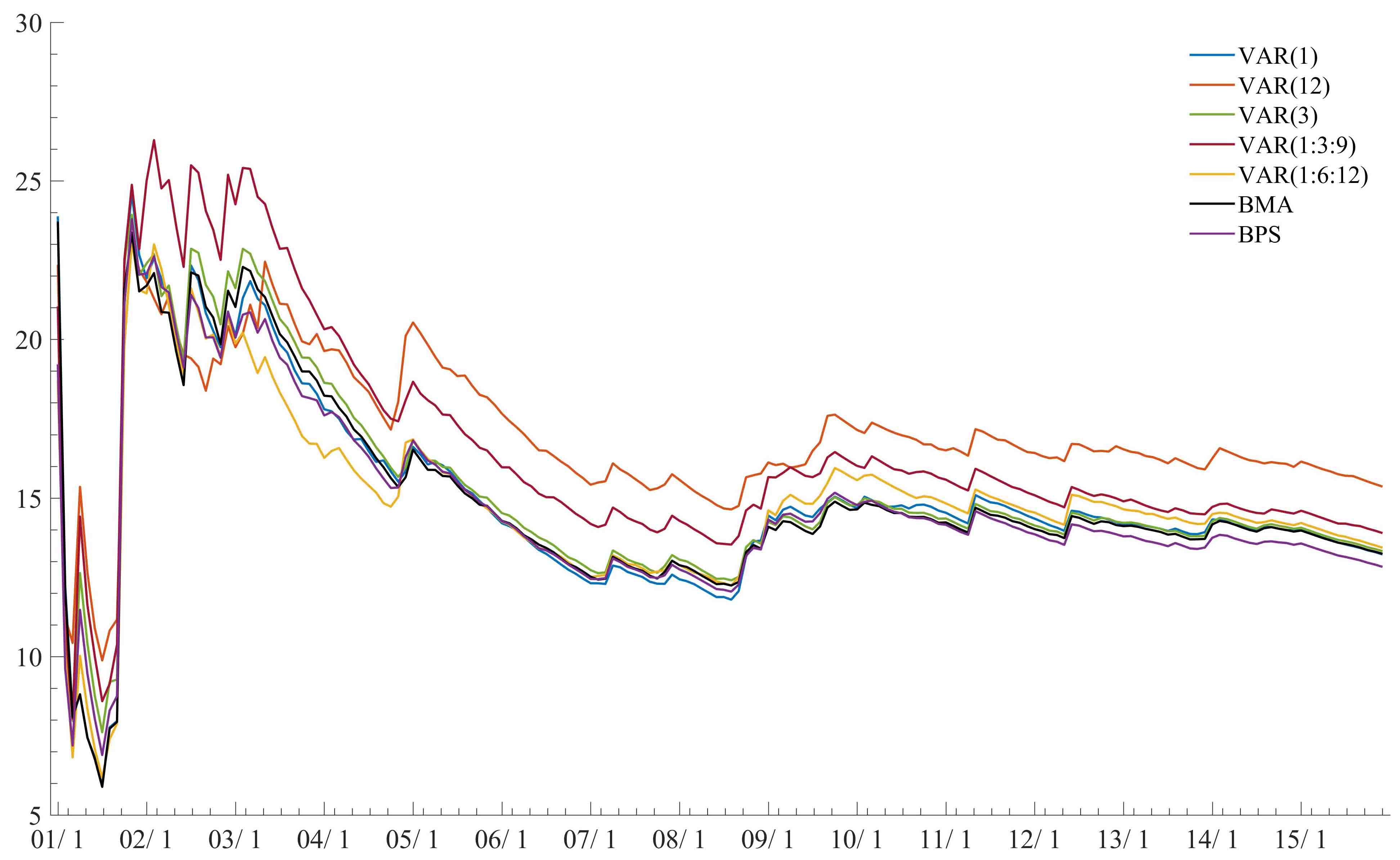}
\caption{US macroeconomic forecasting 2001/1-2015/12: Mean squared 1-step ahead forecast errors MSFE$_{1{:}t}(1)$ of investment $(i)$ sequentially revised at each of the $t=\seq1{180}$ months.
\label{1mse5}}
\end{figure}

\begin{figure}[htbp!]
\centering
\includegraphics[width=0.75\textwidth]{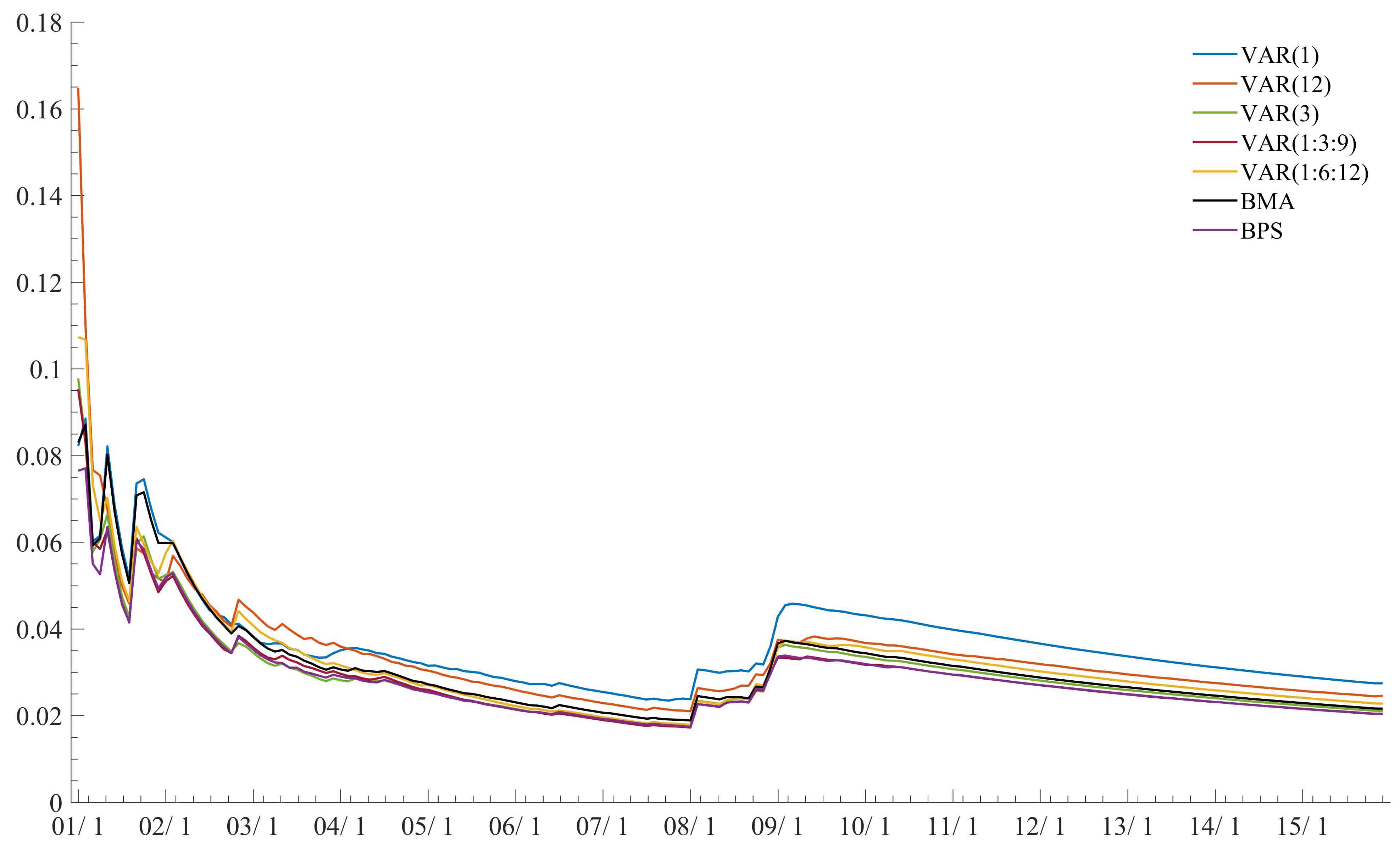}
\caption{US macroeconomic forecasting 2001/1-2015/12: Mean squared 1-step ahead forecast errors MSFE$_{1{:}t}(1)$ of interest rate $(r)$ sequentially revised at each of the $t=\seq1{180}$ months.
\label{1mse6}}
\end{figure}

\clearpage

\begin{figure}[htbp!]
\centering
\includegraphics[width=0.75\textwidth]{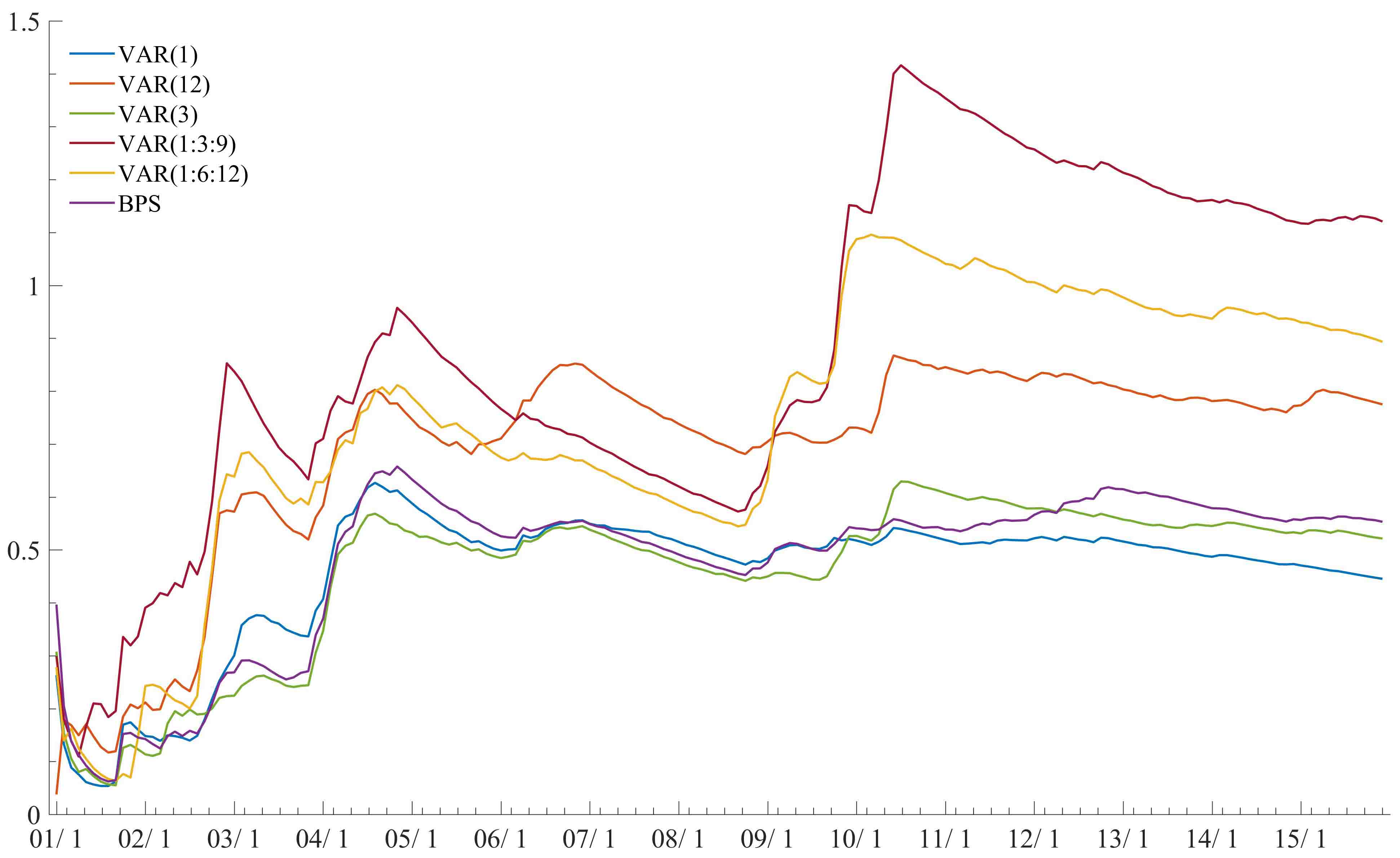}
\caption{US macroeconomic forecasting 2001/1-2015/12: Mean squared 12-step ahead forecast errors MSFE$_{1{:}t}(12)$ of wage $(w)$ sequentially revised at each of the $t=\seq1{180}$ months.
\label{12mse2}}
\end{figure}

\begin{figure}[htbp!]
\centering
\includegraphics[width=0.75\textwidth]{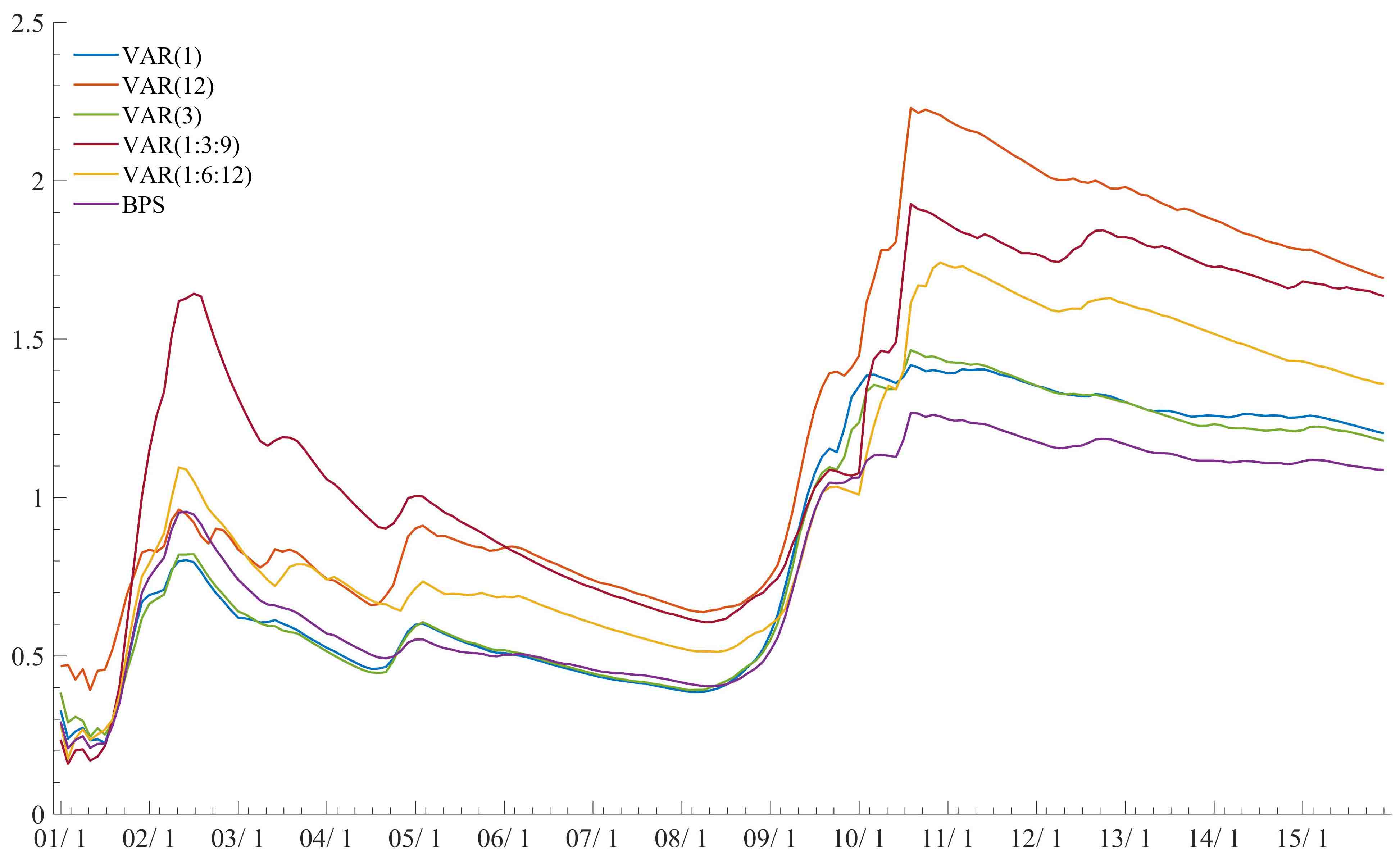}
\caption{US macroeconomic forecasting 2001/1-2015/12: Mean squared 12-step ahead forecast errors MSFE$_{1{:}t}(12)$ of unemployment rate $(u)$ sequentially revised at each of the $t=\seq1{180}$ months.
\label{12mse3}}
\end{figure}

\begin{figure}[htbp!]
\centering
\includegraphics[width=0.75\textwidth]{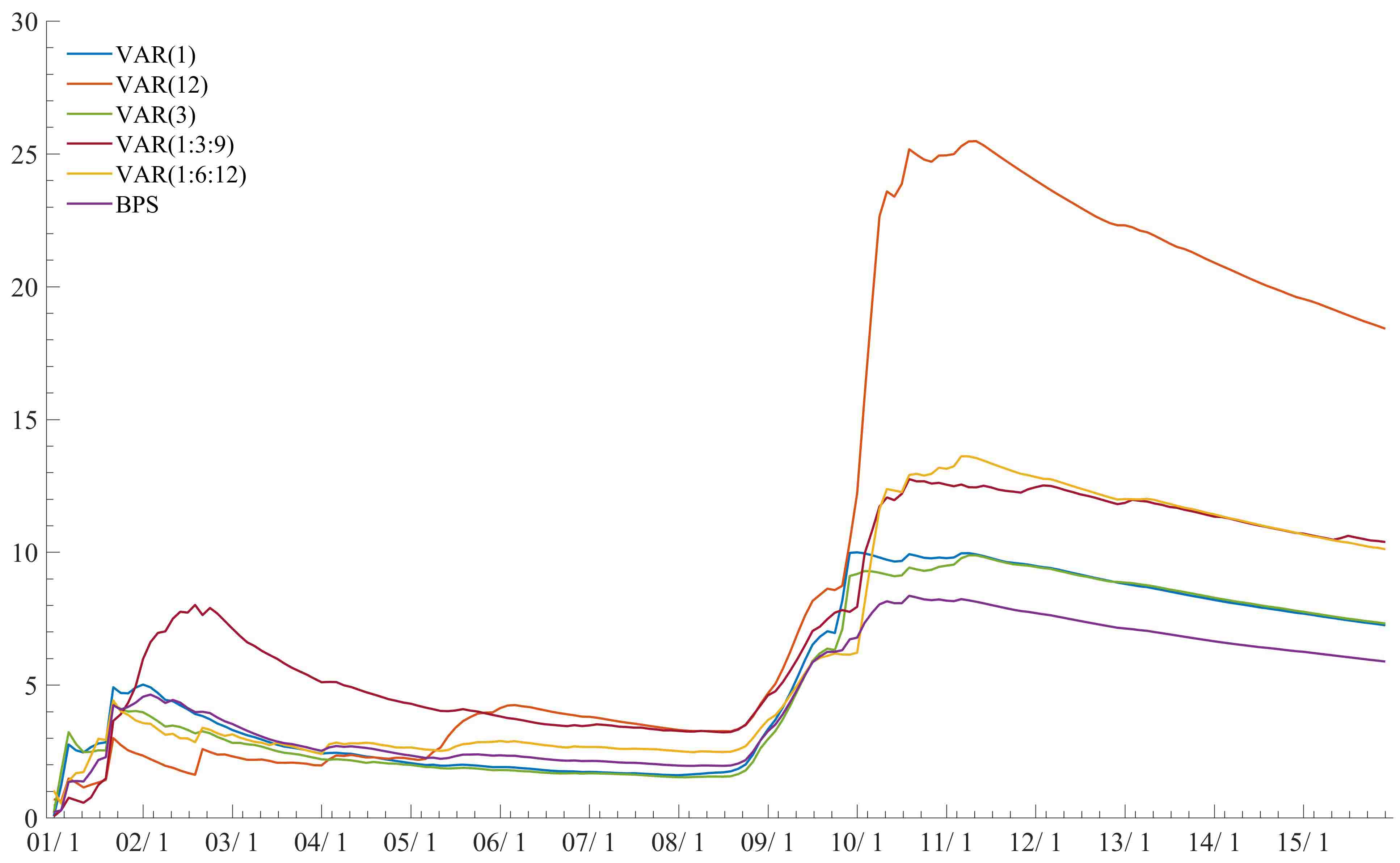}
\caption{US macroeconomic forecasting 2001/1-2015/12: Mean squared 12-step ahead forecast errors MSFE$_{1{:}t}(12)$ of consumption $(c)$ sequentially revised at each of the $t=\seq1{180}$ months.
\label{12mse4}}
\end{figure}

\begin{figure}[htbp!]
\centering
\includegraphics[width=0.75\textwidth]{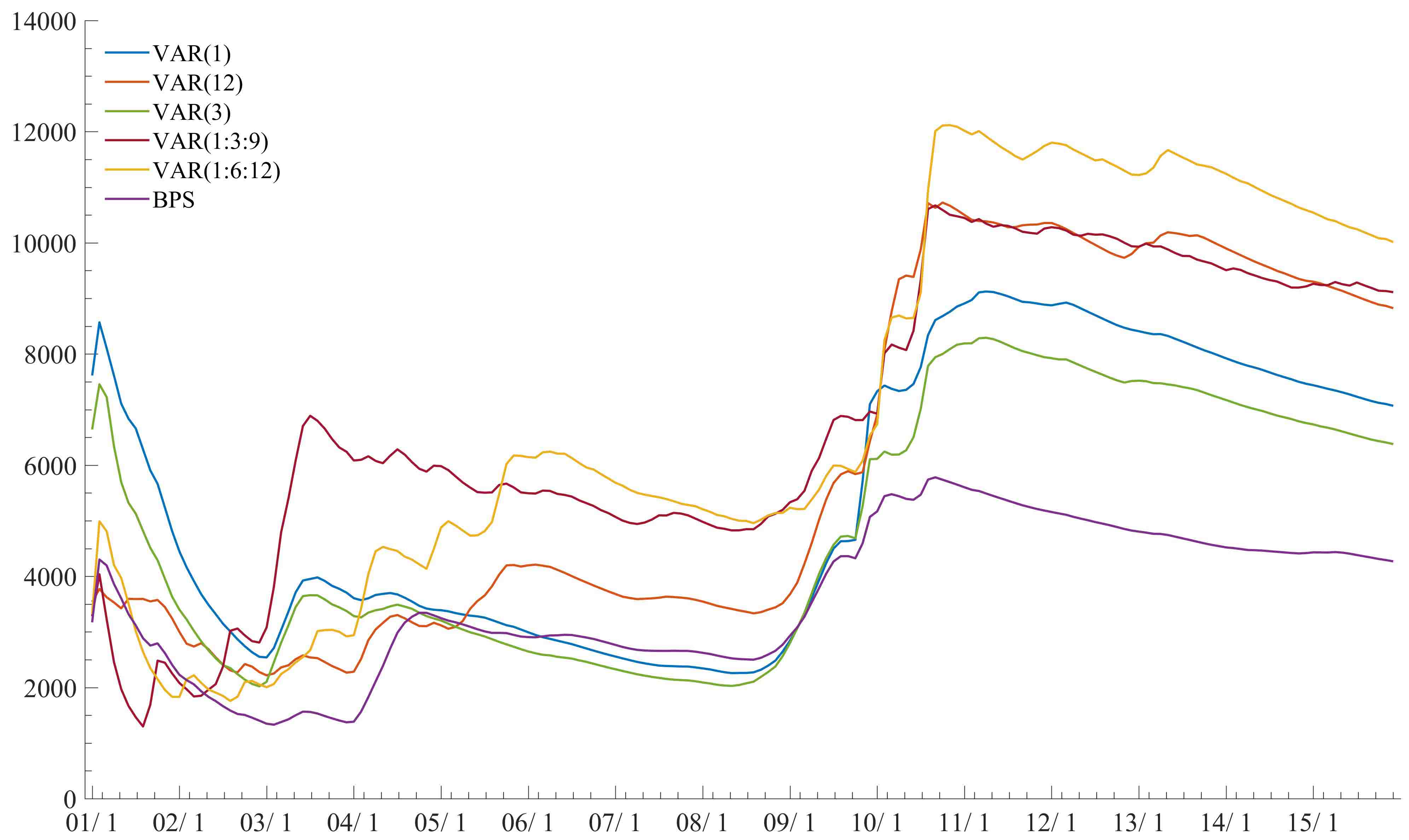}
\caption{US macroeconomic forecasting 2001/1-2015/12: Mean squared 12-step ahead forecast errors MSFE$_{1{:}t}(12)$ of investment $(i)$ sequentially revised at each of the $t=\seq1{180}$ months.
\label{12mse5}}
\end{figure}

\begin{figure}[htbp!]
\centering
\includegraphics[width=0.75\textwidth]{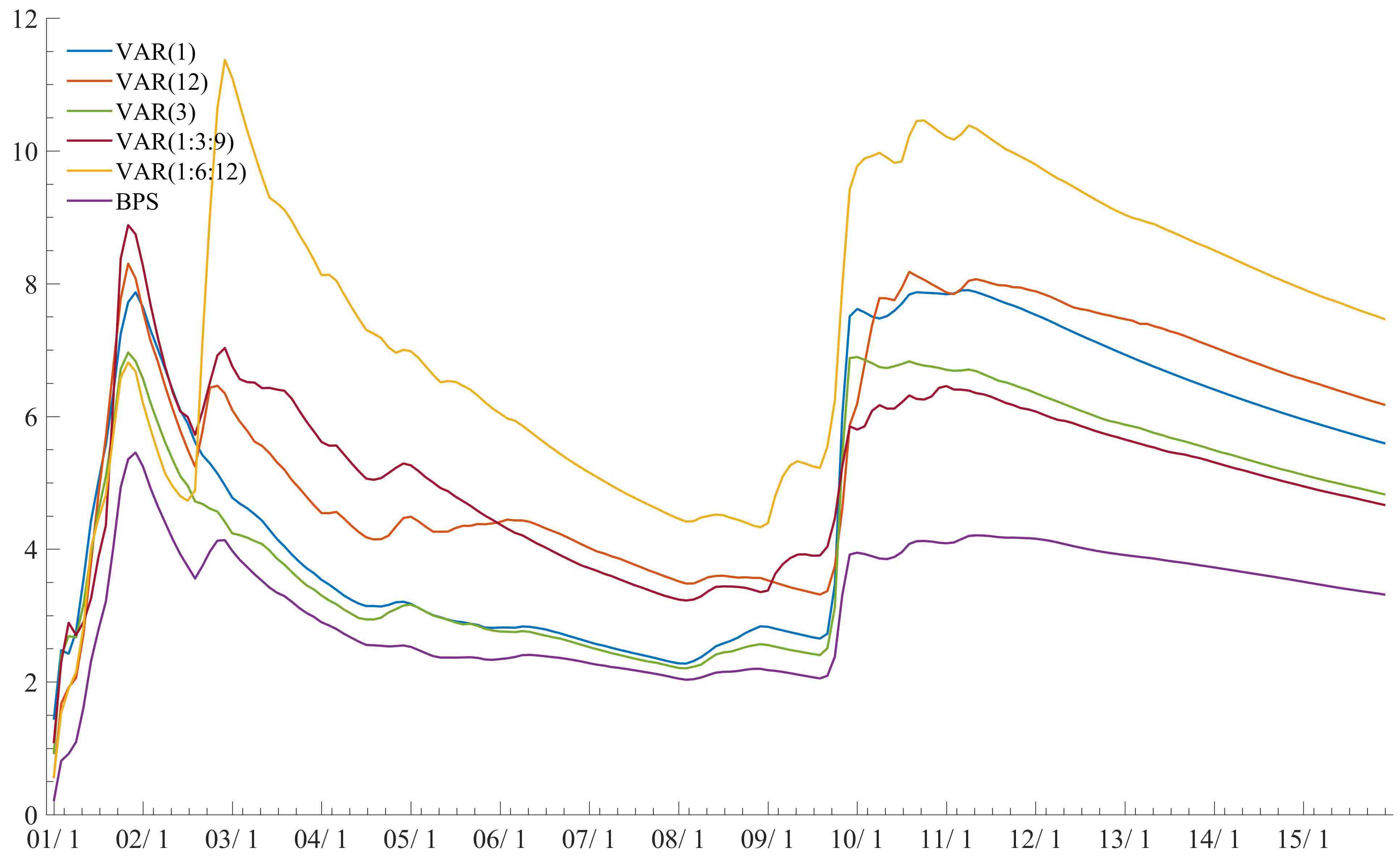}
\caption{US macroeconomic forecasting 2001/1-2015/12: Mean squared 12-step ahead forecast errors MSFE$_{1{:}t}(12)$ of interest rate $(r)$ sequentially revised at each of the $t=\seq1{180}$ months.
\label{12mse12}}
\end{figure}

\clearpage

\begin{figure}[htbp!]
\centering
\includegraphics[width=0.75\textwidth]{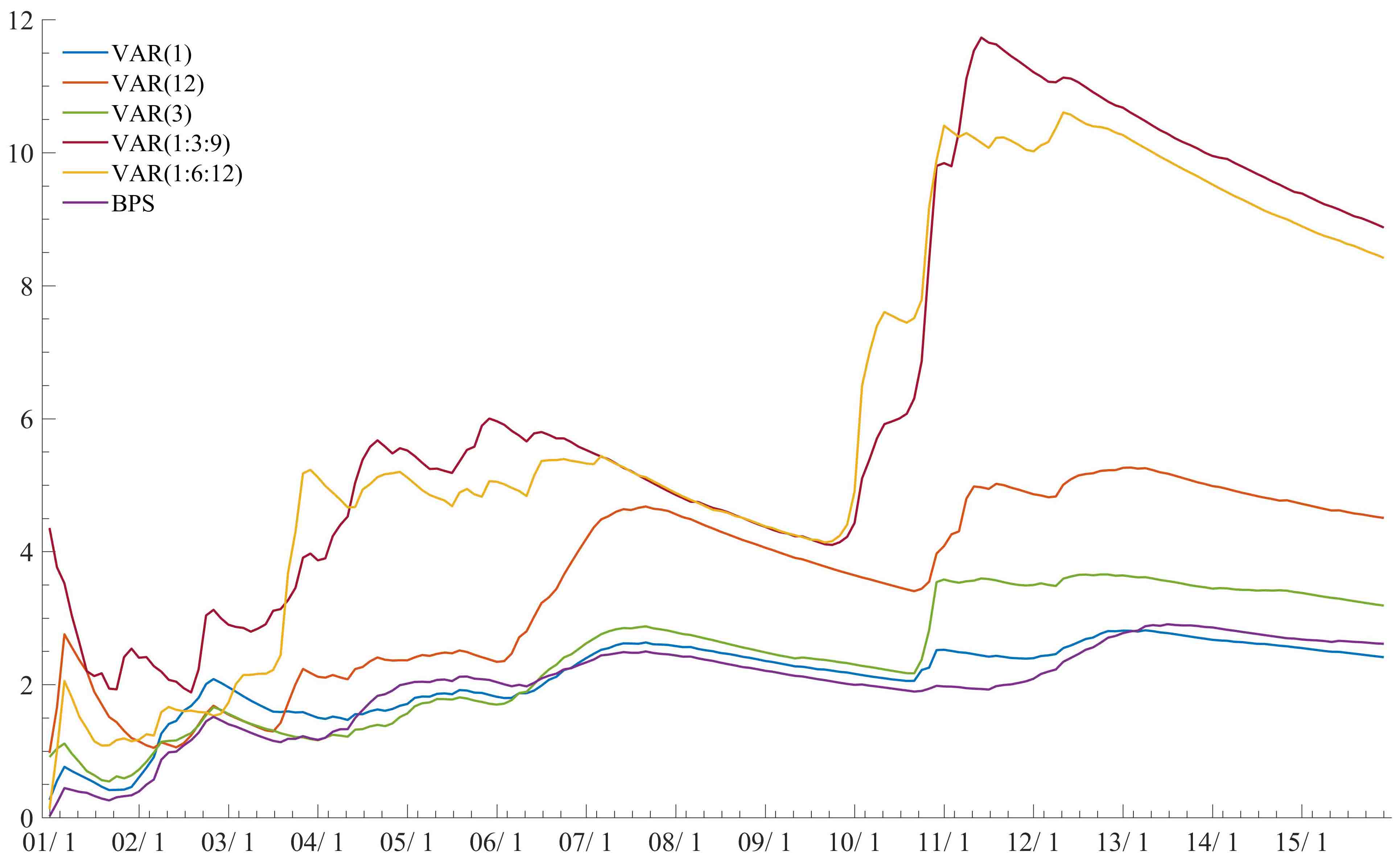}
\caption{US macroeconomic forecasting 2001/1-2015/12: Mean squared 24-step ahead forecast errors MSFE$_{1{:}t}(24)$ of wage $(w)$ sequentially revised at each of the $t=\seq1{180}$ months.
\label{24mse2}}
\end{figure}

\begin{figure}[htbp!]
\centering
\includegraphics[width=0.75\textwidth]{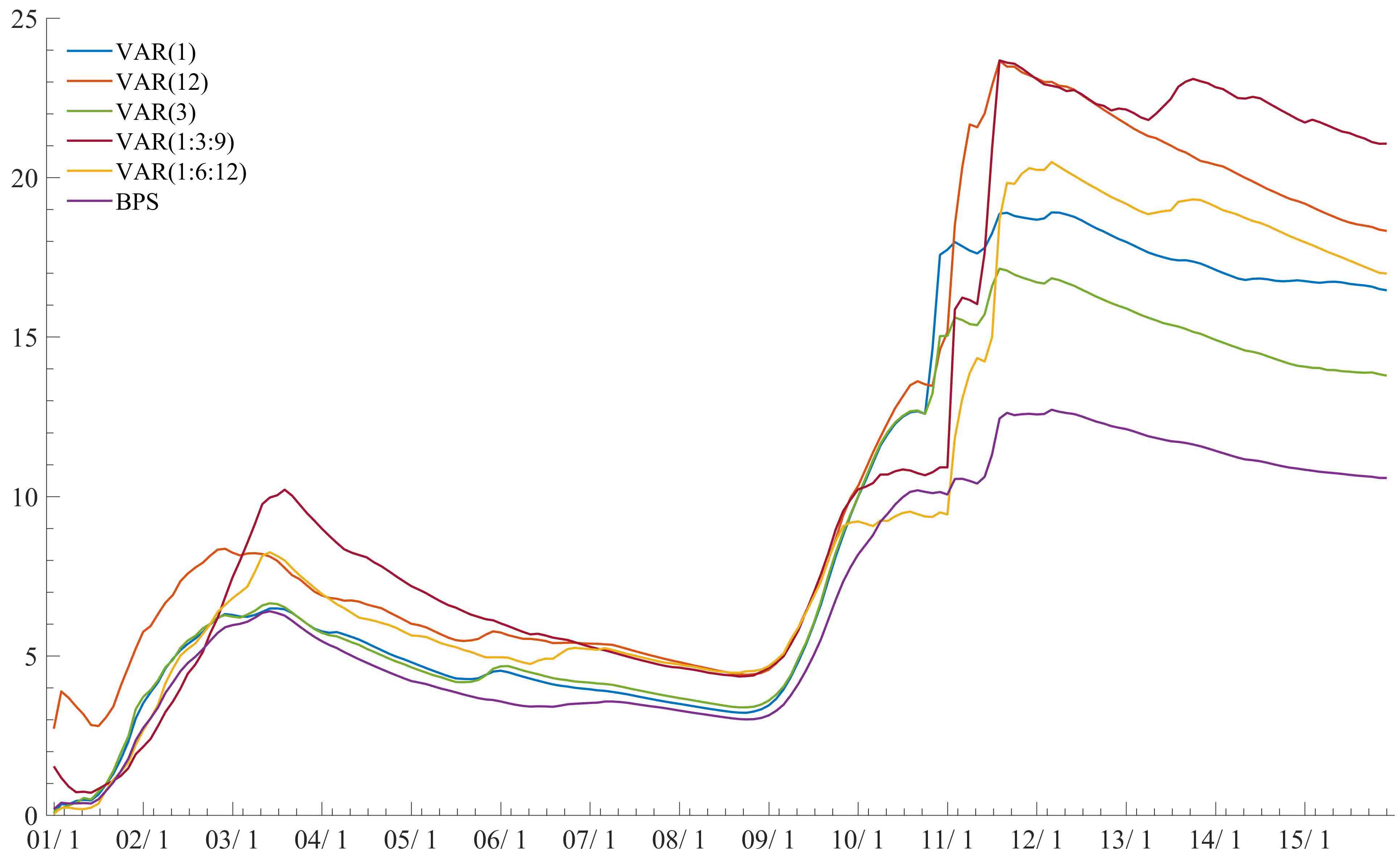}
\caption{US macroeconomic forecasting 2001/1-2015/12: Mean squared 24-step ahead forecast errors MSFE$_{1{:}t}(24)$ of unemployment rate $(u)$ sequentially revised at each of the $t=\seq1{180}$ months.
\label{24mse3}}
\end{figure}

\begin{figure}[htbp!]
\centering
\includegraphics[width=0.75\textwidth]{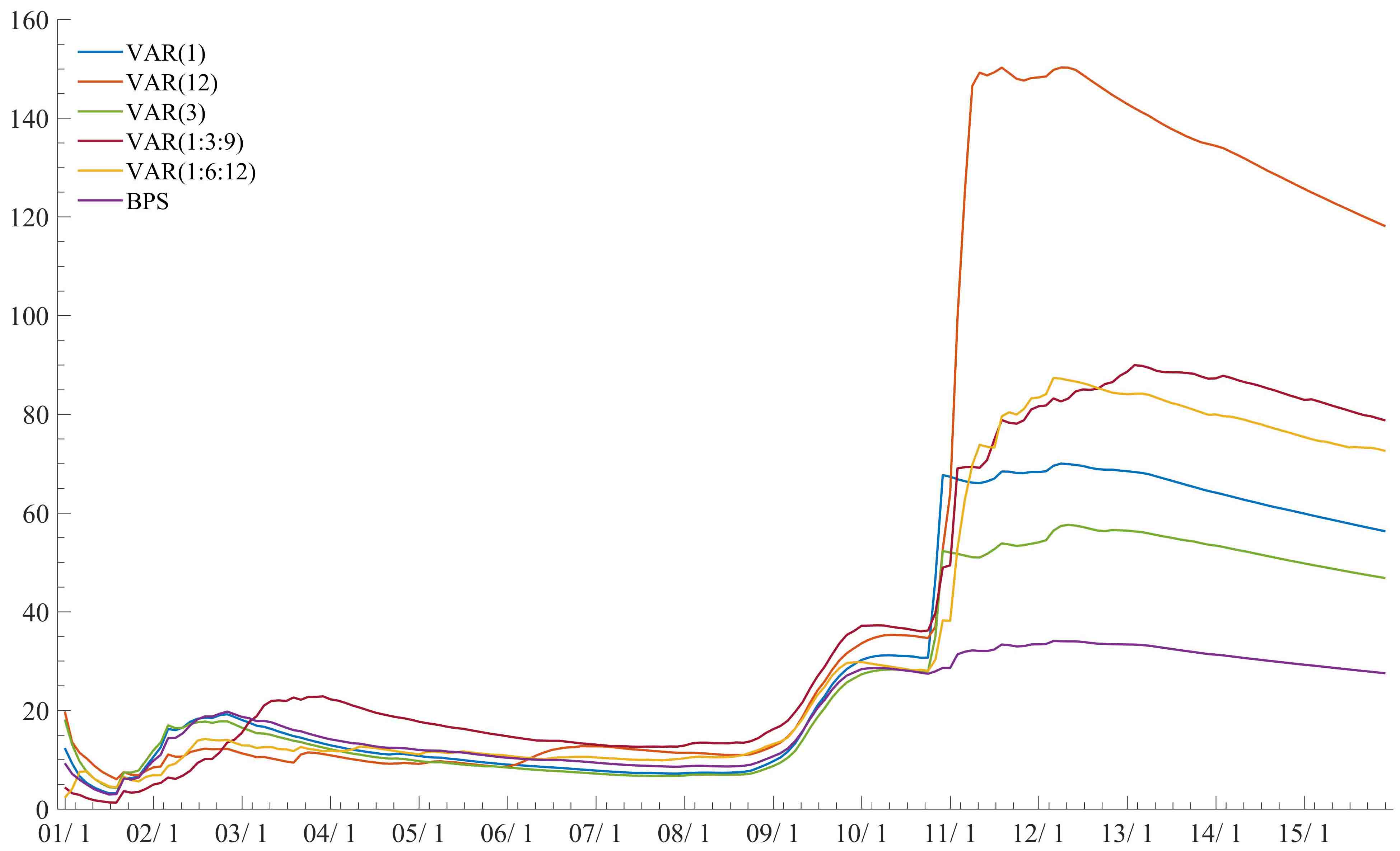}
\caption{US macroeconomic forecasting 2001/1-2015/12: Mean squared 24-step ahead forecast errors MSFE$_{1{:}t}(24)$ of consumption $(c)$ sequentially revised at each of the $t=\seq1{180}$ months.
\label{24mse4}}
\end{figure}

\begin{figure}[htbp!]
\centering
\includegraphics[width=0.75\textwidth]{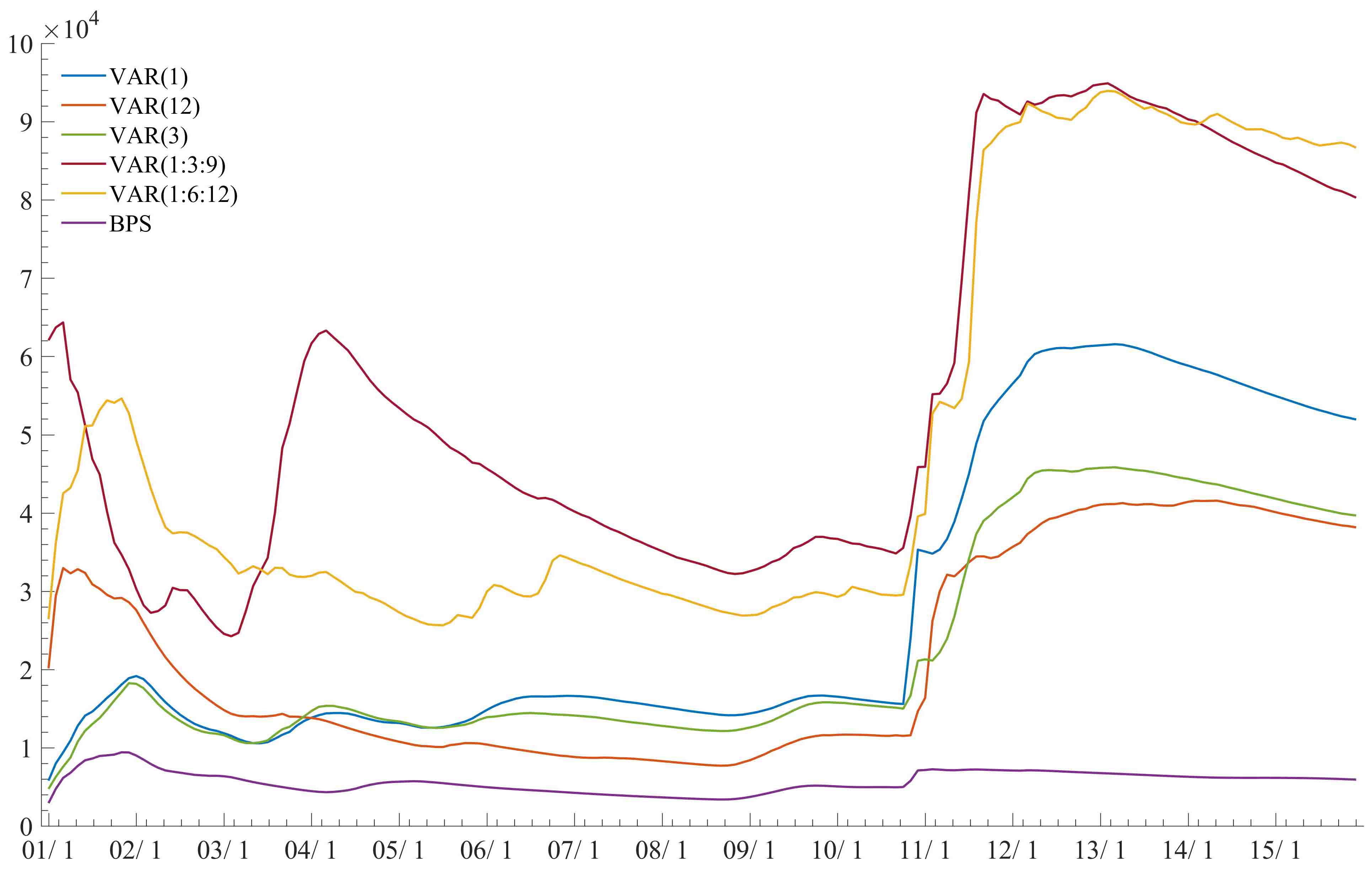}
\caption{US macroeconomic forecasting 2001/1-2015/12: Mean squared 24-step ahead forecast errors MSFE$_{1{:}t}(24)$ of investment $(i)$ sequentially revised at each of the $t=\seq1{180}$ months.
\label{24mse5}}
\end{figure}

\begin{figure}[htbp!]
\centering
\includegraphics[width=0.75\textwidth]{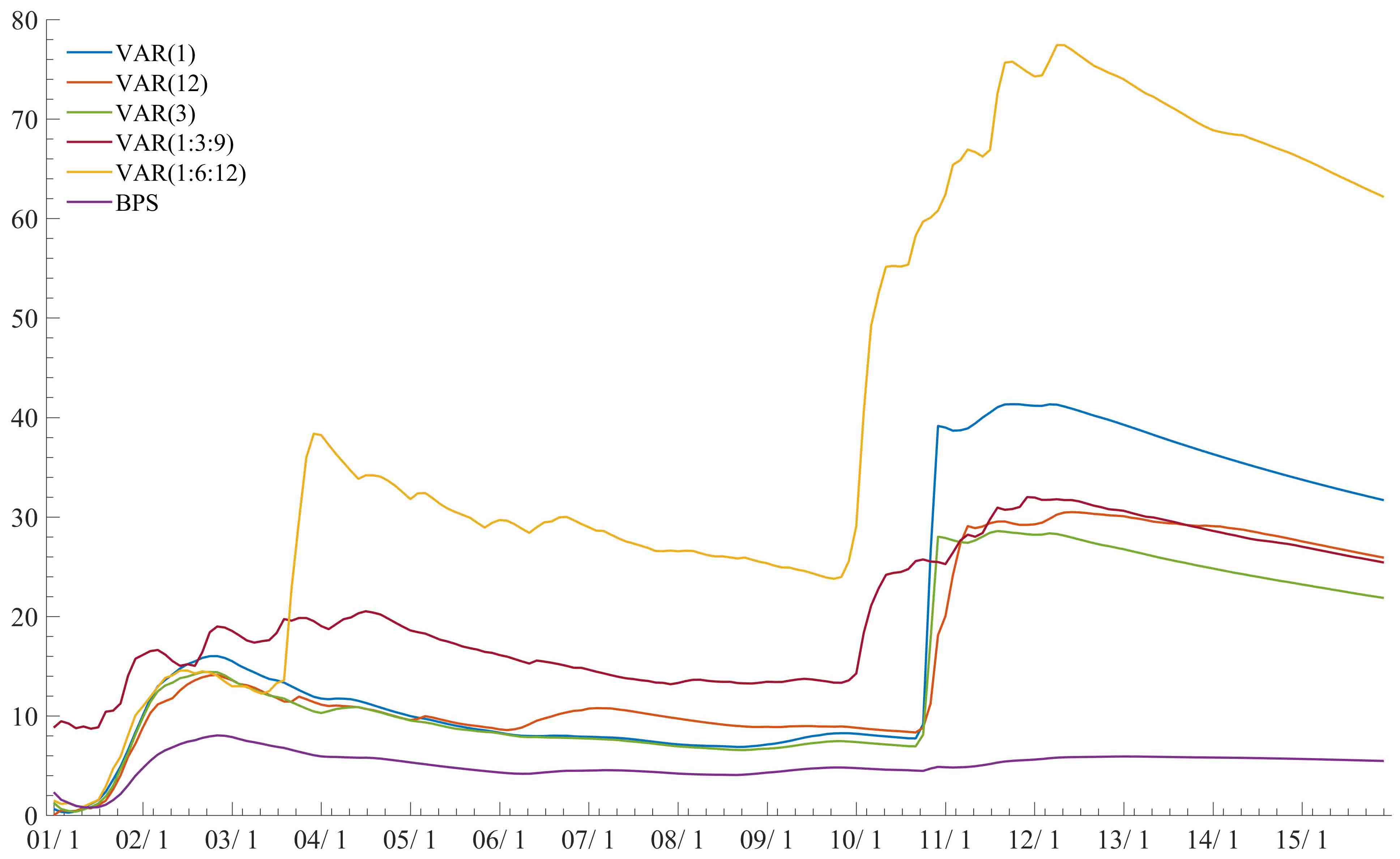}
\caption{US macroeconomic forecasting 2001/1-2015/12: Mean squared 24-step ahead forecast errors MSFE$_{1{:}t}(24)$ of interest rate $(r)$ sequentially revised at each of the $t=\seq1{180}$ months.
\label{24mse24}}
\end{figure}

\clearpage

\begin{figure}[htbp!]
\centering
\includegraphics[width=0.75\textwidth]{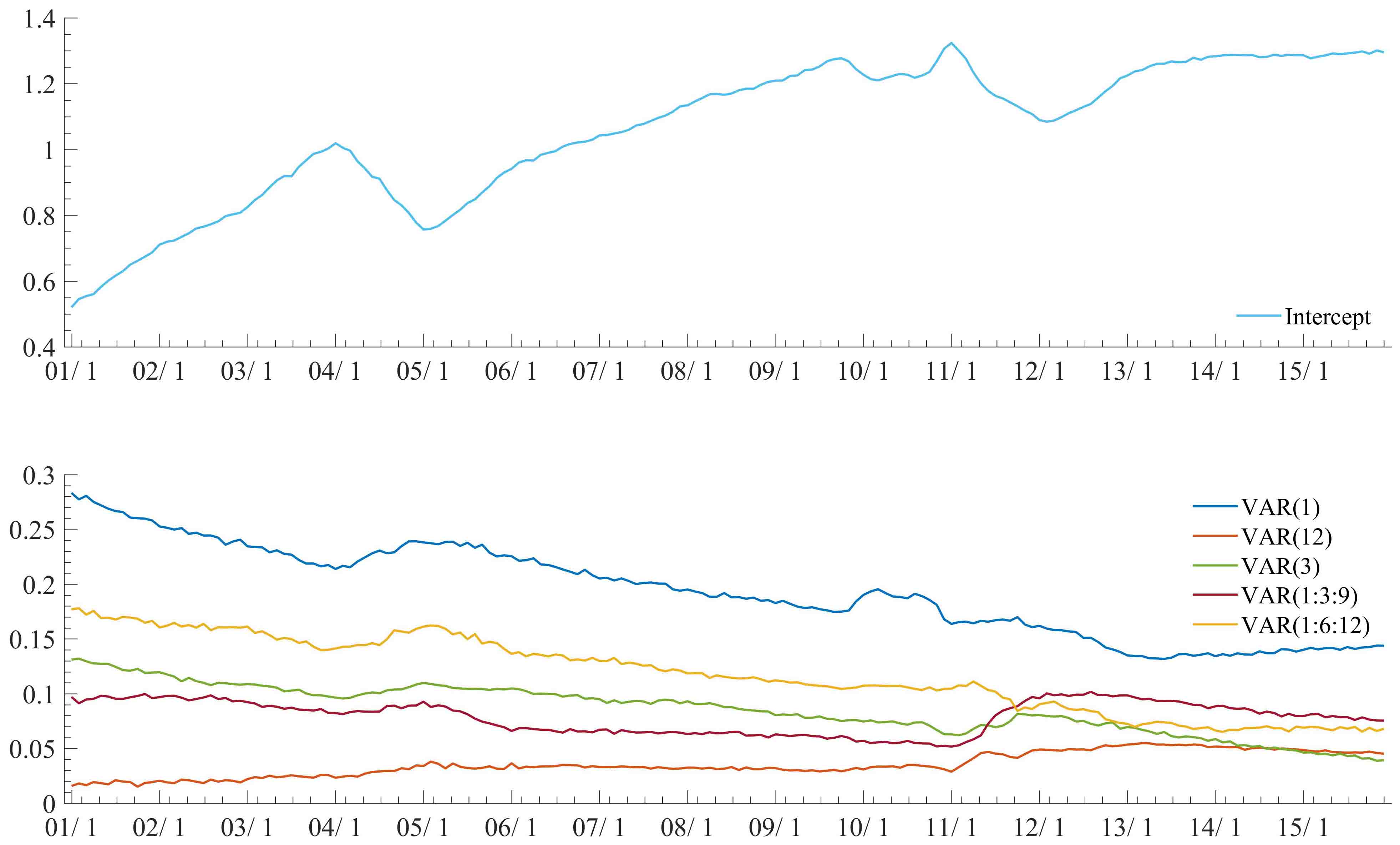}
\caption{US macroeconomic forecasting 2001/1-2015/12:  On-line posterior means of BPS(12) model
 coefficients for inflation $(p)$ sequentially computed at each of the $t=\seq1{180}$ months.
\label{12coeff1}}
\end{figure}

\begin{figure}[htbp!]
\centering
\includegraphics[width=0.75\textwidth]{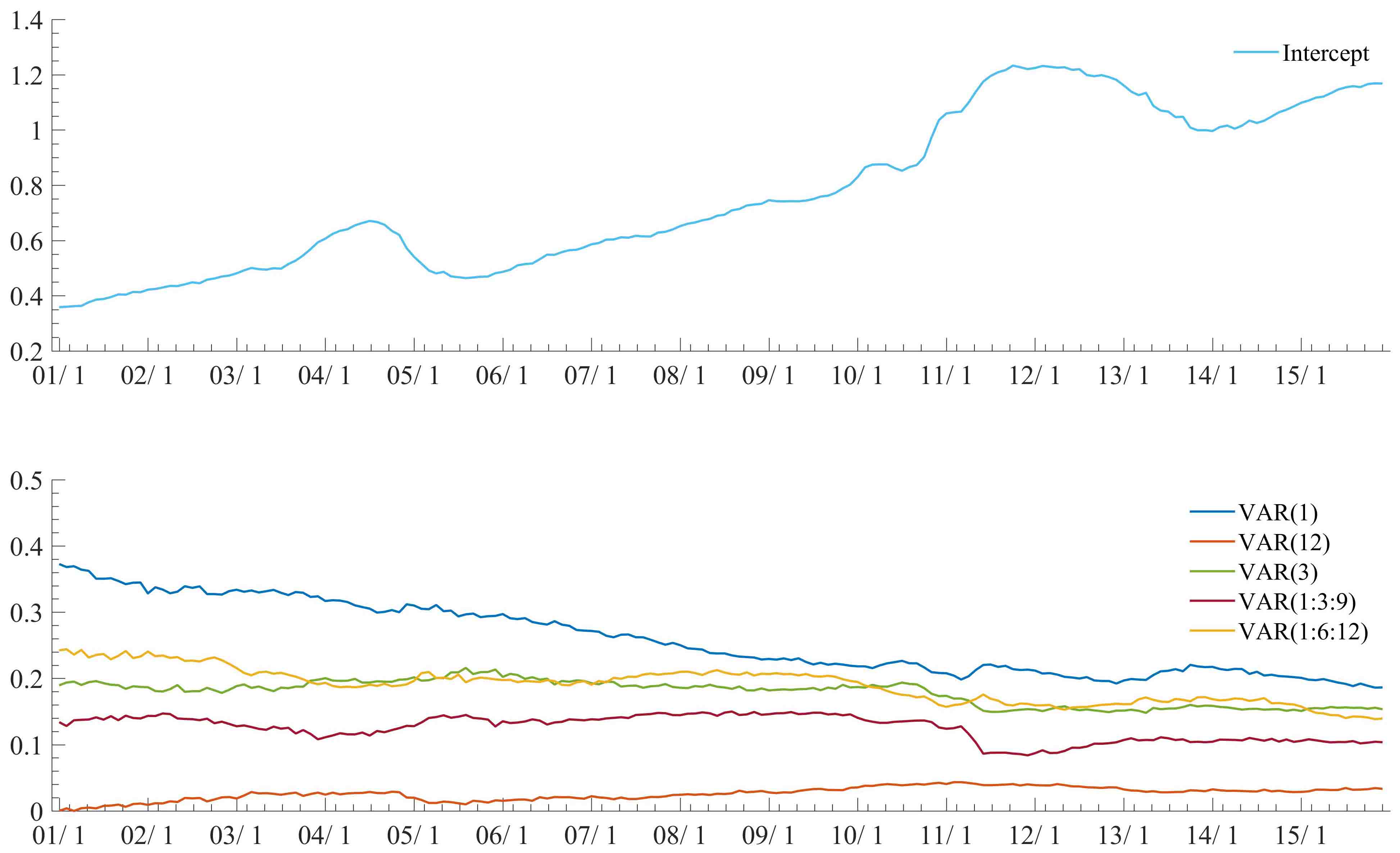}
\caption{US macroeconomic forecasting 2001/1-2015/12:  On-line posterior means of BPS(12)  model
 coefficients for wage $(w)$sequentially computed at each of the $t=\seq1{180}$ months.
\label{12coeff2}}
\end{figure}

\begin{figure}[htbp!]
\centering
\includegraphics[width=0.75\textwidth]{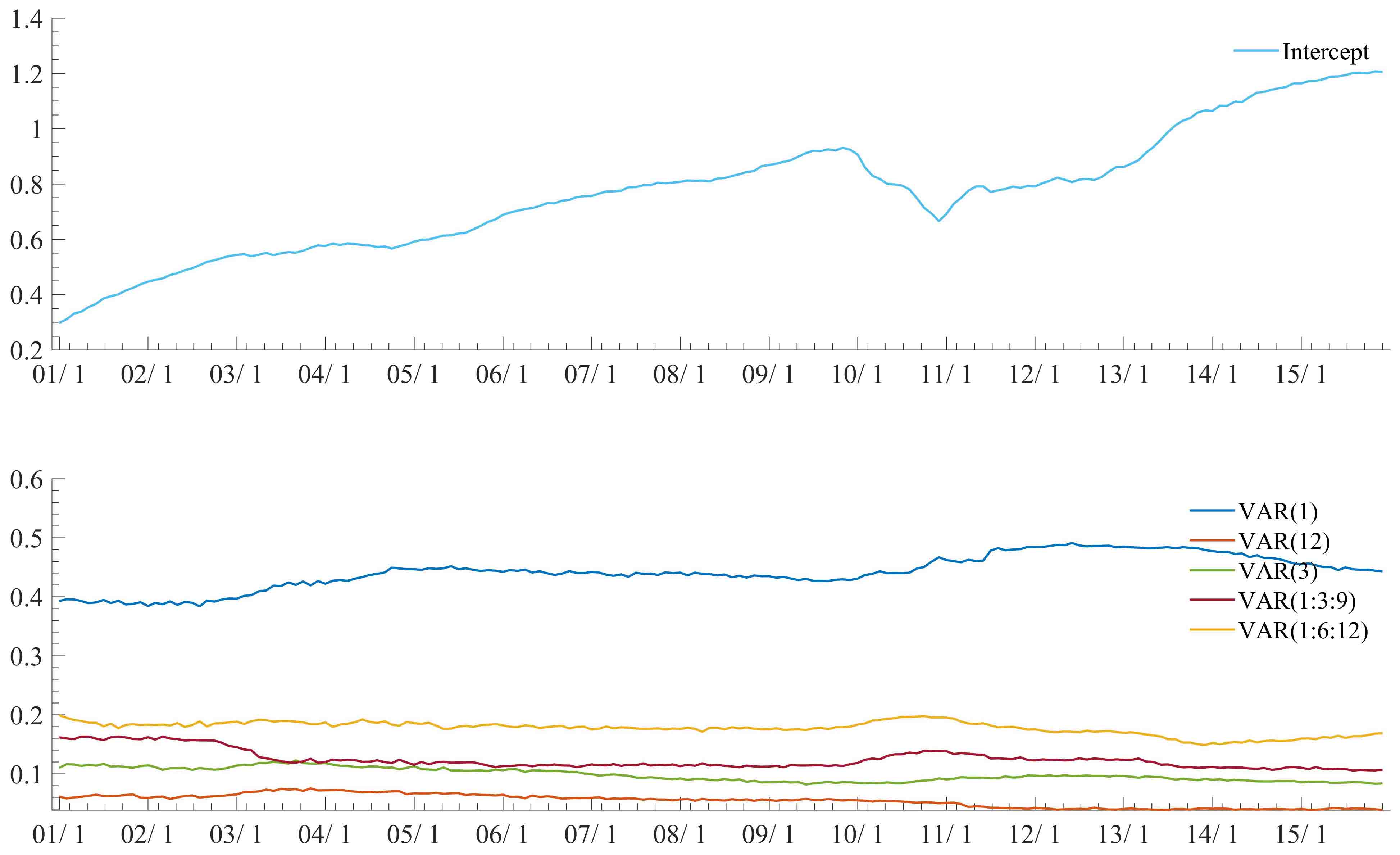}
\caption{US macroeconomic forecasting 2001/1-2015/12:  On-line posterior means of BPS(12)  model
 coefficients for unemployment $(u)$ sequentially computed at each of the $t=\seq1{180}$ months.
\label{12coeff3}}
\end{figure}

\begin{figure}[htbp!]
\centering
\includegraphics[width=0.75\textwidth]{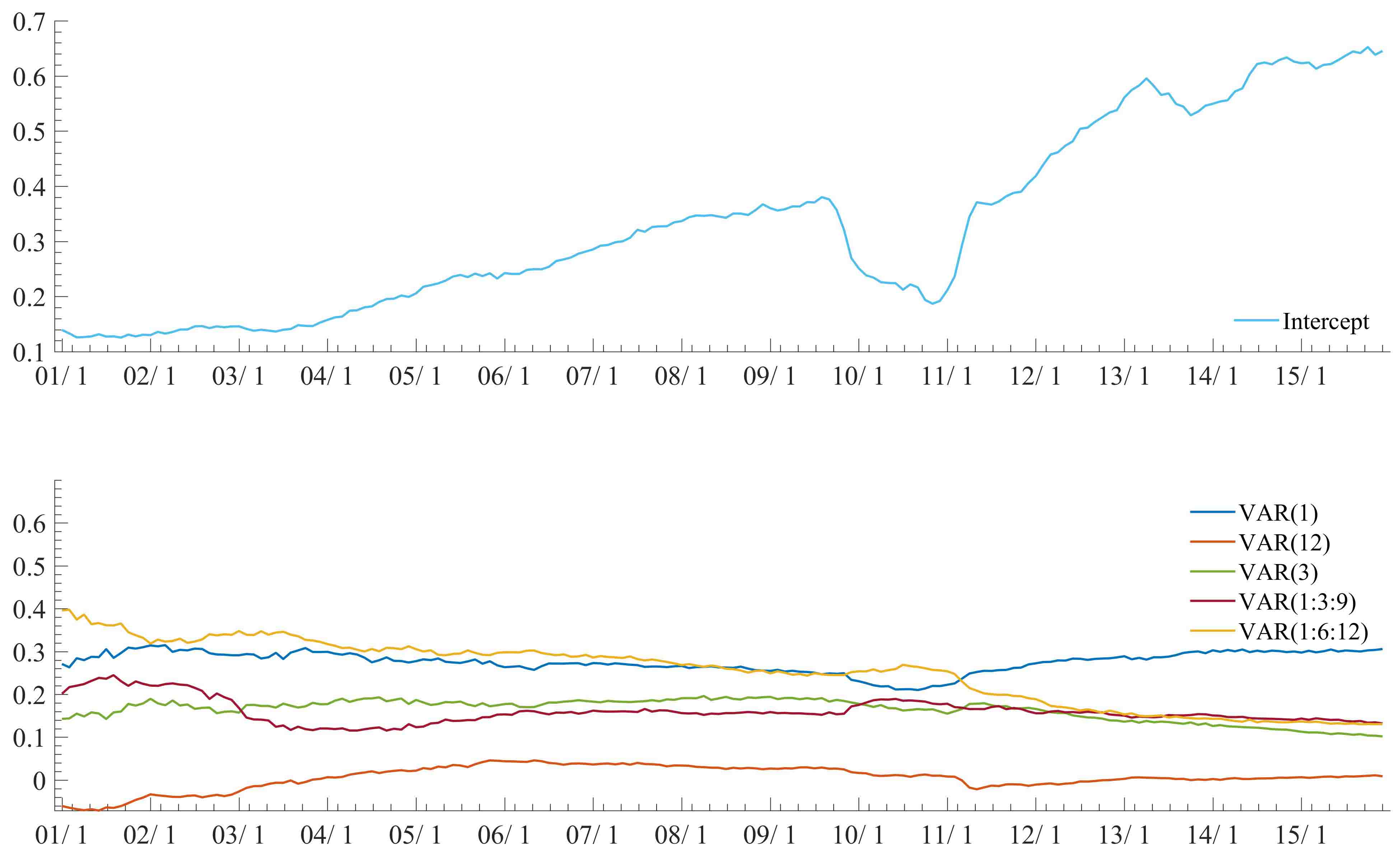}
\caption{US macroeconomic forecasting 2001/1-2015/12:  On-line posterior means of BPS(12)  model
 coefficients for consumption $(c)$ sequentially computed at each of the $t=\seq1{180}$ months.
\label{12coeff4}}
\end{figure}

\begin{figure}[htbp!]
\centering
\includegraphics[width=0.75\textwidth]{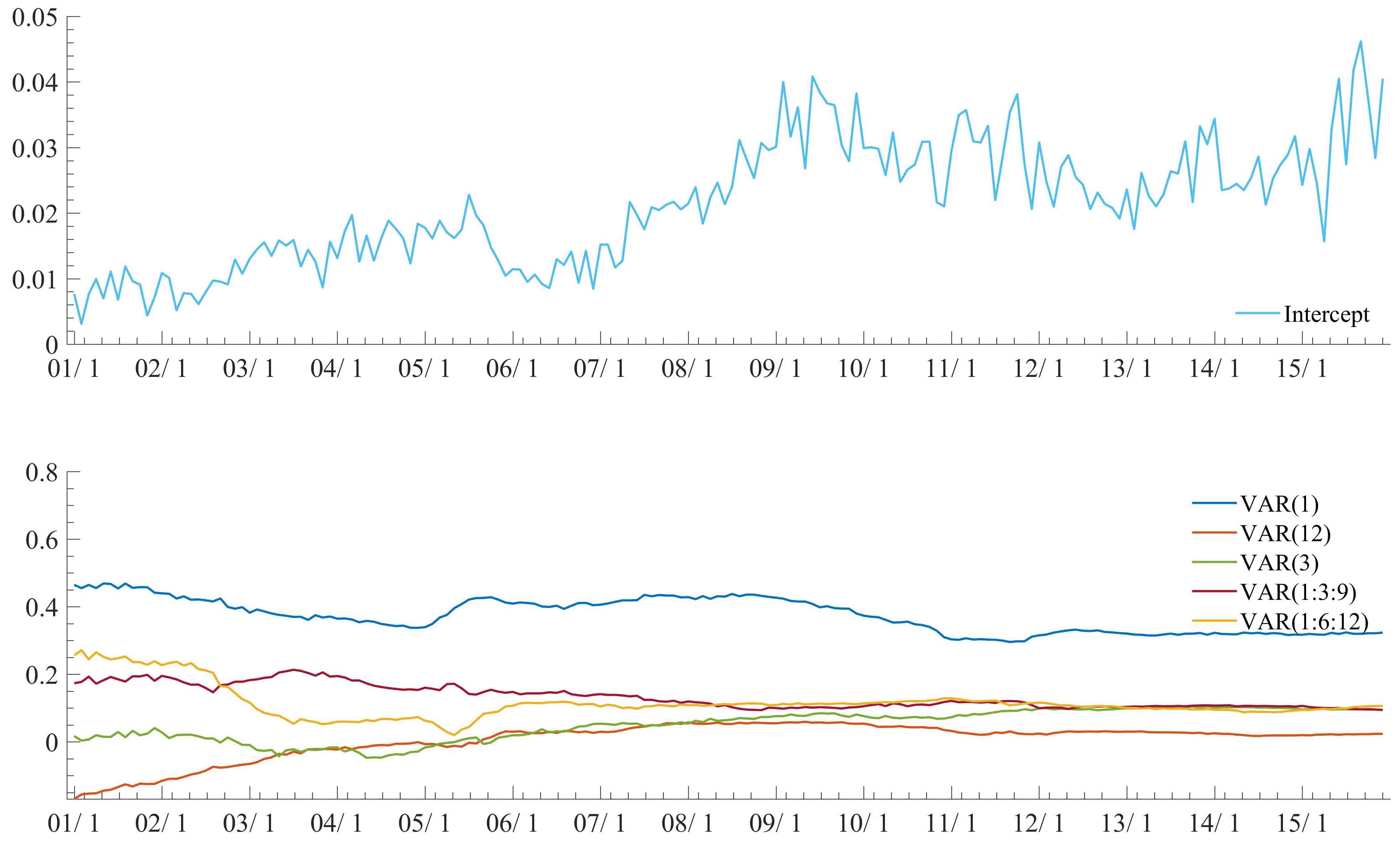}
\caption{US macroeconomic forecasting 2001/1-2015/12:  On-line posterior means of BPS(12)  model
 coefficients for investment $(i)$ sequentially computed at each of the $t=\seq1{180}$ months.
\label{12coeff5}}
\end{figure}

\begin{figure}[htbp!]
\centering
\includegraphics[width=0.75\textwidth]{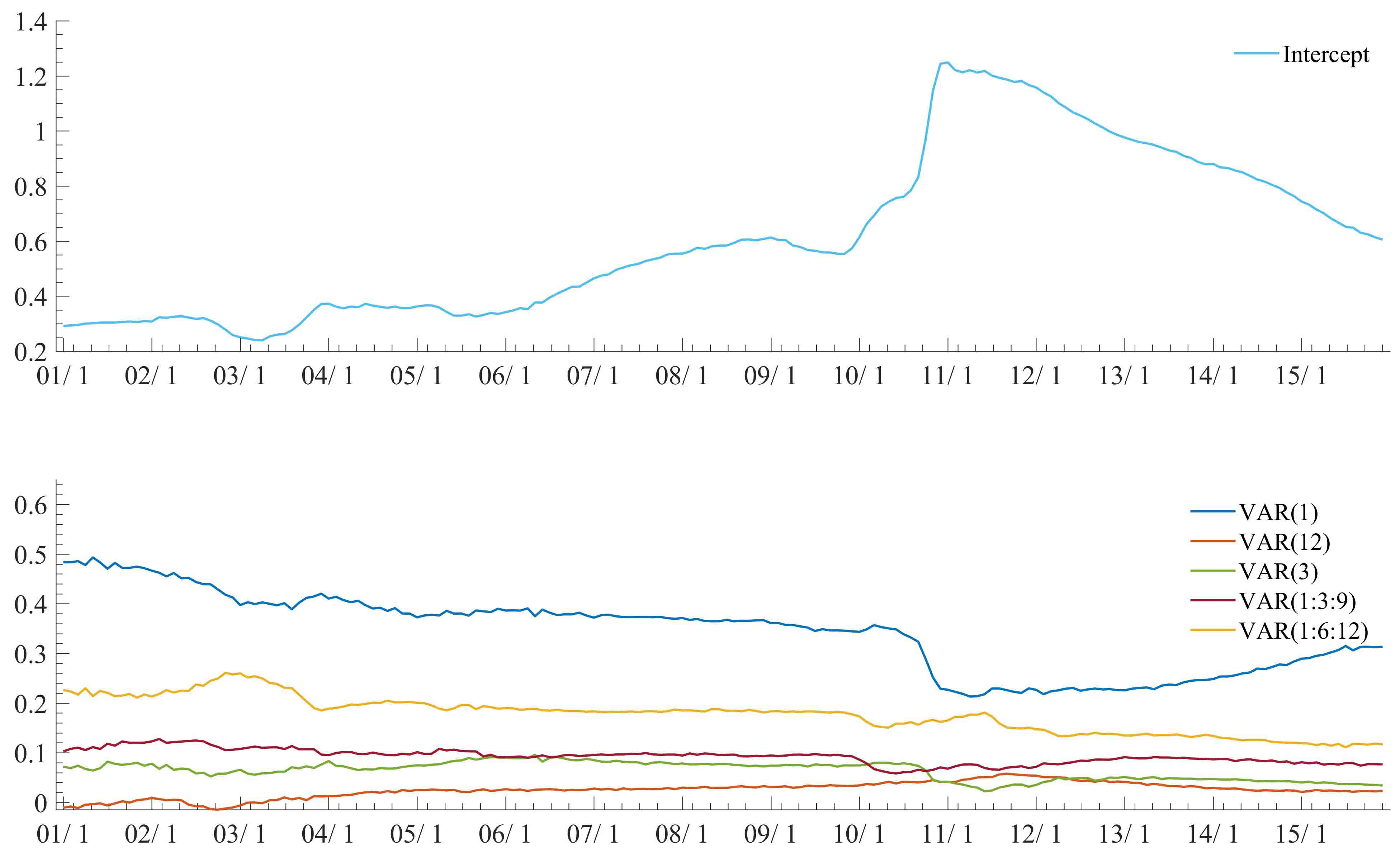}
\caption{US macroeconomic forecasting 2001/1-2015/12:  On-line posterior means of BPS(12)  model
 coefficients for interest rate $(r)$ sequentially computed at each of the $t=\seq1{180}$ months.
\label{12coeff6}}
\end{figure}
\clearpage

\begin{figure}[htbp!]
\centering
\includegraphics[width=0.75\textwidth]{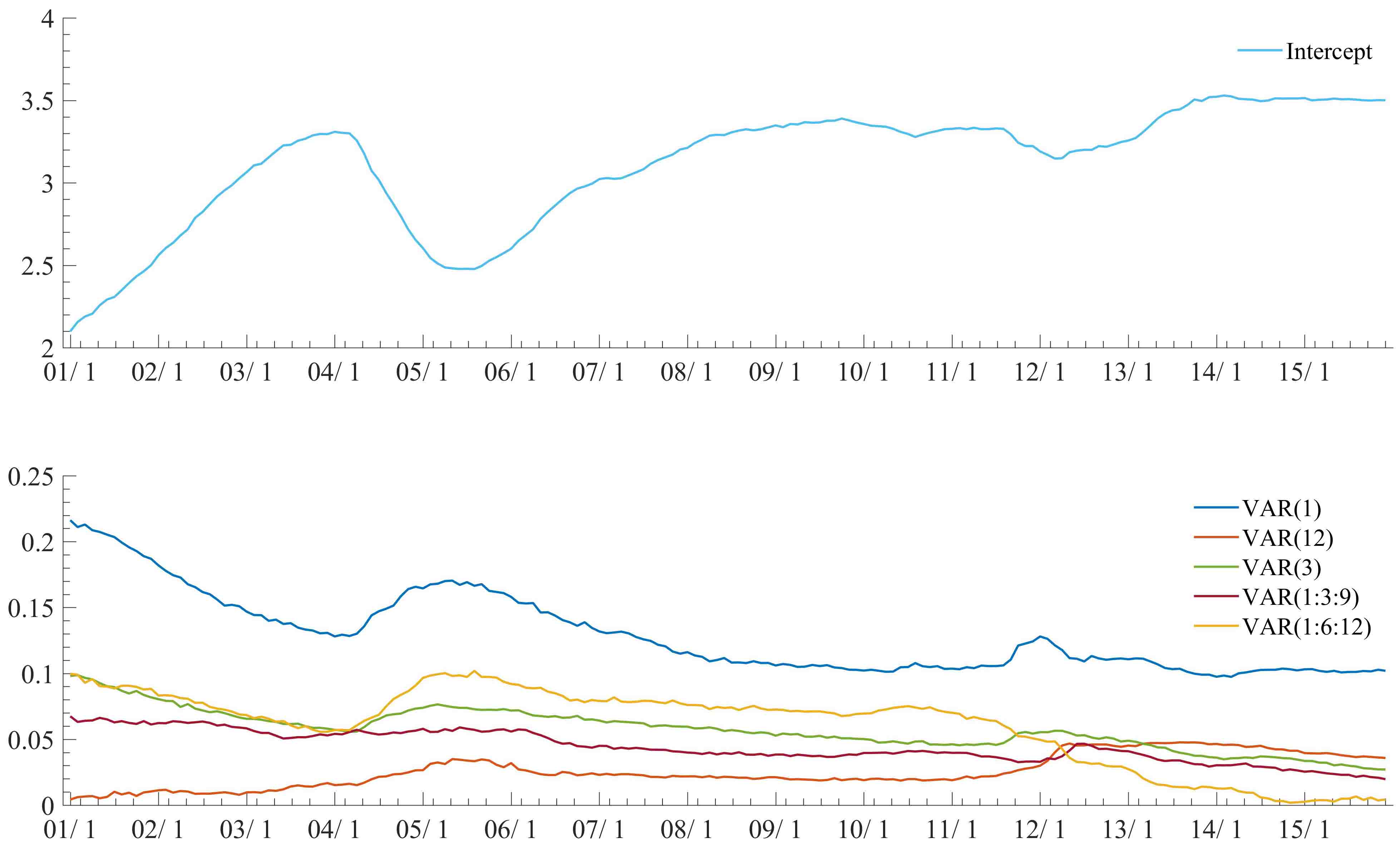}
\caption{US macroeconomic forecasting 2001/1-2015/12:  On-line posterior means of BPS(24) model
 coefficients for inflation $(p)$ sequentially computed at each of the $t=\seq1{180}$ months.
\label{24coeff1}}
\end{figure}

\begin{figure}[htbp!]
\centering
\includegraphics[width=0.75\textwidth]{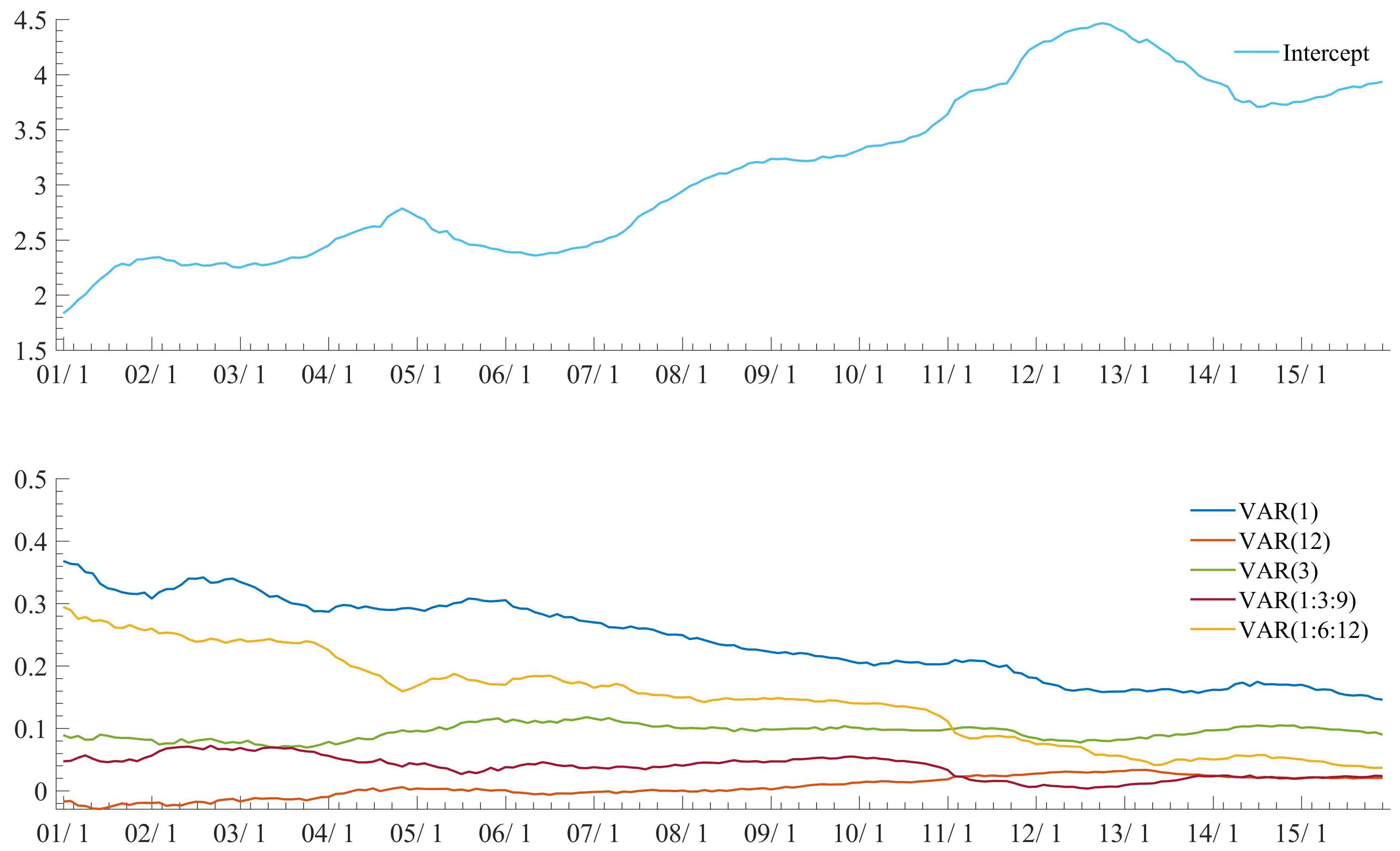}
\caption{US macroeconomic forecasting 2001/1-2015/12:  On-line posterior means of BPS(24)  model
 coefficients for wage $(w)$sequentially computed at each of the $t=\seq1{180}$ months.
\label{24coeff2}}
\end{figure}

\begin{figure}[htbp!]
\centering
\includegraphics[width=0.75\textwidth]{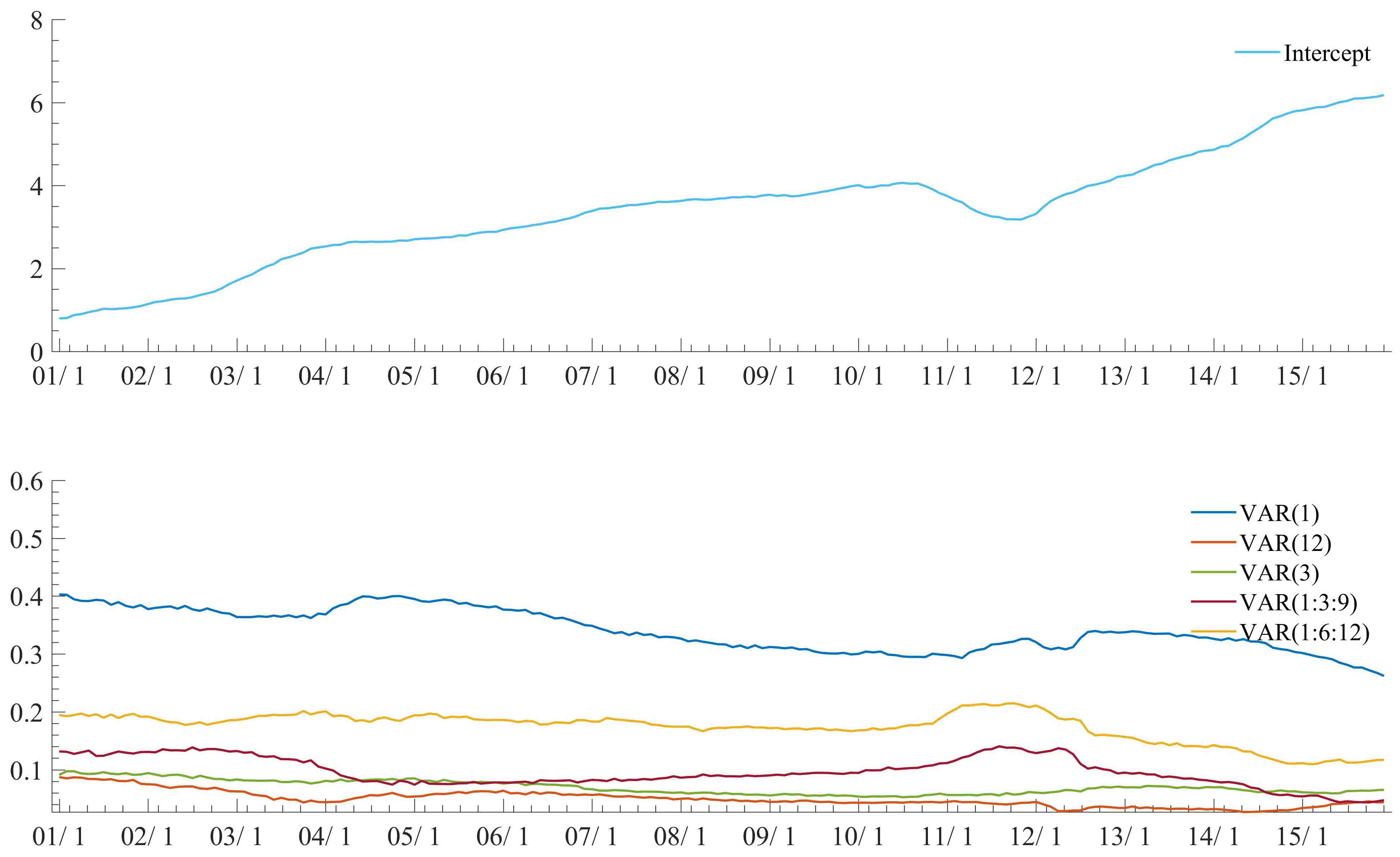}
\caption{US macroeconomic forecasting 2001/1-2015/12:  On-line posterior means of BPS(24)  model
 coefficients for unemployment $(u)$ sequentially computed at each of the $t=\seq1{180}$ months.
\label{24coeff3}}
\end{figure}

\begin{figure}[htbp!]
\centering
\includegraphics[width=0.75\textwidth]{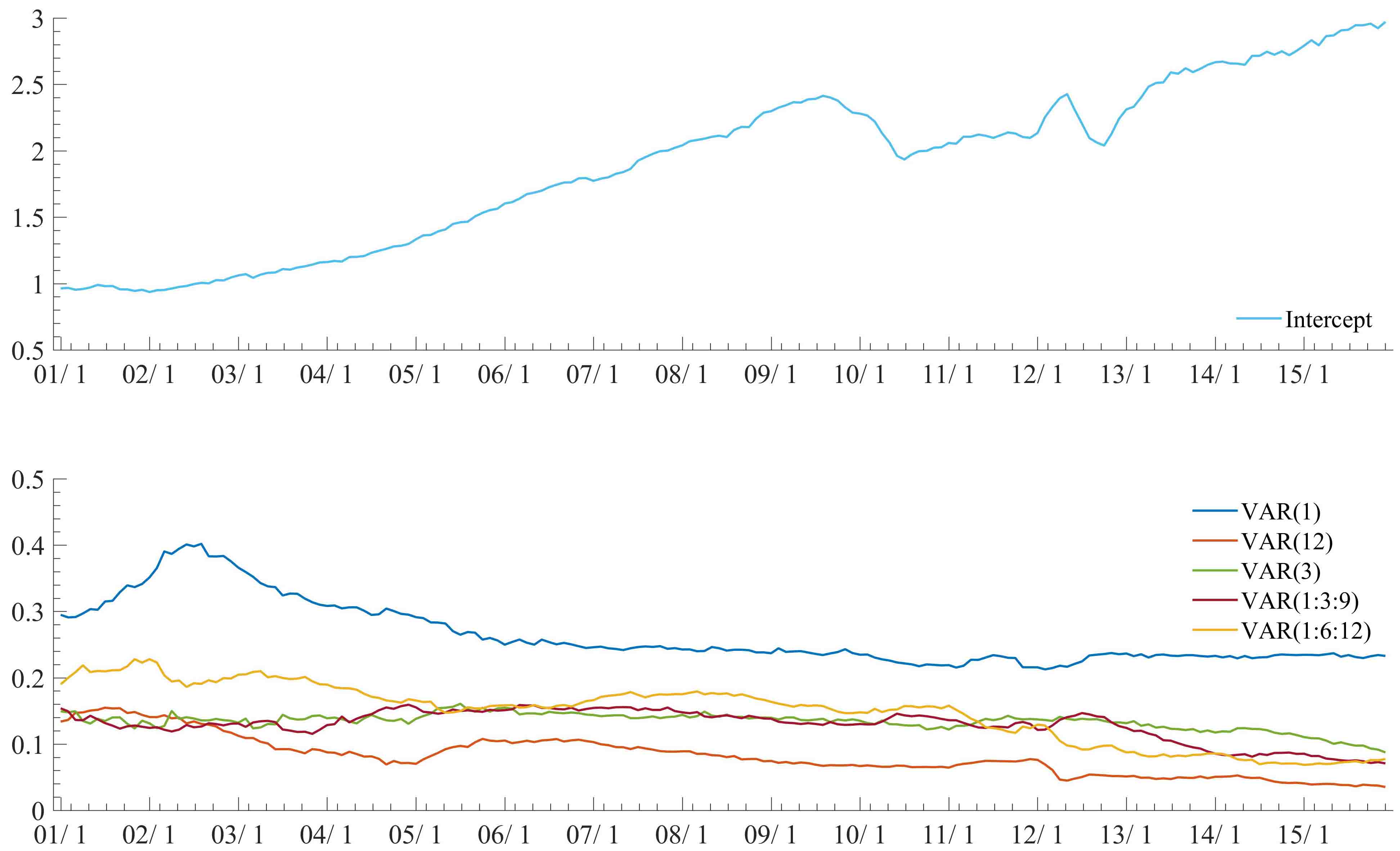}
\caption{US macroeconomic forecasting 2001/1-2015/12:  On-line posterior means of BPS(24)  model
 coefficients for consumption $(c)$ sequentially computed at each of the $t=\seq1{180}$ months.
\label{24coeff4}}
\end{figure}

\begin{figure}[htbp!]
\centering
\includegraphics[width=0.75\textwidth]{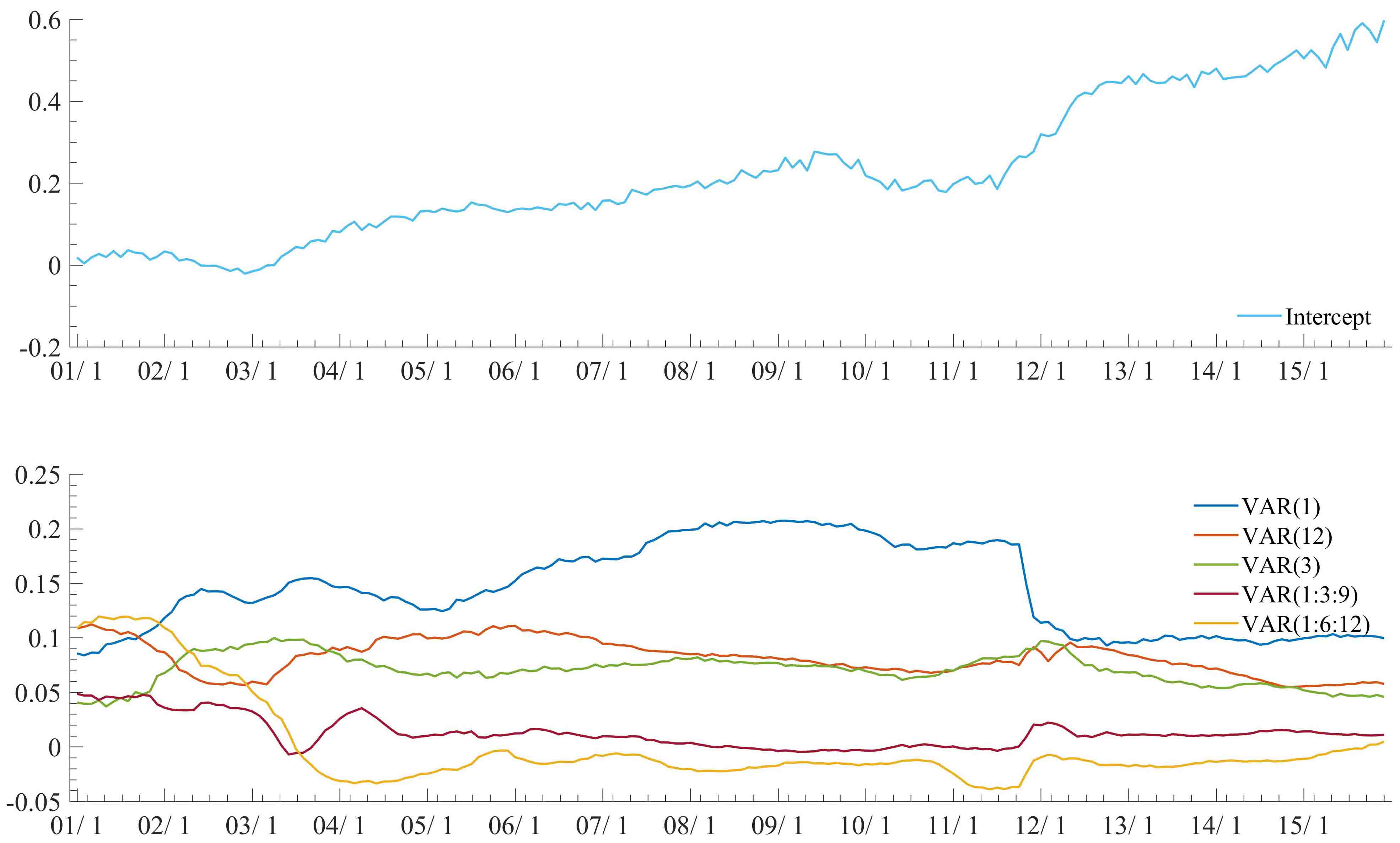}
\caption{US macroeconomic forecasting 2001/1-2015/12:  On-line posterior means of BPS(24)  model
 coefficients for investment $(i)$ sequentially computed at each of the $t=\seq1{180}$ months.
\label{24coeff5}}
\end{figure}

\begin{figure}[htbp!]
\centering
\includegraphics[width=0.75\textwidth]{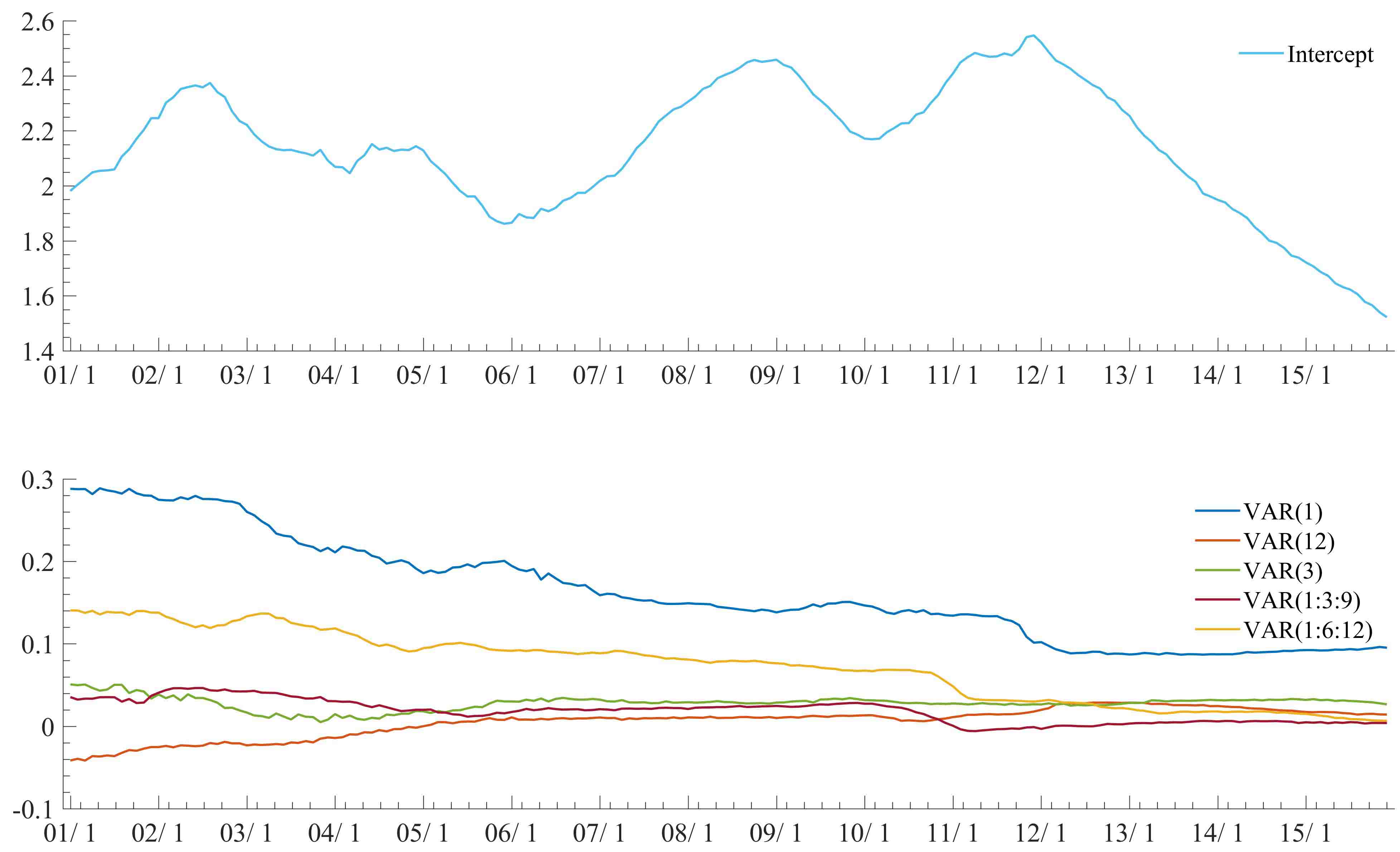}
\caption{US macroeconomic forecasting 2001/1-2015/12:  On-line posterior means of BPS(24)  model
 coefficients for interest rate $(r)$ sequentially computed at each of the $t=\seq1{180}$ months.
\label{24coeff6}}
\end{figure}


\end{document}